\documentclass[11pt,a4paper]{article}
\usepackage{jheppub}
\usepackage{amsmath,amsthm,amssymb}
\usepackage{comment}
\usepackage[utf8]{inputenc}
\usepackage[T1]{fontenc}
\usepackage[rgb,dvipsnames,svgnames,x11names]{xcolor}
 
\usepackage{tikz}
 
\usepackage[mathscr]{eucal}
\usepackage{mathtools}
\usepackage{lie-hasse}
\usepackage{longtable}
\DeclareMathAlphabet{\mathbbold}{U}{bbold}{m}{n}
\usepackage{hyperref}
\hypersetup{
    colorlinks=true,                 
    linkcolor=black,
    linktoc=all,                     
    citecolor=blue,
    filecolor=pink,                  
    urlcolor=blue                 
}
\theoremstyle{plain}

\numberwithin{theorem}{section}

\theoremstyle{definition}

\DeclareMathOperator{\Tr}{Tr}

\begin{document}
\title{Non-Coulomb \(Q\)-cohomology and quantum lifting}

\author[a]{Tanay Kibe,}
\author[b,c,d]{Sam van Leuven}
\author[b]{and Kayleigh Mathieson}

\affiliation[a]{Instituto Balseiro, Centro 
At{\'o}mico Bariloche, S.C. de Bariloche, 8400, 
R{\'i}o Negro, Argentina}
\affiliation[b]{Mandelstam Institute for Theoretical Physics, School of Physics, University of the Witwatersrand, Johannesburg 2050, South Africa
}
\affiliation[c]{DSI-NRF Centre of Excellence in Mathematical and Statistical Sciences (CoE-MaSS),  South Africa}
\affiliation[d]{National Institute for Theoretical and Computational Sciences (NITheCS), Gauteng, South Africa}

\emailAdd{tanay.kibe@ib.edu.ar}
\emailAdd{sam.vanleuven@wits.ac.za}
\emailAdd{kayleigh.mathieson.wits@gmail.com}

\date{\today}

\abstract{
We study quantum lifting of classical $Q$-cohomology classes in four-dimensional $\mathcal N=4$ super-Yang--Mills theory.
Certain powers of the gaugino bilinear $\mathcal S=\Tr(\lambda_+\lambda_-)$ in orthogonal theories provide non-Coulomb monotones: multi-graviton classes that vanish upon Cartan restriction.
For $SO(7)$, we identify $\mathcal S^4$ as a superconformal descendant of a known lifted monotone class, explaining the latter's non-Coulomb character.
We explicitly construct a fortuitous lifting partner for $\mathcal S^4$ that is, by contrast, visible on the Coulomb branch.
This construction provides a strategy for finding analogous partners for higher powers at higher ranks.
Our main result establishes restrictions on quantum lifting in the multi-trace sector of the classical gauge-group families.
We show that the one-loop correction $Q_1$ to the supercharge annihilates every monotone class in classical cohomology, while every $[Q_1]$ image has vanishing Cartan restriction.
Thus every non-zero one-loop pairing has a fortuitous source and a non-Coulomb target, and Coulomb monotones survive at this order at finite rank.
In the BMN sector, our results imply that non-Coulomb monotones also survive at one loop.
A crucial step shows that dressing any classical class by a sufficiently high power of a scalar discriminant yields a multi-graviton class, or zero in cohomology when the original class is non-Coulomb.
We interpret this annihilation as a partial no-hair property.
}

\maketitle

\section{Introduction}\label{sec:intro}

A central problem in four-dimensional $\mathcal N=4$ super-Yang--Mills theory is to identify its $1/16$-BPS spectrum at strong coupling.
Through the AdS/CFT correspondence, this spectrum includes the microstates of supersymmetric black holes in $AdS_5$ \cite{Maldacena:1997re,Witten:1998qj,Gutowski:2004ez,Kunduri:2006ek}.
For $SU(N)$ at large $N$, these black holes have an entropy of order $N^2$, which requires the existence of an exponentially large number of BPS states at strong coupling.

A well-understood part of this spectrum consists of graviton operators, which belong to short gauge theory multiplets and match precisely with the Kaluza--Klein multiplets of supergravity on $S^5$ \cite{Witten:1998qj}.
These alone cannot account for black hole entropy, and the challenge is to identify the additional states required.
A natural tool is the superconformal index, a protected signed count of BPS states \cite{Kinney:2005ej,Romelsberger:2005eg}.
However, its early large $N$ evaluation matches the supergraviton index, so that it does not appear to detect any additional contributions corresponding to black hole microstates \cite{Kinney:2005ej}.

An alternative approach constructs BPS operators directly through the cohomology of a nilpotent supercharge $Q$ at leading nontrivial order in weak coupling \cite{Grant:2008sk}.
In this language, BPS operators are described by $Q$-closed operators modulo $Q$-exact ones.
The subsequent analysis in \cite{Chang:2013fba} constructs explicit representatives of graviton cohomology classes and shows that their products span the cohomology in the infinite $N$ limit, where trace relations trivialize.
Additional classes must therefore arise through finite-rank trace relations, anticipating the mechanism now termed fortuity \cite{Chang:2024zqi}.
Explicit examples nevertheless remained elusive for some time.

To connect this weak-coupling construction to black hole microstates at strong coupling, one must determine which states remain BPS as the coupling increases.
The cohomology of the classical, or tree-level, supercharge, denoted below by $Q_0$, describes the states that remain BPS in the spectrum of the one-loop dilatation operator.
A non-renormalization conjecture states that the spectrum receives no further corrections \cite{Kinney:2005ej,Grant:2008sk,Chang:2013fba}, and was extended to all loop orders more recently in \cite{Chang:2022mjp}.
That argument, however, assumes that the loop-corrected supercharge continues to act on composite operators by the graded Leibniz rule.
Quantum contractions generate additional terms that invalidate this assumption \cite{Budzik:2023xbr,Budzik:2025zvu} and can lift classical cohomology classes, as recent results demonstrate \cite{Chang:2025mqp,Choi:2025bhi,Kim:2026rnx} and as we discuss in more detail below.

The index, instead, is protected independently of this conjecture.
Its earlier agreement with the supergraviton index does not exclude black hole contributions, which could be obscured by boson--fermion cancellations.
Following an observation in \cite{Hosseini:2017mds}, the evaluation of the index at complex chemical potentials does turn out to exhibit the required growth at large $N$ and reproduces the entropy of supersymmetric $AdS_5$ black holes \cite{Cabo-Bizet:2018ehj,Choi:2018hmj,Benini:2018ywd}.
Further analyses demonstrate this growth directly in index coefficients at finite $N$ \cite{Murthy:2020scj,Agarwal:2020zwm}.

Subsequent to these advances, the operator approach yielded the first classical $SU(2)$ cohomology classes beyond multi-gravitons in \cite{Chang:2022mjp}.
Explicit representatives, infinite families of such non-gravitons, higher rank examples and graviton dressings were discovered soon after in \cite{Choi:2022caq,Choi:2023znd,Choi:2023vdm,deMelloKoch:2024pcs}.\footnote{See also \cite{deMelloKoch:2026BMNEveryN} for a recent construction of fortuitous BMN classes for every $SU(N)$ with $N\geq3$.}
The role of trace relations is captured by the distinction between \emph{monotone} and \emph{fortuitous} classes, defined relative to a family of CFTs, obtained in the $\mathcal{N}=4$ case by changing the rank of the gauge group \cite{Chang:2024zqi}.
Namely, a monotone class admits a representative whose $Q$-closedness does not rely on trace relations and therefore remains closed throughout the family.
On the other hand, a fortuitous class has no such representative, and its closedness relies on trace relations that necessarily hold only at particular ranks.\footnote{This terminology was recently refined in the context of the conifold theory \cite{Choi:2026faq}, where certain apparently fortuitous operators have been classified as \emph{generalized monotones}, while ``truly'' fortuitous operators are called \emph{strongly fortuitous}. We find some evidence that a similar refinement of the notion of fortuity may be necessary in the $\mathcal{N}=4$ theory as well, see, \textit{e.g}, the end of Section \ref{ssec:Q1-pairing} and Section \ref{sec:discussion}.}
The proposed connection between fortuitous states and black hole microstates has inspired further studies of their dynamics and analogous phenomena in other models.\footnote{Related work constructs the weak-coupling BPS operator associated with the first black hole cohomology, studies the rank dependence of its anomalous dimension and the spectrum of nearby non-BPS states \cite{Chang:2023decoding,Budzik:2023vtr}. Complementary approaches organize finite-rank state spaces through trace relations and gauge invariants \cite{Lee:2023iil,deMelloKoch:2025ngs,deMelloKoch:2025cec,deMelloKoch:2025qeq,deMelloKoch:2025rkw,deMelloKoch:2025eqt}, while \cite{Chen:2024BPSChaos,Johnson:2026plw,Chen:2026vml,Belin:2026ssc} study BPS dynamics and chaos. Fortuity and its relation to state complexity have also been explored beyond supersymmetry in a simple quark model \cite{Fliss:2026rad}.
Other settings include relevant deformations of $\mathcal N=4$ SYM \cite{Choi:2025pqr}, the conifold \cite{Choi:2026faq}, supersymmetric SYK \cite{Chang:2024lxt,Chryssanthacopoulos:2026mni}, a solvable single-matrix quantum mechanics \cite{Chen:2025sum}, D1--D5 \cite{Chang:2025rqy,Hughes:2025car,Chang:2025wgo,Belin:2026vfd,Giusto:2026rpl,Chae:2026vgn}, ABJM \cite{Belin:2025ABJM,Behan:2025ABJM,Behan:2026ABJM}, and the ABJ vector model with a higher-spin dual \cite{Kim:2025vup}.
}

As alluded to above, recent results for the $S$-dual $SO(7)$ and $USp(6)$ theories invalidate the non-renormalization conjecture.
A comparison of truncated indices has exhibited an additional fortuitous class in the $SO(7)$ BMN sector, which is absent in the corresponding $USp(6)$ sector \cite{Gadde:2025yoa}.
A subsequent comparison of the full classical cohomologies shows that the mismatch also includes a monotone class \cite{Chang:2025mqp}.
The latter is non-Coulomb, namely, it vanishes when all letters are restricted to the Cartan subalgebra of the gauge symmetry.
Although the two classes cancel in the protected index, the mismatch is incompatible with coupling independence of the classical $Q_0$-cohomology if $S$-duality holds.
And indeed, shortly thereafter, the additional pair was shown to lift at next-to-leading order through the generalized Konishi anomaly \cite{Choi:2025bhi}.
It follows that even a multi-graviton class, albeit of non-Coulomb type, can be lifted from the quantum $Q$-cohomology.

Building on the holomorphic-twist approach to quantum corrected supercharges \cite{Budzik:2023xbr}, the analysis in \cite{Budzik:2025zvu} determines the complete one-loop correction $Q_1$ in the $\mathcal{N}=4$ theory, including its action on derivative letters.
Nilpotence makes $[Q_1]$ a differential on $Q_0$-cohomology, although it need not obey a Leibniz rule.
Its non-zero images describe pairs of classical classes that are lifted at one-loop.\footnote{We use ``one-loop lifting'' for lifting by the first quantum correction to the supercharge, distinct from the one-loop dilatation spectrum described by classical $Q_0$-cohomology.}
Such lifts also take place among fortuitous classes in the $SU(2)$ BMN sector, even though the threshold class and several graviton dressings of it and its descendants are unlifted \cite{Kim:2026rnx}.
These developments raise the question of which classical classes can participate in quantum lifting.

In this paper, we investigate the role of non-Coulomb classes in these pairings.
Our starting point is a simple family of monotones built from the gaugino bilinear,
\begin{equation}\label{eq:Sn-intro}
    \mathcal S=\Tr(\lambda_+\lambda_-),
    \qquad [\mathcal S^m]\neq0,
    \qquad r<m<h,
\end{equation}
where $r$ and $h\equiv h^\vee$ are the rank and dual Coxeter number of the gauge algebra.
The non-vanishing of these powers in $Q_0$-cohomology follows from classical chiral-ring results \cite{Cachazo:2002ry,Witten:2003ye,Etingof:2003dd,Etingof:2003fy}.
On the Cartan, however, $\mathcal S$ is a sum of only $r$ nilpotent bilinears, so $\mathcal S^m$ vanishes there for $m>r$.
The powers in \eqref{eq:Sn-intro} therefore form a tower of non-Coulomb monotones in $SO(N)$ theories with $N\geq7$, as also observed in \cite{Choi:2025pqr}.
Note that the tower is empty for the special unitary and symplectic families, where $h-1=r$.

We study these classes using field restrictions, following the idea of the Cartan-based non-$Q$-exactness argument in \cite{Witten:2003ye}.
Instead of restricting to the Cartan, we retain a small set of mutually commuting root generators for $\lambda_+$ and the corresponding negative-root generators for $\lambda_-$.\footnote{SvL thanks Chi-Ming Chang for sharing this idea with him.}
This extends the method to the non-Coulomb powers in \eqref{eq:Sn-intro}.
Comparing the restricted powers of $\mathcal S$ with the similarly restricted $Q_0$-exact operators establishes their non-exactness and gives a concrete picture of these classes beyond the Cartan.
We give explicit examples for $\mathfrak{so}_7$, $\mathfrak{so}_8$ and $\mathfrak{so}_9$, and report results through $\mathfrak{so}_{12}$.
Related restrictions also enter constructions of black hole cohomologies \cite{deMelloKoch:2024pcs}.

For $SO(7)$, we show that $\mathcal S^4$ is a superconformal descendant of the non-Coulomb monotone identified in \cite{Chang:2025mqp,Choi:2025bhi}, which we call the Chang--Lin--Choi--Lee (CLCL) monotone.
Its known lifting therefore implies that $\mathcal S^4$ lifts as well.
We independently construct a fortuitous source $F_4$ and verify
\begin{equation}\label{eq:intro-quantum-pairing}
    [Q_1][F_4]=c[\mathcal S^4]\,,
    \qquad c\neq0\,.
\end{equation}
The construction uses an explicit $\mathfrak{so}_7$ trace relation given in Appendix~\ref{app:trace-relations}, following the approach suggested in \cite{Choi:2023vdm,Chang:2024zqi}.
Our method extends to higher powers and ranks, and lifting the first operator in the non-Coulomb tower implies the lifting of the entire tower.
While $\mathcal S^4$ vanishes on the Cartan, its partner $F_4$ has a non-zero Cartan restriction even modulo monotone classes.
Fortuity therefore does not imply non-Coulombness.\footnote{The argument for the opposite conclusion in \cite{Chang:2025mqp} uses the vanishing of the associated trace relation $Q_0F$ on the Cartan.
This does not, however, imply that its preimage $F$ also vanishes there, and $F_4$ provides a concrete counterexample.}

Finally, we establish two general restrictions on quantum pairings in the multi-trace sector of the classical gauge-group families, including operators with derivatives.
First, we show that every monotone class is annihilated by $[Q_1]$, and second, that every $[Q_1]$ image has vanishing Cartan restriction.
Consequently, every non-zero one-loop pairing has a fortuitous source and a non-Coulomb target.
This answers the question raised in \cite{Choi:2025bhi}: two monotones cannot pair, while a fortuitous source may pair with either a non-Coulomb monotone or another fortuitous class with vanishing Cartan restriction.
In addition, together with a simple grading argument, this excludes the lifting of non-Coulomb BMN monotones, including those recently discovered for $SU(3)$ and $SU(4)$ \cite{Kim:2026bhc}; see also the online computational report \cite{Fortuity:2026summary} for additional results.
The same argument shows that fortuitous operators in the BMN sector can only lift by failing to be $[Q_1]$-closed.
Instead, a Coulomb monotone is both $[Q_1]$-closed and non-exact, and therefore survives at this order.
For these classes, we thus establish at finite rank the protection expected from the large-$N$ graviton picture.

A key step to obtain the second result is to show that any $Q_0$-closed multi-trace operator $F$ satisfies, at fixed rank,
\begin{equation}\label{eq:intro-discriminant-dressing}
    \Delta^kF=G+Q_0H,
\end{equation}
for sufficiently large $k$, where $G$ is a multi-graviton expression, $H$ is any multi-trace and $\Delta$ is the discriminant of the characteristic polynomial of a scalar.
We derive this identity by choosing a diagonal gauge for one scalar, working where its eigenvalues are distinct and temporarily allowing division by their differences.
This lets us ``integrate out'' the off-diagonal fields from the $Q_0$ cohomology by removing $Q_0$ pairs.\footnote{A related construction appears in \cite{Choi:2025pqr}, where an adjoint chiral multiplet is eliminated from cohomology after it is made massive by an explicit superpotential deformation.
Here we work in the undeformed theory, with scalar eigenvalue differences playing the role of mass parameters for the off-diagonal pairs.
This is similar to giving the scalar a generic vacuum expectation value, but we keep its eigenvalues as variables and take no low-energy limit.
Division by operators also underlies the notion of generalized monotones introduced in \cite{Choi:2026faq}, and we discuss the possible relation to our construction in Section~\ref{sec:discussion}.}
The remaining variables are closed replacements for the diagonal components, constructed from gravitons with rational scalar coefficients.
Consequently, every class in the enlarged cohomology has a ``rational multi-graviton'' representative.
Clearing the denominators with powers of $\Delta$ then gives the polynomial identity \eqref{eq:intro-discriminant-dressing} in the original theory, valid also when eigenvalues coincide.
When $F$ has vanishing Cartan restriction, the remaining expression vanishes and we may take $G=0$.
Thus sufficiently many dressings by $\Delta$ annihilate any non-Coulomb class in classical cohomology, whether monotone or fortuitous, analogous to the partial no-hair results for specific graviton dressings in \cite{Choi:2023znd}.
For a fortuitous operator with a non-vanishing Cartan restriction, such as $F_4$, the multi-graviton $G$ is instead non-zero.
In this case, the dressing makes the class monotone but does not annihilate it.

The rest of this paper is organized as follows.
Section~\ref{sec:rev-class-coh} reviews classical cohomology and fixes conventions, and Section~\ref{sec:index} presents the refined-index evidence.
Section~\ref{sec:nC-from-S} studies the gaugino powers and their restrictions.
Section~\ref{sec:quantum-lift} constructs the $SO(7)$ pairing and explains the higher-rank strategy, while Section~\ref{sec:gen-patterns} establishes the general restrictions on quantum lifting.
The appendices collect Lie algebra conventions, supporting chiral-ring arguments, explicit trace identities and details of the quantum pairing.

\section{Classical \texorpdfstring{$Q$}{Q}-cohomologies}\label{sec:rev-class-coh}

In this section, we fix the conventions for the classical supercharge and its cohomology.
We follow the conventions of Chang and Yin \cite{Chang:2013fba}, with the classical supercharge $Q\equiv Q^4_-$.
The letters that enter the $1/16$-BPS complex of $Q$ are denoted by
\begin{equation}
    \phi^n\equiv \Phi^{4n}\,,\qquad
    \psi_n\equiv-i\Psi_{n+}\,,\qquad
    \lambda_{\dot\alpha}\equiv\bar\Psi^4_{\dot\alpha}\,,\qquad
    f\equiv-iF_{++}\,,\qquad
    D_{\dot\alpha}\equiv D_{+\dot\alpha}\,.
\end{equation}
Each of these letters saturates the BPS bound $E=2J_L^3+q_1+q_2+q_3$ of $Q$, with the fields and their covariant derivatives taking values in the complexified gauge algebra $\mathfrak g$. 
Their corresponding free-theory quantum numbers are collected in Table~\ref{tab:n=4-letters} and include the scaling dimension $E$, the Cartan generators $J_L^3$ and $J_R^3$ of the Lorentz group $SU(2)_L\times SU(2)_R$, acting on undotted and dotted spinor indices, respectively, and the Cartan generators $q_1,q_2,q_3$ of the $SO(6)_R$ R-symmetry, which rotate the three complex scalars so that $\phi^m$ has $q_n=\delta_{mn}$.

\begin{table}[hbt]
\begin{centering}
\bgroup
\def\arraystretch{1.3}
\begin{tabular}{|c|c|c|c|c|c|}
\hline
Letters & $E$ & $J_L^3$ & $J_R^3$ & $q_n$ & $i(y_1,y_2,y_3;p,q)$ \tabularnewline
\hline
$f$ & $2$ & $1$ & $0$ & $0$ & $y_1 y_2 y_3$ \tabularnewline
\hline
$\lambda_{\pm}$ & $\frac{3}{2}$ & $0$ & $\pm\frac{1}{2}$ & $\frac{1}{2}$ & $-p, \, - q$ \tabularnewline
\hline
$\psi_m$ & $\frac{3}{2}$ & $\frac{1}{2}$ & $0$ & $\frac{1}{2}-\delta_{mn}$ & $-y_2y_3,\,-y_3y_1,\,-y_1y_2$ \tabularnewline
\hline
$\phi^m$ & $1$ & $0$ & $0$ & $\delta_{mn}$ & $y_1,\,y_2,\,y_3$ \tabularnewline
\hline
$D_+$ & $1$ & $\frac{1}{2}$ & $\frac{1}{2}$ & $0$ & $p$ \tabularnewline
\hline
$D_-$ & $1$ & $\frac{1}{2}$ & $-\frac{1}{2}$ & $0$ & $q$ \tabularnewline
\hline
$D_{\dot\alpha}\lambda^{\dot\alpha}=0$ & $\frac{5}{2}$ & $\frac{1}{2}$ & $0$ & $\frac{1}{2}$ & $pq$ \tabularnewline
\hline
\end{tabular}
\egroup
\par\end{centering}
\caption{The $1/16$-BPS letters of the free $\mathcal N=4$ theory in the Chang--Yin charge convention, which each transform in the adjoint of $\mathfrak g$, and $m,n=1,2,3$. The last column records the contribution $(-1)^Fp^{n_+}q^{n_-}y_1^{n_1}y_2^{n_2}y_3^{n_3}$ of each letter to the single-letter index \eqref{eq:gidx-letter}, with $(n_+,n_-,n_1,n_2,n_3)$ a grading specified in \eqref{eq:mu-grading}. The last row contains the free gaugino equation of motion, which contributes as a constraint in the index.}
\label{tab:n=4-letters}
\end{table}

At tree level, the letters obey
\begin{equation}\label{eq:derivativesantisymm}
    [D_{\dot\alpha},D_{\dot\beta}]
    =\epsilon_{\dot\alpha\dot\beta}f\,,\qquad
    D_{\dot\alpha}\lambda^{\dot\alpha}=[\phi^n,\psi_n]\,,
\end{equation}
where, as in \cite{Chang:2013fba}, the Yang--Mills coupling is absorbed into the fields, so that the classical supercharge below is independent of $g_{\rm YM}$ and its loop corrections are organized in powers of $g_{\rm YM}^2$.
The classical supercharge acts on the undifferentiated letters as
\begin{equation}\label{Eq:SCalgebra}
    [Q,\phi^m]=0\,,\qquad
    \{Q,\lambda_{\dot\alpha}\}=0\,,\qquad
    \{Q,\psi_m\}=-i\epsilon_{mnp}[\phi^n,\phi^p]\,,\qquad
    [Q,f]=i[\phi^n,\psi_n]\,,
\end{equation}
and on a covariant derivative as
\begin{equation}
    [Q,D_{\dot\alpha}\xi]
    =-i[\lambda_{\dot\alpha},\xi]+D_{\dot\alpha}\bigl([Q,\xi]\bigr)\,,
\end{equation}
with graded brackets when $\xi$ is fermionic. 
By \eqref{eq:derivativesantisymm}, antisymmetric combinations of derivative letters can be expressed in terms of existing letters, so when constructing independent letters, it suffices to only retain symmetrized derivatives, with the gaugino index included in the symmetrization whenever the derivatives act on $\lambda_{\dot{\alpha}}$. 
We can introduce the auxiliary commuting variables $z^{\dot\alpha}$, which allow us to collect the fields and their symmetrized covariant derivatives into the generating fields \cite{Grant:2008sk,Chang:2013fba} 
\begin{equation}
    \begin{aligned}
        \phi^m(z)&=\sum_{n=0}^{\infty}\frac{1}{n!}
        (z^{\dot\alpha}D_{\dot\alpha})^n\phi^m\,,
        &\psi_m(z)&=\sum_{n=0}^{\infty}\frac{1}{n!}
        (z^{\dot\alpha}D_{\dot\alpha})^n\psi_m\,,\\
        \lambda(z)&=\sum_{n=0}^{\infty}\frac{1}{(n+1)!}
        (z^{\dot\alpha}D_{\dot\alpha})^n
        (z^{\dot\beta}\lambda_{\dot\beta})\,,
        &f(z)&=\sum_{n=0}^{\infty}\frac{1}{n!}
        (z^{\dot\alpha}D_{\dot\alpha})^nf\,.
    \end{aligned}
\end{equation}
These generating fields are unconstrained, except for the condition $\lambda(0)=0$, and their transformations under $Q$ are given by
\begin{equation}\label{eq:superalgebra-gener-fields}
    \begin{aligned}
    &[Q,\phi^m(z)]=-i[\lambda(z),\phi^m(z)]\,,\\
    &[Q,f(z)]=-i[\lambda(z),f(z)]+i[\phi^n(z),\psi_n(z)]\,,\\
    &\{Q,\psi_m(z)\}=-i\{\lambda(z),\psi_m(z)\}
    -i\epsilon_{mnp}[\phi^n(z),\phi^p(z)]\,,\\
    &\{Q,\lambda(z)\}=-i\lambda(z)^2\,.
    \end{aligned}
\end{equation}
The action of $Q$ on each of the derivative letters can be determined by expanding each of these expressions in powers of $z$. In the above, the generating field $\lambda(z)$ should not be confused with either physical gaugino component, which we denote as
\begin{equation}\label{eq:lambda-pm-definition}
    \lambda_+\equiv\lambda_{\dot +}\,,\qquad
    \lambda_-\equiv\lambda_{\dot -}
\end{equation}
throughout the rest of the paper.

The generating fields can be further organized into a single superfield by introducing three auxiliary Grassmann variables $\theta_{m}$ \cite{Chang:2013fba}
\begin{equation}\label{eq:chang-yin-superfield}
    \Psi(z,\theta)=-i\left[
    \lambda(z)+2\theta_n\phi^n(z)
    +\epsilon^{mnp}\theta_m\theta_n\psi_p(z)
    +4\theta_1\theta_2\theta_3f(z)\right]\,.
\end{equation}
By collectively writing the auxiliary variables as $\mathcal Z=(z,\theta)$, the $Q$-transformations and constraint can be expressed in the compact form
\begin{equation}
    \{Q,\Psi(\mathcal Z)\}=\Psi(\mathcal Z)^2\,,\qquad
    \Psi(0)=0\,.
\end{equation}
For the classical gauge algebras considered below, the gauge-invariant operators can be constructed from traces of products of $z$- and $\theta$-derivatives of $\Psi$, evaluated at the origin $\mathcal{Z} = 0$. However, for the gauge algebra $\mathfrak{g}=\mathfrak{so}_{2r}$, these trace invariants must also be supplemented by Pfaffian-type contractions. 

\subsubsection*{Graviton representatives and monotone classes}

With $\partial_\pm\equiv\partial/\partial z^{\dot\pm}$ and $\partial_i\equiv\partial/\partial\theta_i$, the single-trace graviton representatives of \cite{Chang:2013fba} can be written as
\begin{equation}\label{eq:single-trace-graviton-representatives}
    \mathcal G=\left.
    \partial_+^{a_+}\partial_-^{a_-}\prod_{i=1}^3\partial_i^{b_i}\,
    \Tr\operatorname{Sym}_{\mathrm{gr}}\!\left[
    (\partial_+\Psi)^{k_+}(\partial_-\Psi)^{k_-}
    \prod_{i=1}^3(\partial_i\Psi)^{l_i}
    \right]\right|_{\mathcal Z=0}\,,
\end{equation}
where $a_\pm,l_i\geq0$ and $b_i,k_\pm\in\{0,1\}$ and $\operatorname{Sym}_{\mathrm{gr}}$ symmetrizes the factors inside the trace with a minus sign for each exchange of two fermionic factors.
These expressions are $Q$-closed prior to imposing (rank-dependent) trace relations.

A $Q$-cohomology class is represented by a $Q$-closed operator modulo $Q$-exact terms, and hence is non-trivial when its representative is not $Q$-exact.
The result we need is that, before imposing trace relations at finite rank, every single-trace $Q$-cohomology class is a linear combination of these graviton classes, and every multi-trace class is a linear combination of their products \cite{Chang:2013fba}.\footnote{These statements are established in mathematical terms in Proposition~5 and equation~(12) of \cite{Chang:2026bil} respectively for $\mathfrak{g}=\mathfrak{gl}_{\infty}$. In particular, it identifies the full infinite-rank cohomology with the multi-particle Fock space built from the single-trace classes, with the appropriate bosonic and fermionic statistics. See also remark 3 for the analogous statements in the orthogonal and symplectic cases.}
When the infinite-rank limit is taken at fixed charges, no trace relations enter, and the graviton description of the classical multi-trace cohomology is complete.

As a brief aside, for orthogonal and symplectic algebras, transpose identities for the Lie algebra generators (see, \textit{e.g.}, \eqref{eq:so(n)-bil-form} and \eqref{eq:sp(n)-bil-form}) identify a trace with its reversed word, up to signs.
On the (anti-)symmetrized traces in \eqref{eq:single-trace-graviton-representatives}, reversing a word of $L$ letters gives the factor $(-1)^L$, so only even $L$ single-traces survive.
The even-length graviton generators and their products therefore give the corresponding description in the orthogonal and symplectic multi-trace sectors \cite{Gadde:2025yoa}.
This statement does not include the additional Pfaffian-type invariants of $\mathfrak{so}_{2r}$.

Following \cite{Chang:2024zqi}, a non-zero finite-rank class in the multi-trace sector is called monotone if it admits a representative that is $Q$-closed before rank-dependent trace relations are imposed.
Monotone classes are consequently the non-zero finite-rank images of the multi-graviton classes just described, since trace relations can make some of these classes dependent or $Q$-exact.
The classes that have no such representative are called fortuitous and their closedness relies on trace relations.
They are proposed in \cite{Chang:2024zqi} as the operators dual to black-hole microstates.

It is convenient to grade gauge-invariant operators by the total numbers of auxiliary $z$- and $\theta$-derivatives in their superfield expressions \eqref{eq:chang-yin-superfield},
\begin{equation}\label{eq:mu-grading}
    \mu=(n_+,n_-,n_1,n_2,n_3) \, ,
\end{equation}
where $n_\pm$ count the $z^{\dot\pm}$ and $n_m$ the $\theta_m$ derivatives of $\Psi$.
Under this grading, each gaugino $\lambda_\pm$ carries a single $z^{\dot\pm}$ derivative, the scalar $\phi^m$ the single $\theta_m$ derivative, the fermion $\psi_m$ the complementary pair $\theta_n\theta_p$ of derivatives, and $f$ all three $\theta_{1,2,3}$ derivatives, while a covariant derivative $D_{\dot\alpha}$ adds one $z^{\dot\alpha}$ derivative to the letter it acts on.
For words built from the undifferentiated gauginos $\lambda_\pm$ alone, $n_\pm$ count their respective multiplicities, which we will refer to as their bidegree in Section~\ref{sec:nC-from-S}.

Two features of this grading will be used in the following.
First, $Q$ preserves $\mu$ while raising the number of letters by one, so that each $\mu$-sector is itself a $Q$-complex, graded by letter number.
Second, every letter carries at least one derivative of $\Psi$, which implies that the letter number is bounded by the total grading,
\begin{equation}\label{eq:letter-bound}
    L\leq\ |\mu|\ \equiv\ n_++n_-+n_1+n_2+n_3 \, ,
\end{equation}
with equality precisely when every letter carries a single derivative, which happens when a word is built using just the $\lambda_\pm$ and $\phi^m$ letters.
It follows that a fixed $\mu$-sector is a finite complex whose top piece, at $L=|\mu|$, is annihilated by $Q$.

\subsubsection*{Letter-based truncations}

Let $\mathcal C_{\mathcal N=4}$ denote the full Chang--Yin complex, with $\mathcal C_\lambda\subset\mathcal C_{\mathcal N=4}$ the subcomplex of words built from $\lambda(z)$ alone, \textit{i.e.}, from the gauginos and their covariant derivatives.\footnote{
Classical (and loop-corrected) cohomologies involving these letters have also been analyzed recently in the context of $\mathcal N=1$ gauge theories using the holomorphic twist formalism \cite{Budzik:2023xbr}.}
Since $\phi^m$, $\psi_m$ and $f$ each carry at least one $\theta$-derivative, $\mathcal C_\lambda$ is the sum of the $\mu$-sectors with $n_1=n_2=n_3=0$.
As $Q$ preserves $\mu$, $\mathcal C_\lambda$ is a direct summand of $\mathcal C_{\mathcal N=4}$ as a complex.
In particular, a non-trivial pure-gaugino class in the $Q$-cohomology of $\mathcal C_\lambda$ cannot become exact by allowing other $\mathcal N=4$ letters, since any $Q$-preimage would have to lie in the same $\mu$-sector.

The truncation to $\mathcal C_\lambda$ is an example of the letter-based truncations considered in \cite{Gadde:2025yoa}.
The best-known one is the BMN truncation \cite{Choi:2023znd,Choi:2023vdm}, that discards $\lambda_{\dot\alpha}$ and $D_{\dot\alpha}$ and keeps $\phi^m$, $\psi_m$ and $f$, and which by the same argument consists of the $\mu$-sectors with $n_+=n_-=0$.
In terms of the superfield, the BMN and pure-gaugino truncations restrict $\Psi(\mathcal Z)$ to $z=0$ and to $\theta=0$, respectively, and are complementary in this sense.
Their indices are the specializations $p=q=0$ and $y_1=y_2=y_3=0$ of the classical refined index defined in Section~\ref{sec:index}.
In \cite{Gadde:2025yoa}, a mismatch between the BMN indices of the S-dual $SO(7)$ and $USp(6)$ theories was used to detect non-graviton cohomologies, and in Section~\ref{sec:index} we compare the pure-gaugino indices of $SO(2r+1)$ and $USp(2r)$ in the same spirit.

\section{Index evidence for non-Coulomb classes}\label{sec:index}

We compare the classical refined indices of the $S$-dual $SO(2r+1)$ ($B_r$) and $USp(2r)$ ($C_r$) theories in the pure-gaugino truncation $\mathcal C_\lambda$, which keeps the gauginos and their covariant derivatives. 
The two theories have the same rank and Weyl group, and their common graviton representatives give identical monotone sectors on the Coulomb branch \cite{Gadde:2025yoa,Chang:2025mqp}.
Discrepancies between their classical spectra can nevertheless arise from non-Coulomb classes, whether monotone or fortuitous, and from fortuitous classes on the Coulomb branch.
In the full theory, these possibilities are illustrated by the non-Coulomb CLCL monotone \cite{Chang:2025mqp,Choi:2025bhi} and the Coulomb-visible fortuitous class $F_4$ of Section~\ref{ssec:Q1-pairing} (and its related BMN class \cite{Gadde:2025yoa}).
Here, we use index discrepancies to probe such differences between the classical spectra.

Using the grading $\mu$ of Section~\ref{sec:rev-class-coh}, we can define the classical refined index \cite{Chang:2013fba}
\begin{equation}
 \mathcal I^{\mathrm{cl}}_G(y_1,y_2,y_3;p,q)
 =\operatorname{Tr}_{H_Q(\mathcal C_{\mathcal N=4})}
 (-1)^F p^{n_+}q^{n_-}y_1^{n_1}y_2^{n_2}y_3^{n_3}.
 \label{eq:gidx-refined}
\end{equation}
Here, $Q$ denotes the classical interacting supercharge.\footnote{Its cohomology describes the states that remain BPS under the one-loop dilatation operator.
Accordingly, this cohomology is also called one-loop cohomology \cite{Gadde:2025yoa}, and the corresponding refined index is called the one-loop index in \cite{Gaikwad:2025ugk}.
Our terminology refers instead to the order of the supercharge used to define the cohomology.}
The single-letter index, with the adjoint gauge character suppressed, is
\begin{equation}
 i(y_1,y_2,y_3;p,q)=1-\frac{(1-y_1)(1-y_2)(1-y_3)}{(1-p)(1-q)}.
 \label{eq:gidx-letter}
\end{equation}
Since $Q$ preserves the $\mu$-grading, the classical refined index can be computed on the free BPS spectrum, without determining the classical cohomology explicitly.
Further quantum corrections to the supercharge need not preserve the full $\mu$-grading, so this five-fugacity index is not protected at the quantum level.
The fully protected superconformal index can be recovered by imposing $y_1y_2y_3=pq$.

By setting $y_1=y_2=y_3=0$, we can select the pure-gaugino truncation $\mathcal C_\lambda$ introduced in Section~\ref{sec:rev-class-coh}.
Its single letter index is
\begin{equation}
 i_\lambda(p,q)=\frac{-p-q+pq}{(1-p)(1-q)}\,,
 \label{eq:gidx-gaugino-letter}
\end{equation}
where the free gaugino equation of motion accounts for the positive $pq$ term.
We write $\mathcal I_G(p,q)=\mathcal I^{\mathrm{cl}}_G(0,0,0;p,q)$ and $\mathcal I_G(t)=\mathcal I_G(t,t)$, so $t$ counts $n_++n_-$. 
As usual, we project onto gauge invariant states by integrating the multi-letter index over the gauge group with the Haar measure \cite{Kinney:2005ej}
\begin{equation}
 \mathcal I_G(p,q)=\int_G d\mu(g)\,
 \exp\!\left[\sum_{\ell\geq1}\frac{i_\lambda(p^\ell,q^\ell)}{\ell}
 \chi_{\mathrm{adj}}(g^\ell)\right]\,,
 \label{eq:gidx-haar}
\end{equation}
where the adjoint character is given by
\begin{equation}
  \chi_{\mathrm{adj}}(z)=r+\sum_{\alpha\in\Phi}z^\alpha\,,
  \qquad z^\alpha=\prod_{k=1}^r z_k^{c_k}\,,
  \qquad \alpha=\sum_{k=1}^r c_ke_k\,.
  \label{eq:gidx-character}
\end{equation}
Here, $\Phi$ is the root system of $\mathfrak{g}$ and $c_k$ are the coordinates of the root systems listed in \eqref{eq:classical-roots}.
The constant term $r$ counts the Cartan generators. 
In order to express the integrand in product form, we introduce the notation
\begin{equation}
 (u;p,q)_\infty=\prod_{m,n\geq0}(1-up^m q^n)\,,
 \qquad (u;p)_\infty=\prod_{m\geq0}(1-up^m)\,.
 \label{eq:gidx-products}
\end{equation}
with $|p|,|q| < 1$, which enables us to express the integral as
\begin{equation}
 \mathcal I_G(p,q)
 =\frac{\bigl[(p;p)_\infty(q;q)_\infty(pq;p,q)_\infty\bigr]^r}{|W|}
 \oint_{\mathbb T^r}\prod_{k=1}^r\frac{dz_k}{2\pi i z_k}
 \prod_{\alpha\in\Phi}(z^\alpha;p,q)_\infty\,,
 \label{eq:gidx-contour}
\end{equation}
where each $z_k$ is integrated around the unit circle and $|W|=2^r r!$ is the order of the Weyl group, which is the same for $B_r$ and $C_r$.
Factors with $m=n=0$ in the double products contain the Vandermonde determinants associated with the Haar measure.

We evaluate \eqref{eq:gidx-contour} as a power series after setting $p=q=t$.
Denoting the $SO(2r+1)$ and $USp(2r)$ indices by $\mathcal I_{B_r}$ and $\mathcal I_{C_r}$, we set $\Delta_r=\mathcal I_{B_r}-\mathcal I_{C_r}$.
At rank three, the two expansions first differ at order $t^8$:\footnote{This discrepancy does not conflict with $S$-duality, since the refined index of classical cohomology is not generally protected against further quantum corrections.
The full protected superconformal index requires $y_1y_2y_3=pq$, which is not satisfied by the pure-gaugino specialization at non-zero $p,q$. 
We note that the fully protected specialization $p=0$ gives the one-fermion, one-derivative index studied recently for $U(N)$ in \cite{Eleftheriou:2026yct}, where an exact evaluation follows from a $q$-Dyson identity.
Analytic evaluations of other index integrals through residues were also developed in \cite{vanLeuven:2025gwr}.
Adapting such methods to the two-fugacity orthogonal and symplectic indices could help in deriving the discrepancies below at arbitrary rank.}
\begin{align}
 \mathcal I_{SO(7)}(t)
 &=1+t^2+2t^3+4t^4+6t^5+12t^6+18t^7+33t^8
 \nonumber\\
 &\quad+52t^9+87t^{10}+136t^{11}+225t^{12}+348t^{13}+O(t^{14})\,,
 \label{eq:gidx-so7-series}\\
 \mathcal I_{USp(6)}(t)
 &=1+t^2+2t^3+4t^4+6t^5+12t^6+18t^7+32t^8
 \nonumber\\
 &\quad+50t^9+84t^{10}+132t^{11}+217t^{12}+332t^{13}+O(t^{14})\,,
 \label{eq:gidx-sp6-series}
\end{align}
so that
\begin{equation}\label{eq:gidx-delta3-series}
     \Delta_3(t)=t^8+2t^9+3t^{10}+4t^{11}+8t^{12}+16t^{13}+O(t^{14})\,.
\end{equation}
The first discrepancies are summarized for ranks $1\leq r\leq6$ in Table~\ref{tab:gidx-thresholds}.
\begin{table}[!ht]
 \centering
 \renewcommand{\arraystretch}{1.18}
 \begin{tabular}{c c c }
 \hline
 $r$ & Groups & First discrepancy \\
 \hline
 1 & $SO(3),\ USp(2)$   & none \\
 2 & $SO(5),\ USp(4)$   & none  \\
 3 & $SO(7),\ USp(6)$   & $t^8$    \\
 4 & $SO(9),\ USp(8)$   & $t^{10}$  \\
 5 & $SO(11),\ USp(10)$ & $t^{12}$\\
 6 & $SO(13),\ USp(12)$ & $t^{14}$ \\
 \hline
 \end{tabular}
 \caption{First discrepancies between the gaugino indices $\mathcal I_{B_r}$ and $\mathcal I_{C_r}$.
 For $r=1,2$, equality holds to all orders by the Lie-algebra isomorphisms.}
 \label{tab:gidx-thresholds}
\end{table}

For each of the tested ranks $3\leq r\leq6$ we find
\begin{equation}
 \Delta_r(t)=t^{2r+2}+2t^{2r+3}+O(t^{2r+4}).
 \label{eq:gidx-threshold-pattern}
\end{equation}
The observed discrepancy has a simple cohomological interpretation.
The gaugino bilinear $\mathcal S=\Tr(\lambda_+\lambda_-)$ has weight $pq$, or $t^2$ on the diagonal.
As we will review in detail in Section~\ref{sec:nC-from-S}, its powers satisfy $\mathcal S^h=0$ and $\mathcal S^{h-1}\neq0$ in classical $Q$-cohomology, where $h$ is the dual Coxeter number.
Since $h(B_r)=2r-1$ and $h(C_r)=r+1$, the power $\mathcal S^{r+1}$ represents a non-zero class for $SO(2r+1)$ but is trivial in cohomology for $USp(2r)$ when $r\geq3$.
On the rank-$r$ Cartan, however, $\mathcal S$ is a sum of only $r$ independent gaugino pairs, so $\mathcal S^{r+1}$ vanishes under restriction.
The surviving $SO(2r+1)$ class is therefore non-Coulomb, as recently noted in \cite{Choi:2025pqr} as well.
The weight $t^{2r+2}$ of $\mathcal S^{r+1}$ coincides with the first discrepancy in the computed indices.
The next coefficient is consistent with its two first derivatives $D_{\dot\pm}\mathcal S^{r+1}$, each of weight $t^{2r+3}$.
At order $t^{2r+4}$, the coefficient of $\Delta_r$ is $3$ for $r=3$ and $4$ for $4\leq r\leq6$.
These values are consistent with the three second derivatives of $\mathcal S^{r+1}$ and, for $r\geq4$, the additional class
$\mathcal S^{r+2}$.
The latter is non-trivial in the $SO(2r+1)$ cohomology precisely when $r\geq4$.

These matches do not determine all the contributing classes, since bosonic and fermionic contributions can cancel in the index. 
For the same reason, agreement below the first discrepancy does not establish agreement of the cohomologies. 
In the following section, we establish the relevant non-Coulomb classes directly.

\section{Non-Coulomb classes from the gaugino bilinear}\label{sec:nC-from-S}

In this section, we explain and execute a strategy to determine the classical $Q$-cohomology classes built from the gaugino components $\lambda_+$ and $\lambda_-$ for a general classical gauge algebra.

As explained in Section~\ref{sec:rev-class-coh}, the gauginos and their covariant derivatives form a subcomplex of the full classical $Q$-complex.
Expanding \eqref{eq:superalgebra-gener-fields} to lowest order in $z$, one finds
\begin{equation}
    \{Q,\lambda_{\dot\alpha}\}=0\,,\qquad
    \{Q,D_{\dot\alpha}\lambda_{\dot\beta}\}
    \propto\{\lambda_{\dot\alpha},\lambda_{\dot\beta}\}\,.
\end{equation}
Without derivatives, the $Q$-closed operators are therefore the gauge-invariant functions of the Grassmann-odd, $\mathfrak g$-valued matrices $\lambda_+$ and $\lambda_-$, while the $Q$-exact ones form the invariant part $I^{\mathfrak g}$ of the ideal $I$ generated by the components of
\begin{equation}\label{eq:gaugino-anticommutator-ideal}
    \{\lambda_+,\lambda_+\}\,,\qquad
    \{\lambda_-,\lambda_-\}\,,\qquad
    \{\lambda_+,\lambda_-\}\,.
\end{equation}

The CDSW conjecture \cite{Cachazo:2002ry} states that this classical chiral ring is generated by the gaugino bilinear $\mathcal{S}=\Tr(\lambda_+\lambda_-)$, with the nilpotence relation
\begin{equation}\label{eq:cdsw-conj}
    \mathcal{S}^{h-1}\neq0\mod I^{\mathfrak{g}}\,,\qquad
    \mathcal{S}^{h}=0\mod I^{\mathfrak{g}}\,,
\end{equation}
where $h\equiv h^\vee$ is the dual Coxeter number of $\mathfrak g$.
Witten gave a particularly simple proof of $\mathcal{S}^{h-1}\neq0$ for $\mathfrak{sl}_n$ and $\mathfrak{sp}_{2n}$ and a proof of $\mathcal{S}^{h}=0$ for all classical Lie algebras \cite{Witten:2003ye}, while his proof of $\mathcal{S}^{h-1}\neq0$ for $\mathfrak{so}_n$ is less direct and uses instanton calculations.
Etingof and Kac developed a uniform algebraic approach and proved the $G_2$ case \cite{Etingof:2003dd}, while Etingof subsequently obtained a purely algebraic proof for all classical Lie algebras \cite{Etingof:2003fy}.

Every power $\mathcal S^m$ is $Q$-closed without the use of trace relations, and by \eqref{eq:cdsw-conj} the powers with $m<h$ are non-trivial monotone classes in the terminology of \cite{Chang:2024zqi}.
On the Cartan subalgebra, $\mathcal S$ reduces to a sum of $r$ nilpotent terms, with $r$ the rank of $\mathfrak g$, so that $\mathcal S^m$ vanishes there for $m>r$.
The powers with $r<m<h$ thus form a tower of non-Coulomb monotones which is empty for $\mathfrak{sl}_n$ and $\mathfrak{sp}_{2n}$, where $h-1=r$, but not for $\mathfrak{so}_n$ with $n\geq7$.
This was recently observed in \cite{Choi:2025pqr} as well.

We develop a restriction method to establish the non-triviality of $\mathcal S^{h-1}$, and with it of the whole tower, by placing $\lambda_+$ and $\lambda_-$ on a few basis elements of $\mathfrak g$.
The support is chosen such that $\mathcal S^{h-1}$ survives while the restricted invariant ideal becomes simple enough to test non-exactness.
The Cartan restriction recovers Witten's non-vanishing argument for $\mathfrak{sl}_n$ and $\mathfrak{sp}_{2n}$.
We work out non-Coulomb restrictions explicitly for $\mathfrak{so}_7$, $\mathfrak{so}_8$ and $\mathfrak{so}_9$, and indicate the generalization to higher rank.

\subsection{The gaugino chiral ring}\label{ssec:gaugino-chiral-ring}

Let $\mathfrak g$ be a complex simple Lie algebra of rank $r$ and dimension $d$.
We work in the normalized Cartan--Weyl basis $T_a=(H_i,E_\alpha)$ of Appendix~\ref{app:cw-basis}, with $i=1,\dots,r$ and $\alpha$ running over the roots $\Phi$ of $\mathfrak g$, in which the invariant form is the identity on the Cartan subalgebra and the dual basis for $\mathfrak g^*$ is $H^i=H_i$ and $E^\alpha=E_{-\alpha}$ such that  $\langle T_a,T^b\rangle=\delta_a{}^b$.
We expand
\begin{equation}
    \lambda_+=x^aT_a\,,\qquad
    \lambda_-=y_aT^a\,,
\end{equation}
with Grassmann-odd coefficients $x^a$ and $y_a$, $a=1,\ldots,d$.
Let $R$ be the algebra of polynomials in the $x^a$ and $y_a$, that is, the exterior algebra on $\mathfrak g\oplus\mathfrak g^*$, and let $I\subset R$ be the ideal generated by the components of \eqref{eq:gaugino-anticommutator-ideal}.
The classical chiral ring is then the invariant quotient \cite{Witten:2003ye}
\begin{equation}\label{eq:invariant-ring-quotient}
    (R/I)^{\mathfrak g}\simeq R^{\mathfrak g}/I^{\mathfrak g}\,,\qquad I^{\mathfrak g}=I\cap R^{\mathfrak g}\,.
\end{equation}
We write $R_{(n_+,n_-)}$ and $I_{(n_+,n_-)}$ for the parts of bidegree $(n_+,n_-)$ in the numbers of $x$ and $y$ factors, which is the grading \eqref{eq:mu-grading} of Section~\ref{sec:rev-class-coh} restricted to pure-gaugino operators.
The three generators in \eqref{eq:gaugino-anticommutator-ideal} are copies of the adjoint in bidegrees $(2,0)$, $(0,2)$ and $(1,1)$.

In the following multi-trace description, we specialize to the classical Lie algebras.
For $\mathfrak{sl}_n$ and $\mathfrak{sp}_{2n}$, gauge invariants of $\lambda_\pm$ are spanned by multi-traces, where traces of single letters vanish.
For $\mathfrak{so}_n$, multi-traces span the $O(n)$-invariant sector.
For testing the non-exactness of $\mathcal S^p$, it is enough to work in this multi-trace sector.\footnote{For $\mathfrak{so}_{2r}$, the additional single-epsilon, or Pfaffian-type, invariants are odd under an orientation-reversing $O(n)$ transformation, whereas $\mathcal S^p$ is even.
Since $Q$ commutes with this transformation, any $Q$-exact expression for $\mathcal S^p$ can be replaced by its even part.
Products with an even number of epsilon tensors reduce to multi-traces by $\epsilon_{i_1\cdots i_{2r}}\epsilon_{j_1\cdots j_{2r}}=\det(\delta_{i_a j_b})$ \cite{Witten:2003ye}.
Therefore, in the $O(n)$ even sector, there are no additional exact expressions beyond the multi-trace sector.}
Two kinds of single traces lie in the ideal.
A trace in which two $\lambda_+$ or two $\lambda_-$ are adjacent contains $\lambda_\pm^2=\tfrac12\{\lambda_\pm,\lambda_\pm\}$, and the remaining, alternating traces satisfy
\begin{equation}\label{eq:alternating-traces}
    \mathcal T_m\equiv\Tr\bigl((\lambda_+\lambda_-)^m\bigr)
    =\Tr\bigl(\{\lambda_+,\lambda_-\}(\lambda_+\lambda_-)^{m-1}\bigr)\mod\lambda_+^2\,,
    \qquad m>1\,,
\end{equation}
while $\mathcal T_1=\mathcal S$.
Hence every multi-trace other than a power of $\mathcal S$ lies in $I^{\mathfrak g}$, and these traces span (the multi-trace part of) the invariant ideal.

It follows that the quotient in \eqref{eq:invariant-ring-quotient} is spanned by the powers $\mathcal S^p$ in the multi-trace sector.
Their non-triviality depends on whether trace relations express them as combinations of these other multi-traces.
Therefore, the non-trivial part of the CDSW conjecture lies in the nilpotence order $h$, especially when it exceeds the rank, as it does for $\mathfrak{so}_n$.

\subsection{Support restrictions}\label{ssec:supp-restr}

For subsets $A$ and $B$ of the basis labels, the support restriction $\rho_{A,B}$ keeps only the coefficients $x^a$ with $a\in A$ and $y_b$ with $b\in B$,
\begin{equation}\label{eq:support-restriction-fields}
    \rho_{A,B}:\qquad
    \lambda_+\longmapsto\sum_{a\in A}x^aT_a\,,
    \qquad
    \lambda_-\longmapsto\sum_{b\in B}y_bT^b\,.
\end{equation}
Note that we allow different restrictions of different letters, so that the restriction should not be viewed as a restriction at the level of the Lie algebra.
Since we have the elementary implication that $X\in I^{\mathfrak g}$ implies $\rho_{A,B}(X)\in\rho_{A,B}(I^{\mathfrak g})$, we also have 
\begin{equation}\label{eq:restriction-contrapositive}
    \rho_{A,B}(X)\notin\rho_{A,B}\bigl(I^{\mathfrak g}\bigr)
    \quad\Longrightarrow\quad
    X\notin I^{\mathfrak g}\,.
\end{equation}
A well-chosen, and possibly non-uniform, restriction can thus demonstrate that an operator is non-trivial in $Q$-cohomology, and this will be our key tool in the following.

For powers of $\mathcal S$ it is natural to keep paired variables, $\rho_J\equiv\rho_{J,J}$ for a single set of labels $J$, which we call a support,
\begin{equation}\label{eq:paired-support}
    \lambda_+\big|_J=\sum_{a\in J}x^aT_a\,,\qquad
    \lambda_-\big|_J=\sum_{a\in J}y_aT^a\,.
\end{equation}
We always take $J$ to consist of positive roots, so that $\lambda_-$ takes values in the corresponding negative roots, and the restriction $\rho_J$ then matches the pairing in $\mathcal S$.
Writing $\Tr(XY)=c_{\mathrm{tr}}\langle X,Y\rangle$ for the relation between the trace and the normalized invariant form, we have
\begin{equation}\label{eq:S-paired-support}
    \mathcal{S}\big|_J=c_{\mathrm{tr}}\sum_{a\in J}x^a y_a\,,
    \qquad
    \mathcal S^p\big|_J=c_{\mathrm{tr}}^p p!\sum_{\substack{J'\subset J\\|J'|=p}}\ \prod_{a\in J'}x^ay_a\,.
\end{equation}
In particular, $\mathcal S^{h-1}$ survives only on supports with at least $h-1$ elements.
On a minimal support, with $|J|=h-1$, it is proportional to the top form of $J$,
\begin{equation}\label{eq:minimal-support-monomial}
    \mathcal{S}^{h-1}\big|_J\propto\prod_{a\in J}x^a y_a\,.
\end{equation}
In this case, we note that the restriction test \eqref{eq:restriction-contrapositive} only succeeds when the restricted ideal vanishes in this bidegree.

When no single minimal support achieves this, several can be combined.
For a family of minimal supports
\begin{equation}\label{eq:chosen-subsets}
    \mathcal F=\{J_1,\dots,J_s\}\,,\qquad |J_1|=\dots=|J_s|=h-1\,,
\end{equation}
let $J$ be their union and $\pi_{\mathcal F}$ the projection onto the top forms of $J_1,\dots,J_s$, so that $\pi_{\mathcal F}\rho_J(X)=\rho_{J_1}(X)+\dots+\rho_{J_s}(X)$ in bidegree $(h-1,h-1)$.
In analogy with \eqref{eq:restriction-contrapositive}, we again have
\begin{equation}\label{eq:projected-restriction-test}
    \pi_{\mathcal F}\rho_J(\mathcal{S}^{h-1})
    \notin
    \pi_{\mathcal F}\rho_J\bigl(I^{\mathfrak g}_{(h-1,h-1)}\bigr)\quad \Longrightarrow\quad \mathcal S^{h-1}\notin I^{\mathfrak g}\,.
\end{equation}
We note that not all $h-1$-cardinality subsets of the union $J$ are always necessary to check \eqref{eq:projected-restriction-test}, hence our definition of $\pi_\mathcal{F}$.

The simplest support is the Cartan, or Coulomb, restriction $J_C=\{1,\ldots,r\}$, which places $\lambda_\pm$ in the Cartan subalgebra.
All commutators then vanish, so that the entire ideal restricts to zero, while $\mathcal S^r|_{J_C}\propto x^1\cdots x^ry_1\cdots y_r\neq0$.
For $\mathfrak{sl}_n$ and $\mathfrak{sp}_{2n}$, where $h-1=r$, this recovers Witten's simple proof of the non-vanishing statement $\mathcal S^{h-1}\neq0$ in the CDSW conjecture.

\subsection{Simplifying the \texorpdfstring{$Q$}{Q}-exact ideal}\label{ssec:strong-comm-vanish-ideal}

The restricted ideal simplifies on supports whose generators commute.
We call $J$ a commuting support if
\begin{equation}\label{eq:lower-lower-comm}
    [T_a,T_b]=0\,,\qquad a,b\in J\,,
\end{equation}
which in the Cartan--Weyl basis implies $[T^a,T^b]=0$, but not $[T_a,T^b]=0$, since $T^b=T_{-b}$ for root generators.
On such a support, $\lambda_+^2$ and $\lambda_-^2$ restrict to zero, and with them every multi-trace containing a non-alternating trace.
By Section~\ref{ssec:gaugino-chiral-ring}, the restricted multi-trace part of $I_{(p,p)}^{\mathfrak g}$ is then spanned by the restrictions of the alternating multi-traces
\begin{equation}\label{eq:alternating-multitraces}
    \mathcal T_\nu=\mathcal T_{\nu_1}\cdots\mathcal T_{\nu_s}\,,\qquad
    \nu=(\nu_1,\dots,\nu_s)\,,
\end{equation}
with $\nu$ a partition of $p$ other than $(1,\ldots,1)$, which corresponds to $\mathcal S^p$.
We will call their span the residual ideal at bidegree $(p,p)$.

The residual ideal vanishes altogether if one also imposes the lower--upper condition
\begin{equation}\label{eq:lower-upper-comm}
    [T_a,T^b]=0\,,\qquad a\neq b\in J\,,
\end{equation}
in which case we call $J$ strongly commuting.
For root generators the two conditions state that $\alpha\pm\beta\notin\Phi$ for all distinct $\alpha,\beta\in J$, which is the definition of \emph{strongly orthogonal} roots. 
To see that the residual ideal vanishes, note that on such a support 
\begin{equation}
    \{\lambda_+,\lambda_-\}|_J=\sum_{\alpha\in J\cap\Phi}x^\alpha y_\alpha[E_\alpha,E^\alpha]\,,
\end{equation}
so that $\lambda_+^2|_J=0$ turns \eqref{eq:alternating-traces} into
\begin{equation}
    \mathcal T_m\big|_J=\sum_{\alpha\in J\cap\Phi}x^\alpha y_\alpha\Tr\bigl([E_\alpha,E^\alpha]A_\alpha\bigr)\,, \qquad A_\alpha=(\lambda_+\lambda_-)^{m-1}\big|_{J\setminus\{\alpha\}}\,,\qquad m>1\,.
\end{equation}
Here the restriction to $J\setminus\{\alpha\}$ in $A_\alpha$ is due to Grassmann statistics.
By assumption, each of the generators in $J\setminus\{\alpha\}$ commute with $E_\alpha$ and $E^\alpha$. 
Cyclicity of trace then gives
\begin{equation}
\Tr\bigl([E_\alpha,E^\alpha]A_\alpha\bigr)=0\,,
\end{equation}
and therefore $\mathcal T_m\big|_J=0$.
A strongly commuting support with $h-1$ elements therefore proves that $\mathcal S^{h-1}\neq 0$ is non-trivial in cohomology, similar to the Cartan restriction.

However, the cardinality of a strongly commuting support is limited by the rank of $\mathfrak{g}$.
Indeed, strongly orthogonal roots are mutually orthogonal and hence linearly independent, and Cartan generators commuting with them lie in their orthogonal complement, so that $|J|\leq r$.
This suffices for $\mathfrak{sl}_n$ and $\mathfrak{sp}_{2n}$, where $h-1=r$, but not for $\mathfrak{so}_n$ with $n\geq7$.

\subsection{Commuting supports for \texorpdfstring{$\mathfrak{so}_n$}{so(n)}}\label{ssec:comm-suports-son}

For $\mathfrak{so}_n$ with $n\geq7$, we have $h-1>r$.
In order for the support set $J$ to have cardinality $h-1$, we therefore have to relax the strongly commuting condition.
Instead, we only impose the commuting condition \eqref{eq:lower-lower-comm} on $J$.
This allows much larger supports, since commuting roots can span abelian subalgebras larger than the rank, and the maximal such dimensions are listed in Table~\ref{tab:classical-lie-data}.
The challenge is then to find supports large enough to see $\mathcal{S}^{h-1}$, while the residual ideal remains small or structured enough to show transparently that $\mathcal{S}^{h-1}$ lies outside it.

\begin{table}[t]
\centering
\renewcommand{\arraystretch}{1.8}
\begin{tabular}{c|c|c|c|c}
\(\mathfrak g\)
& \(\operatorname{rank}(\mathfrak g)\)
& \(h\)
& \(\max |\Sigma|\)
& \(\max \dim \mathfrak a\)
\\
\hline
\(A_r=\mathfrak{sl}_{r+1}\)
& \(r\)
& \(r+1\)
& \(\left\lfloor \frac{r+1}{2}\right\rfloor\)
& \(\left\lfloor \frac{(r+1)^2}{4}\right\rfloor\)
\\
\hline
\(B_r=\mathfrak{so}_{2r+1}\)
& \(r\)
& \(2r-1\)
& \(r\)
& \(\max\left\{2r-1, \frac{r(r-1)}{2}+1\right\}\)
\\
\hline
\(C_r=\mathfrak{sp}_{2r}\)
& \(r\)
& \(r+1\)
& \(r\)
& \(\frac{r(r+1)}{2}\)
\\
\hline
\(D_r=\mathfrak{so}_{2r}\)
& \(r\)
& \(2r-2\)
&
\(\begin{cases}
r, & r\text{ even},\\
r-1, & r\text{ odd}
\end{cases}\)
& \(\frac{r(r-1)}{2}\)
\end{tabular}
\caption{Classical simple Lie algebras, their ranks, dual Coxeter numbers \(h\), maximal cardinalities of strongly orthogonal root sets \(\Sigma\subset\Phi\), and maximal dimensions of abelian subalgebras \(\mathfrak a\subset\mathfrak g\), where \(r\geq2\) for \(B_r\) and \(r\geq4\) for \(D_r\). For the final column, see, \textit{e.g.}, \cite{Burde:2012abelian}.}
\label{tab:classical-lie-data}
\end{table}

Before turning to our strategy, let us note a feature of $\mathfrak{so}_n$ and $\mathfrak{sp}_{2n}$ which simplifies the residual ideal.
The generators preserve a non-degenerate bilinear form, a symmetric form $\Xi$ for $\mathfrak{so}_n$ and an antisymmetric form $\Omega$ for $\mathfrak{sp}_{2n}$,
\begin{equation}\label{eq:so(n)-bil-form}
    X^T \Xi+\Xi X=0\,,\qquad X\in \mathfrak{so}_n\,,
\end{equation}
\begin{equation}\label{eq:sp(n)-bil-form}
    X^T\Omega+\Omega X=0\,,\qquad X\in \mathfrak{sp}_{2n}\,.
\end{equation}
Together with (graded) cyclicity, these transpose identities imply that all even alternating traces vanish identically,
\begin{equation}\label{eq:even-traces}
    \mathcal{T}_{2q}=0\,.
\end{equation}
No such identity exists for $\mathfrak{sl}_n$.

We now propose the following strategy for $\mathfrak{so}_n$:\footnote{Note the similarity with Etingof's strategy as reviewed in Appendix~\ref{app:etinghof-proof}.}
\begin{enumerate}
    \item Restrict to a commuting support, on which every multi-trace containing $\lambda_+^2$ or $\lambda_-^2$ vanishes, so that only the alternating multi-traces \eqref{eq:alternating-multitraces} remain.
    \item Discard the even alternating traces by \eqref{eq:even-traces}, which leaves at bidegree $(h-1,h-1)$ the residual generators
    \begin{equation}
      \bigl\{\;\mathcal{T}_{\nu}\;:\;\nu=(\nu_1,\ldots,\nu_s) \, ,\quad \nu_{i}\ \text{odd}, \, \quad \sum_{i=1}^{s} \nu_i=h-1 ,\quad \nu\neq(1,1,\ldots,1)\;\bigr\} \, ,
      \label{eq:residual-generators}
    \end{equation}
    spanning the residual ideal at this bidegree.
    \item Evaluate the residual generators and $\mathcal{S}^{h-1}$ on a family $\mathcal{F}$ of minimal supports, each comprised of $h-1$ roots, and apply the test \eqref{eq:projected-restriction-test}.
\end{enumerate}

In Appendix~\ref{app:comm-supports} we construct, for each $\mathfrak{so}_n$, a family of commuting supports $J_K(L)$, labeled by a subset $K\subseteq\{1,\dots,r\}$ of size $k$, which we call the \emph{level} of the support, and a collection $L$ of further roots.\footnote{In Appendix~\ref{app:lie-algebra-review}, we review and collect the relevant roots systems. }
They have at least $h-1$ elements, as required for $\mathcal S^{h-1}$ to survive.
For the examples below, we only need to know how to write such a support down, which is conveniently represented by a simple graph.

\subsubsection*{Graph description}

\begin{figure}[t]
\centering
\begin{tikzpicture}[
    vtx/.style={circle,fill=black,inner sep=1.6pt},
    mvtx/.style={rectangle,fill=black,inner sep=2.2pt},
    vl/.style={font=\scriptsize},
    pl/.style={font=\footnotesize},
    lw/.style={line width=0.6pt}
  ]
  \begin{scope}[shift={(0,0)}]
    \draw[lw] (0,0.28) circle (0.28);
    \draw[lw] (0,0) -- (1.3,0);
    \node[mvtx] at (0,0) {};   \node[vtx] at (1.3,0) {};
    \node[vl] at (0,-0.40) {$1$};  \node[vl] at (1.3,-0.40) {$2$};
    \node[pl] at (0.65,-0.95) {(a)};
  \end{scope}
  \begin{scope}[shift={(3.2,0)}]
    \draw[lw] (1.3,0.28) circle (0.28);
    \draw[lw] (0,0) -- (1.3,0);
    \node[vtx] at (0,0) {};   \node[mvtx] at (1.3,0) {};
    \node[vl] at (0,-0.40) {$1$};  \node[vl] at (1.3,-0.40) {$2$};
    \node[pl] at (0.65,-0.95) {(b)};
  \end{scope}
  \begin{scope}[shift={(6.4,0)}]
    \draw[lw] (0,0.28) circle (0.28);
    \draw[lw] (0,0) -- (1.3,0);
    \node[vtx] at (0,0) {};   \node[mvtx] at (1.3,0) {};
    \node[vl] at (0,-0.40) {$1$};  \node[vl] at (1.3,-0.40) {$2$};
    \node[pl] at (0.65,-0.95) {(c)};
  \end{scope}
  \begin{scope}[shift={(9.6,0)}]
    \draw[lw] (0,0.20) circle (0.20);
    \draw[lw] (0,0.34) circle (0.34);
    \node[vtx] at (0,0) {};   \node[mvtx] at (1.3,0) {};
    \node[vl] at (0.12,-0.42) {$1$};  \node[vl] at (1.3,-0.40) {$2$};
    \node[pl] at (0.65,-0.95) {(d)};
  \end{scope}
\end{tikzpicture}
\caption{Choices of support drawn as a marked graph on $K$ at level $k=2$ for $\mathfrak{so}_{9}$, so that there are two vertices and $r-k=2$ loops or edges.
A loop at $i$ carrying the index $l$ stands for the pair $e_{i}\pm e_{l}$, an edge $ij$ carrying $l$ for $e_{i}+e_{l}$ and $e_{j}+e_{l}$, and the marked vertex, drawn as a square, for the short root $e_{i}$. The roots $e_{i}+e_{j}$ are present for every pair of vertices and are not drawn.
Graph (a) is the set \eqref{eq:graph-to-roots-example}.
Graphs (a) and (b) differ by the transposition $1\leftrightarrow2$ of the vertices and therefore describe equivalent supports.
On the other hand, graphs (c) and (d) are inequivalent to (a) and to each other.}
\label{fig:multigraphs}
\end{figure}
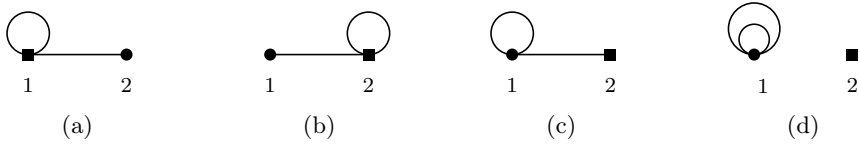

The data specifying $J_K(L)$ can be represented as a graph on the vertex set $K$ with $r-k$ loops and edges in total and, when $n$ is odd, one marked vertex.
The loops and edges are labeled by the $r-k$ indices $l\notin K$, so that each carries its own $l$, which we leave implicit in the graph.
The roots inside $J_K(L)$ are then read off as follows.
Every pair of vertices $i,j$ contributes the root $e_i+e_j$, whether or not the two are joined.
A loop at $i$ carrying the index $l$ contributes the pair $e_i\pm e_l$, an edge joining $i$ and $j$ and carrying $l$ contributes $e_i+e_l$ together with $e_j+e_l$, and a marked vertex $i$ contributes the short root $e_i$.
For instance, the first graph of Figure~\ref{fig:multigraphs}, drawn at level $k=2$ for $\mathfrak{so}_{9}$, gives
\begin{equation}\label{eq:graph-to-roots-example}
    J=\{\,e_1+e_2,\ e_1+e_3,\ e_1-e_3,\ e_1+e_4,\ e_2+e_4,\ e_1\,\} \, ,
\end{equation}
where we labeled the loop by $l=3$ and the edge by $l=4$.
Counting the contributions, the $\binom{k}{2}$ pairs of vertices contribute a single root each, the $r-k$ loops and edges contribute two roots each, and for odd $n$ the marked vertex gives an additional short root. Every such graph therefore yields a support of cardinality
\begin{equation}
    |J|=\binom{k}{2}+2(r-k)+[\,n\text{ odd}\,]=\binom{k}{2}+(n-2k),
\end{equation}
where $n=2r+[\,n\text{ odd}\,]$, and $[\,n\text{ odd}\,]$ equals $1$ for odd $n$ and $0$ for even $n$. The dual Coxeter number for $\mathfrak{so}_n$ in Table~\ref{tab:classical-lie-data} is $n-2$ for both even and odd cases, and hence
\begin{equation}\label{eq:graded-cardinality-main}
    |J|-(h-1)=\frac{(k-2)(k-3)}{2}\geq0 \, .
\end{equation}
Every such support therefore has at least $h-1$ elements, with equality at exactly two levels, $k=2$ and $k=3$.
At the other levels, the members of the family \eqref{eq:chosen-subsets} are drawn from the subsets of size $h-1$.
The extreme level $k=r$ gives the cardinalities $\binom r2+1$ and $\binom r2$ in the $\max|\mathfrak{a}|$ column of Table~\ref{tab:classical-lie-data} for $\mathfrak{so}_{2r+1}$ and $\mathfrak{so}_{2r}$, and $k=1$ gives the remaining entry $2r-1$ for $\mathfrak{so}_{2r+1}$.

Conjugation by an element of $O(n)$ that preserves the Cartan subalgebra permutes the roots by a signed permutation of the $e_i$.
Since the trace invariants are invariant under conjugation by $O(n)$, supports related in this way give equivalent restrictions.
The permutations of the indices in $K$, permutations of the indices $l\notin K$ and sign flips $e_l\to-e_l$ are all of this kind.
The graph records neither the labels $l$ nor these signs, so that two graphs related by a permutation of their vertices describe equivalent supports, as illustrated in Figure~\ref{fig:multigraphs}.
It therefore suffices to consider one representative of each graph isomorphism class. 

\subsection{Examples}\label{ssec:examples}

We now carry out the above strategy for $\mathfrak{so}_n$ with $n=7,8,9$, where there are, respectively,  one, two and three residual generators \eqref{eq:residual-generators}.
For completeness, in the following our convention will be to read off the coefficients of top forms in the ordering where all $x^a$ precede all $y_a$, and we use the Lie algebra conventions specified in Appendix \ref{app:son}.

\subsubsection{\texorpdfstring{$\mathfrak{so}_{7}$}{so(7)}}\label{ssec:example-so7}

Here $h-1=4$, and the residual ideal has the single generator $\mathcal T_3\mathcal S$ in the relevant bidegree.
We take a support at level $k=2$, whose vertex pair contributes $e_1+e_2$ and whose single remaining index $l=3$ labels one loop or edge.
We choose a loop at vertex 2 and mark vertex 1,
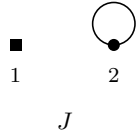
\begin{equation*}
\begin{tikzpicture}[baseline=(bb.base),
    vtx/.style={circle,fill=black,inner sep=1.6pt},
    mvtx/.style={rectangle,fill=black,inner sep=2.2pt},
    vl/.style={font=\scriptsize},
    pl/.style={font=\footnotesize},
    lw/.style={line width=0.6pt}
  ]
  \node (bb) at (0,-0.95) {};
  \draw[lw] (1.3,0.28) circle (0.28);
  \node[mvtx] at (0,0) {};   \node[vtx] at (1.3,0) {};
  \node[vl] at (0,-0.40) {$1$};  \node[vl] at (1.3,-0.40) {$2$};
  \node[pl] at (0.65,-1.00) {$J$};
\end{tikzpicture}
\end{equation*}
which gives the commuting support
\begin{equation}
  J=\{\,e_{1}+e_{2},\;e_{1},\;e_{2}+e_{3},\;e_{2}-e_{3}\,\} \, .
  \label{eq:so7-support}
\end{equation}
Incidentally, this set contains the maximal number, three, of strongly orthogonal roots, namely $e_1$ and $e_2\pm e_3$, while the root $e_1+e_2$ is not strongly orthogonal to the other three.
On this minimal support, $\rho_J(\mathcal S^4)$ is proportional to the top form, while the residual generator vanishes,
\begin{equation}
  \rho_{J}\bigl(\mathcal{T}_{3}\mathcal{S}\bigr)=0 \, ,
  \label{eq:so7-ideal}
\end{equation}
although $\mathcal{T}_{3}$ and $\mathcal{S}$ are separately non-vanishing on $J$.
A single support therefore proves $\mathcal{S}^{4}\notin I^{\mathfrak{so}_7}$, which generalizes Witten's simple argument for $\mathfrak{sl}_{n}$ and $\mathfrak{sp}_{2n}$ to $\mathfrak{so}_{7}$.
The choice of loop instead of edge or the position of the mark is not forced by commutativity, and for example a marking of vertex 2 would give an alternative commuting support.
On it, however, $\mathcal T_3\mathcal S$ does not vanish, and a second support would be needed to prove that $\mathcal{S}^{4}\notin I^{\mathfrak{so}_7}$.
Our choice is therefore the most minimal and hence convenient one.

\subsubsection{\texorpdfstring{$\mathfrak{so}_{8}$}{so(8)}}

Here $h-1=5$, and the residual generators are $\mathcal T_5$ and $\mathcal T_3\mathcal S^2$ at the relevant bidegree. 
Unlike the $\mathfrak{so}_7$ example, the minimal supports considered here do not make both generators vanish. 
We therefore start with two supports at level $k=2$, which will turn out to be sufficient.

At $k=2$, the vertex pair again contributes $e_1+e_2$, while the indices $l=3,4$ label two loops or edges, and there is no marking of vertices since $n$ is even.
We choose the support $J_1$ with a loop at each vertex, and $J_2$ with a loop at vertex 2 and an edge, so that the two graphs are not isomorphic:
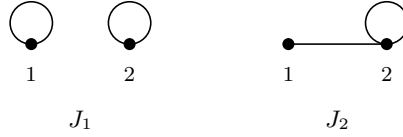
\begin{equation*}
\begin{tikzpicture}[baseline=(bb.base),
    vtx/.style={circle,fill=black,inner sep=1.6pt},
    vl/.style={font=\scriptsize},
    pl/.style={font=\footnotesize},
    lw/.style={line width=0.6pt}
  ]
  \node (bb) at (0,-0.95) {};
  \begin{scope}[shift={(0,0)}]
    \draw[lw] (0,0.28) circle (0.28);
    \draw[lw] (1.3,0.28) circle (0.28);
    \node[vtx] at (0,0) {};   \node[vtx] at (1.3,0) {};
    \node[vl] at (0,-0.40) {$1$};  \node[vl] at (1.3,-0.40) {$2$};
    \node[pl] at (0.65,-1.00) {$J_{1}$};
  \end{scope}
  \begin{scope}[shift={(3.4,0)}]
    \draw[lw] (1.3,0.28) circle (0.28);
    \draw[lw] (0,0) -- (1.3,0);
    \node[vtx] at (0,0) {};   \node[vtx] at (1.3,0) {};
    \node[vl] at (0,-0.40) {$1$};  \node[vl] at (1.3,-0.40) {$2$};
    \node[pl] at (0.65,-1.00) {$J_{2}$};
  \end{scope}
\end{tikzpicture}
\end{equation*}
Explicitly, with the edge of $J_2$ carrying $l=3$,
\begin{equation}
\begin{aligned}
  J_{1}&=\{\,e_{1}+e_{2},\;e_{1}+e_{3},\;e_{1}-e_{3},\;e_{2}+e_{4},\;e_{2}-e_{4}\,\}\, ,\\
  J_{2}&=\{\,e_{1}+e_{2},\;e_{1}+e_{3},\;e_{2}+e_{3},\;e_{2}+e_{4},\;e_{2}-e_{4}\,\}\, ,
\end{aligned}
  \label{eq:so8-supports}
\end{equation}
both of cardinality $5=h-1$.
The support $J_1$ contains the maximal number, four, of strongly orthogonal roots, while $J_2$ contains three.

By direct computation, $\mathcal{T}_{3}\mathcal{S}^{2}$ vanishes on both supports, while $\mathcal{T}_{5}$ does not.
With $\mathcal{F}=\{J_{1},J_{2}\}$, we find
\begin{equation}
  \pi_{\mathcal{F}}\rho_{J}\bigl(\mathcal{S}^{5}\bigr)\propto
  \prod_{a\in J_{1}}x^{a}y_{a}+\prod_{a\in J_{2}}x^{a}y_{a} \, ,
  \qquad
  \pi_{\mathcal{F}}\rho_{J}\bigl(\mathcal{T}_{5}\bigr)\propto
  \prod_{a\in J_{1}}x^{a}y_{a} -4\prod_{a\in J_{2}}x^{a}y_{a}  \, ,
  \label{eq:so8-family}
\end{equation}
where $\pi_\mathcal{F}$ is as in \eqref{eq:projected-restriction-test}.
Since these are linearly independent, it follows that $\mathcal{S}^{5}\notin I^{\mathfrak{so}_8}$.

\subsubsection{\texorpdfstring{$\mathfrak{so}_{9}$}{so(9)}}

Here $h-1=6$, and the residual generators at the relevant bidegree are
\begin{equation}
  \mathcal{T}_{5}\mathcal{S} \, ,\qquad \mathcal{T}_{3}^{2} \, ,\qquad \mathcal{T}_{3}\mathcal{S}^{3} \, .
  \label{eq:so9-generators}
\end{equation}
To separate $\mathcal S^6$ from their span, we take three supports at level $k=2$ and a fourth at level $k=3$.
This turns out to be the minimal number of sets required to demonstrate independence of $\mathcal{S}^6$ from the ideal, as we return to in Section \ref{ssec:higher-rank-supports}.

At level $k=2$, the vertex pair again contributes $e_1+e_2$, the indices $l=3,4$ label two loops or edges, and one vertex is marked since $n$ is odd.
We choose the three graphs
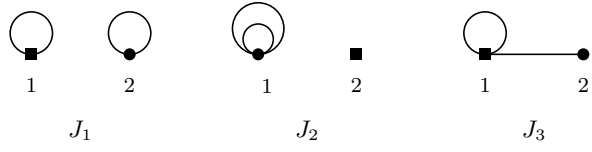
\begin{equation*}
\begin{tikzpicture}[baseline=(bb.base),
    vtx/.style={circle,fill=black,inner sep=1.6pt},
    mvtx/.style={rectangle,fill=black,inner sep=2.2pt},
    vl/.style={font=\scriptsize},
    pl/.style={font=\footnotesize},
    lw/.style={line width=0.6pt}
  ]
  \node (bb) at (0,-0.95) {};
  \begin{scope}[shift={(0,0)}]
    \draw[lw] (0,0.28) circle (0.28);
    \draw[lw] (1.3,0.28) circle (0.28);
    \node[mvtx] at (0,0) {};   \node[vtx] at (1.3,0) {};
    \node[vl] at (0,-0.40) {$1$};  \node[vl] at (1.3,-0.40) {$2$};
    \node[pl] at (0.65,-1.00) {$J_{1}$};
  \end{scope}
  \begin{scope}[shift={(3,0)}]
    \draw[lw] (0,0.20) circle (0.20);
    \draw[lw] (0,0.34) circle (0.34);
    \node[vtx] at (0,0) {};   \node[mvtx] at (1.3,0) {};
    \node[vl] at (0.12,-0.42) {$1$};  \node[vl] at (1.3,-0.40) {$2$};
    \node[pl] at (0.65,-1.00) {$J_{2}$};
  \end{scope}
  \begin{scope}[shift={(6.0,0)}]
    \draw[lw] (0,0.28) circle (0.28);
    \draw[lw] (0,0) -- (1.3,0);
    \node[mvtx] at (0,0) {};   \node[vtx] at (1.3,0) {};
    \node[vl] at (0,-0.40) {$1$};  \node[vl] at (1.3,-0.40) {$2$};
    \node[pl] at (0.65,-1.00) {$J_{3}$};
  \end{scope}
\end{tikzpicture}
\end{equation*}
The fourth support has level $k=3$.
Its three vertices contribute $e_1+e_2$, $e_1+e_3$ and $e_2+e_3$, the remaining index $l=4$ labels an edge between vertices 1 and 2, and vertex 3 is marked:
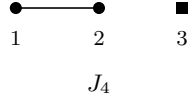
\begin{equation*}
\begin{tikzpicture}[baseline=(bb.base),
    vtx/.style={circle,fill=black,inner sep=1.6pt},
    mvtx/.style={rectangle,fill=black,inner sep=2.2pt},
    vl/.style={font=\scriptsize},
    pl/.style={font=\footnotesize},
    lw/.style={line width=0.6pt}
  ]
  \node (bb) at (0,-0.95) {};
  \begin{scope}[shift={(0,0)}]
    \draw[lw] (0,0) -- (1.1,0);
    \node[vtx] at (0,0) {};   \node[vtx] at (1.1,0) {};   \node[mvtx] at (2.2,0) {};
    \node[vl] at (0,-0.40) {$1$};  \node[vl] at (1.1,-0.40) {$2$};  \node[vl] at (2.2,-0.40) {$3$};
    \node[pl] at (1.1,-1.00) {$J_{4}$};
  \end{scope}
\end{tikzpicture}
\end{equation*}
Explicitly, these supports are
\begin{equation}
\begin{aligned}
  J_{1}&=\{\,e_{1}+e_{2},\;e_{1}+e_{3},\;e_{1}-e_{3},\;e_{2}+e_{4},\;e_{2}-e_{4},\;e_{1}\,\}\, ,\\
  J_{2}&=\{\,e_{1}+e_{2},\;e_{1}+e_{3},\;e_{1}-e_{3},\;e_{1}+e_{4},\;e_{1}-e_{4},\;e_{2}\,\}\, ,\\
  J_{3}&=\{\,e_{1}+e_{2},\;e_{1}+e_{3},\;e_{1}-e_{3},\;e_{1}+e_{4},\;e_{2}+e_{4},\;e_{1}\,\}\, ,\\
  J_{4}&=\{\,e_{1}+e_{2},\;e_{1}+e_{3},\;e_{2}+e_{3},\;e_{1}+e_{4},\;e_{2}+e_{4},\;e_{3}\,\}\, ,
\end{aligned}
  \label{eq:so9-family}
\end{equation}
all of cardinality $6=h-1$.
The largest strongly orthogonal subsets of $J_1,J_2,J_3,J_4$ have four, three, three and two roots, respectively, so that only $J_1$ attains the maximum for $\mathfrak{so}_9$.

On each support, the coefficients of the top form are
\begin{equation}
\renewcommand{\arraystretch}{1.15}
\begin{array}{c|c|rrr}
   & \mathcal{S}^{6} & \mathcal{T}_{5}\mathcal{S} & \mathcal{T}_{3}^{2} & \mathcal{T}_{3}\mathcal{S}^{3}\\\hline
  J_{1} & -46080 &  0 & -288 & -1152\\
  J_{2} & -46080 &  0 & 576 & -4608\\
  J_{3} & -46080 & 720 & 1440 & -2880\\
  J_{4} & -46080 & -960 & 1152 & 3456
\end{array}
  \label{eq:so9-table}
\end{equation}
This $4\times4$ matrix is non-degenerate, so that the column of $\mathcal S^6$ is linearly independent from the other three.
The test \eqref{eq:projected-restriction-test} with $\mathcal F=\{J_1,J_2,J_3,J_4\}$ therefore gives $\mathcal S^6\notin I^{\mathfrak{so}_9}$.

\subsubsection{Higher rank}\label{ssec:higher-rank-supports}

The $\mathfrak{so}_9$ example illustrates why adding more supports at a fixed level need not suffice.
On every level-two support, it can be checked that $\mathcal S^6$ agrees with a specific linear combination of the residual generators, so no family drawn from that level alone can separate it from the ideal.

Applying the same test at higher rank gives the following results, including the preceding examples for comparison:
\begin{equation}
\renewcommand{\arraystretch}{1.2}
\begin{array}{c|c|c|c|c}
  \mathfrak g & h-1 & \#\text{generators} & \text{level }2
  & \text{levels }2\text{ and }3\\\hline
  \mathfrak{so}_7 & 4 & 1 & \text{sufficient} & \text{sufficient}\\
  \mathfrak{so}_8 & 5 & 2 & \text{sufficient} & \text{sufficient}\\
  \mathfrak{so}_9,\dots,\mathfrak{so}_{12} & 6,\dots,9 & 3,4,5,7
  & \text{insufficient} & \text{sufficient}\\
  \mathfrak{so}_{13} & 10 & 9 & \text{insufficient} & \text{insufficient}
\end{array}
  \label{eq:levels-table}
\end{equation}
The number of generators counts the residual ideal generators \eqref{eq:residual-generators}.
Here ``sufficient'' means that the family separates $\mathcal S^{h-1}$ from their span, while ``insufficient'' means that it remains in their span even when all supports at the indicated levels are included.
All supports at levels two and three have cardinality exactly $h-1$, so no passage to subsets is needed.

Thus levels two and three establish $\mathcal S^{h-1}\notin I^{\mathfrak{so}_n}$ through $n=12$, but do not suffice for $\mathfrak{so}_{13}$.
Level four is the next natural enlargement to test, using subsets of cardinality $h-1$ as in \eqref{eq:chosen-subsets}.
More generally, we expect higher levels to be required as the number of residual generators grows.

\section{Quantum lifting of non-Coulomb powers of \texorpdfstring{$\mathcal{S}$}{S}}\label{sec:quantum-lift}

In this section, we show that the non-Coulomb monotone $\mathcal S^4$ of the $\mathcal N=4$ $SO(7)$ theory is lifted by the first quantum correction $Q_1$ to the supercharge.\footnote{Here loop order refers to corrections to the supercharge, not to the one-loop dilatation spectrum described by the classical cohomology in Section~\ref{sec:index}.}
To specify how this happens, we briefly recall that the one-loop supercharge induces a differential on the classical cohomology \cite{Budzik:2023xbr,Budzik:2025zvu}.
Using $\hbar$ as a formal loop-counting parameter, we write
\begin{equation}
    Q=Q_0+\hbar Q_1+\hbar^2Q_2+\mathcal O(\hbar^3)\,,
\end{equation}
where $Q_0$ is the classical supercharge previously denoted by $Q$.
Nilpotence of $Q$ order by order gives
\begin{equation}\label{eq:loop-nilpotence}
    Q_0^2=0\,,\qquad \{Q_0,Q_1\}=0\,,\qquad Q_1^2=-\{Q_0,Q_2\}\,.
\end{equation}
Thus $Q_1$ induces a nilpotent differential $[Q_1]$ on $Q_0$-cohomology, although it need not satisfy a Leibniz rule on composite operators, as we discuss below.
Its cohomology consists of the classes that survive at this order.
 
As we will explain below, $[\mathcal S^4]$ is $[Q_1]$-closed.
Its lifting therefore has to occur through $[Q_1]$-exactness and we will construct a corresponding $Q_0$-closed source $F_4$ which satisfies
\begin{equation}\label{eq:main-result-lift}
    [Q_1][F_4]=c\,[\mathcal S^4]\,,\qquad c\neq0\,,
\end{equation}
where brackets denote $Q_0$ cohomology classes.
We also establish that $[F_4]$ is fortuitous \cite{Chang:2024zqi}, namely, it has no $Q_0$-closed representative in the free trace complex, and is only closed after the $\mathfrak{so}_7$ trace relations are imposed.
 
The quantum lifting of $\mathcal{S}^4$ is closely related to recent works \cite{Chang:2025mqp,Choi:2025bhi}.
Indeed, in these works, another non-Coulomb monotone was found in the $SO(7)$ theory, built from both scalars and gauginos.
We will refer to it as the Chang--Lin--Choi--Lee (CLCL) monotone.
As we show in Appendix~\ref{app:clcl-monotone}, $\mathcal{S}^4$ is a superconformal descendant of this monotone.
Alternatively, one can obtain the CLCL monotone from $\mathcal{S}^4$ by acting with the six superconformal charges which sit in the commutant to $Q_0$,
\begin{equation}\label{eq:clcl-descendant}
    \epsilon_{m_1n_1p_1}\epsilon_{m_2n_2p_2}
    \bigl(\bar S^{m_1}_{\dot\alpha}\bar S^{m_2\dot\alpha}
    \bar S^{n_1}_{\dot\beta}\bar S^{n_2\dot\beta}
    \bar S^{p_1}_{\dot\gamma}\bar S^{p_2\dot\gamma}\bigr)(\mathcal{S}^4)
    \;\propto\; M_1-3M_2 \, .
\end{equation}
This agrees with $(E,J_L^3,J_R^3,q_1,q_2,q_3)=(12,0,0,4,4,4)$ for $\mathcal S^4$ and $(9,0,0,3,3,3)$ for the CLCL monotone.
We view this connection as an explanation of the non-Coulomb character of the CLCL monotone through the known non-Coulomb character of $\mathcal{S}^4$, which we studied at length in Section \ref{sec:nC-from-S}.
The action of the classical commutant charges commutes with the Cartan restriction. 
Therefore, the CLCL representative obtained from $\mathcal S^4$, and the multiplet generated from it, vanish on the Cartan.
 
The $Q_1$ pairing established in \cite{Choi:2025bhi} between the CLCL monotone and the fortuitous BMN operator first constructed in \cite{Gadde:2025yoa} also implies the lifting of $\mathcal S^4$.
Indeed, the classical commutant supercharges $\bar Q_{m\dot\alpha}$ anticommute with $Q_1$ in $Q_0$-cohomology \cite{Budzik:2025zvu}, so applying the same charges that take the CLCL monotone to $\mathcal S^4$ to its fortuitous source produces a partner for $[\mathcal S^4]$.
In the following, we instead give an independent construction that provides a route to partners for other powers of the gaugino bilinear.
In particular, the same strategy can be applied to the non-Coulomb operators $\mathcal S^m$ with $r<m<h$ in $SO(N)$ gauge theories described in Section~\ref{sec:nC-from-S}.
 
In Section~\ref{ssec:rank-sources}, we show that $[Q_1]$ annihilates general monotone classes in the multi-trace sector at finite rank. 
A source with non-zero $[Q_1]$ image must therefore be fortuitous, which is independent motivation from the known pairing for the construction below.
 
Our construction proceeds as follows.
The action of $Q_1$ first fixes the charge sector and letter number in which we have to look for the partner $[F_4]$.
In this sector, we construct the $Q_0$ action and the $\mathfrak{so}_7$ trace relations.
The preimage of a trace relation under $Q_0$ that is not itself the image of a trace relation gives a fortuitous class, which in the present case is unique up to addition of monotone classes.
We then use the four-root support of Section~\ref{ssec:example-so7} to verify \eqref{eq:main-result-lift}.
An additional calculation shows that $[Q_1]$ annihilates all monotone classes in this sector, so that adding such a class to $[F_4]$ does not change its $Q_1$ image in cohomology.

\subsection{Constructing a fortuitous source}\label{ssec:constructing-source}
 
The one-loop supercharge acts on pairs of letters, so obtaining $\mathcal S^4$ requires a pair action that produces two derivative-free gauginos.
The only such entry is the $\psi_m\phi^m$ entry \cite{Budzik:2025zvu}.
In the basis $T_a$ of Section~\ref{ssec:gaugino-chiral-ring}, let $f_{abc}=\langle[T_a,T_b],T_c\rangle$, with color indices raised and lowered by $g_{ab}=\langle T_a,T_b\rangle$ and its inverse.
Using the same invariant form as in Appendix~\ref{app:cw-basis}, the pair action takes the form
\begin{equation}\label{eq:Q1-psi-phi}
    Q_1\bigl((\psi_m)^a(\phi^n)^b\bigr)=-\kappa_Q\,\delta_m^{\,n}\,f^{a}{}_{cd}f^{bc}{}_{e}(\lambda_{\dot\alpha})^d(\lambda^{\dot\alpha})^e\,.
\end{equation}
The overall sign fixes our convention for the non-zero normalization $\kappa_Q$, whose precise value will not be needed.
Note that the $\psi\phi$ pair can sit within a single trace or in two different traces, and all contributions have to be included.
 
Recall the grading \eqref{eq:mu-grading} of Section~\ref{sec:rev-class-coh}.
Since $Q_1$ trades a $\psi\phi$ pair for a pair of gauginos, it preserves the number of letters and shifts $\mu$ by $(1,1,-1,-1,-1)$.
As $\mathcal{S}^4$ has eight letters and grading $(4,4,0,0,0)$, any source for it must have eight letters and lie in the sector
\begin{equation}\label{eq:mu4-sector}
    \mu_4=(3,3,1,1,1) \, .
\end{equation}
The grading of $Q_1$ also shows that it vanishes on every pure-gaugino operator.
Thus $[\mathcal{S}^4]$ is $[Q_1]$-closed, and in the remainder we will demonstrate it is exact as well.

\subsubsection{Classical source map}\label{ssec:class-source-map}
 
As announced above, we obtain $[F_4]$ as a $Q_0$-preimage of a trace relation that is not the $Q_0$ image of another trace relation.
As a first step, we construct the action of $Q_0$ on the $\mu_4$ sector of the free-trace complex, in which finite-rank trace relations are not yet imposed.
We recall from Section~\ref{sec:rev-class-coh} the relevant $Q_0$ actions,
\begin{equation}\label{eq:D-action}
   \{Q_0,\lambda_{\dot\alpha}\}=0\,,\quad  \{Q_0,\psi_m\}=-i\epsilon_{mnp}[\phi^n,\phi^p] \, ,
    \quad
    Q_0\bigl(D_{\dot\alpha}\xi\bigr)=-i[\lambda_{\dot\alpha},\xi]+D_{\dot\alpha}\bigl([Q_0,\xi]\bigr) \, ,
\end{equation}
and that $Q_0$ raises the letter number by one while preserving $\mu$.
 
Since $|\mu_4|=9$, the bound \eqref{eq:letter-bound} limits words in this sector to at most nine letters.
The class $[F_4]$ is built from eight letters and its $Q_0$ image from nine, so the spaces we need sit at the top of this bound.
Moreover, $\mathcal{S}^4$ is a singlet under both the $SU(3)_R$ R-symmetry and the $SU(2)_R$ part of the Lorentz group, and $Q_1$ commutes with both, so $F_4$ may be taken to be a singlet as well.
Throughout the construction we therefore restrict attention to singlets under both groups.
 
An eight-letter word with $|\mu_4|=9$ carries one superfield derivative more than it has letters, so one of its letters carries two derivatives.
This means that either one of the letters is the fermion $\psi_m$, or a covariant derivative is placed on a gaugino or on a scalar letter.
A nine-letter word saturates \eqref{eq:letter-bound} and is therefore built from $\lambda_\pm$ and $\phi^m$ alone, which at $\mu_4$ means six gauginos and one scalar of each flavor.
Writing $\mathcal{C}_\infty^{\mathrm{src}}$ and $\mathcal{C}_\infty^{\mathrm{tar}}$ for the corresponding free multi-trace complexes, we find by explicit enumeration
\begin{equation}\label{eq:mu4-letter-content}
\begin{aligned}
    \mathcal C_\infty^{\mathrm{src}}&:\quad \lambda^6\psi_m\phi^m\ \oplus\ D_{\dot\alpha}\bigl(\lambda^5\phi^1\phi^2\phi^3\bigr)\,,
    & &\dim\mathcal C_\infty^{\mathrm{src}}=54+258=312\,,\\
    \mathcal C_\infty^{\mathrm{tar}}&:\quad \lambda^6\phi^1\phi^2\phi^3\,, & &\dim\mathcal C_\infty^{\mathrm{tar}}=85\,.
\end{aligned}
\end{equation}
Here the derivative notation denotes all placements of a single covariant derivative on an individual gaugino or scalar.
The dimensions are those of the spaces of singlets under $SU(3)_R$ and $SU(2)_R$.
We denote the two families of the source by $\mathcal C_\psi$ and $\mathcal C_\partial$, so that $\mathcal C_\infty^{\mathrm{src}}=\mathcal C_\psi\oplus\mathcal C_\partial$.
To calculate the dimensions we have imposed the transpose relation \eqref{eq:so(n)-bil-form}, which holds uniformly in the rank, but, as already mentioned, not yet the $\mathfrak{so}_7$ trace relations, to which we turn below.
The derivative words of $\mathcal C_\partial$ are annihilated by $Q_1$, since their image would have to lie at $\mu=(4,4,0,0,0)$, which contains no scalars, while an action on two letters cannot remove all three.
The complex $\mathcal C_\partial$ is however needed when imposing that $Q_0F_4$ be a trace relation, since by \eqref{eq:D-action} their $Q_0$ images can cancel those of the first family.
This observation will guide our choice of representative in Section~\ref{ssec:Q1-pairing}.
 
By the above, we have that
\begin{equation}\label{eq:terminal-D}
    \mathcal{C}_\infty^{\mathrm{src}}
    \ \xrightarrow{\ \ D\,\equiv\,Q_0\ \ }\ 
    \mathcal{C}_\infty^{\mathrm{tar}}
    \ \xrightarrow{\ \ Q_0\ \ }\ 0 \, .
\end{equation}
Constructing the explicit matrix $D$ which implements this map, one finds
\begin{equation}\label{eq:terminal-free-counts}
    \operatorname{rank}D=85=\dim\mathcal{C}_\infty^{\mathrm{tar}} \, ,
\end{equation}
so that every element of the target has a preimage.
Different preimages differ by an element of the $227$-dimensional kernel of $D$.
These elements are sources closed in the free-trace complex. 
Those that remain non-$Q_0$-exact after evaluation at finite rank represent the monotone classes in this sector.

\subsubsection{Including \texorpdfstring{$\mathfrak{so}_7$}{so(7)} trace relations}\label{ssec:including-trace-relations}
 
Let $\pi_7$ evaluate these expressions on $\mathfrak{so}_7$-valued letters, satisfying \eqref{eq:so(n)-bil-form}, and define
\begin{equation}
    \mathcal{I}_7^{\mathrm{src}}\equiv\ker\bigl(\pi_7|_{\mathcal{C}_\infty^{\mathrm{src}}}\bigr)\,,\qquad \mathcal{I}_7^{\mathrm{tar}}\equiv\ker\bigl(\pi_7|_{\mathcal{C}_\infty^{\mathrm{tar}}}\bigr)\,.
\end{equation}
These are the $\mathfrak{so}_7$ trace relations in the two spaces.
Since $Q_0$ acts letter by letter, independently of the matrix size, it is compatible with $\pi_7$, so that $D\mathcal{I}_7^{\mathrm{src}}\subset\mathcal{I}_7^{\mathrm{tar}}$.
An explicit computation gives\footnote{Evaluating the source and target bases on randomly sampled $\mathfrak{so}_7$ matrices, retaining the Grassmann nature of the fermionic letters, bounds the first two dimensions from above, see Appendix~\ref{app:finding-so7-trace-identities}. Explicit relations built from the vanishing eight-index antisymmetrizer in seven dimensions supply the matching lower bounds and the third dimension, see Appendix~\ref{app:antisymmetrizer-ids}.}
\begin{equation}\label{eq:terminal-identity-counts}
    \dim\mathcal{I}_7^{\mathrm{src}}=3\,,\qquad \dim\mathcal{I}_7^{\mathrm{tar}}=4\,,\qquad \dim D\mathcal{I}_7^{\mathrm{src}}=3\,.
\end{equation}
Thus the quotient $\mathcal I_7^{\mathrm{tar}}/D\mathcal I_7^{\mathrm{src}}$ is one dimensional.
As a representative of this class, we choose the explicit relation
\begin{equation}\label{eq:main-explicit-If}
    I_{\mathrm f}=-3\Pi_\epsilon\mathcal B\bigl(\lambda_+,\lambda_-,\lambda_-,\phi^1,\lambda_+,\lambda_-,\lambda_+\phi^2,\phi^3\bigr)\,,
\end{equation}
where $\mathcal B$ is the eight-index antisymmetrizer contraction \eqref{eq:app-B-contraction} and $\Pi_\epsilon=\mathscr P_0\mathscr A_\epsilon$ performs the Lorentz and flavor singlet projections defined in Appendix~\ref{app:antisymmetrizer-ids}.
Its vanishing for $\mathfrak{so}_7$, and its independence from the $Q_0$ images of the source relations, are established there, see \eqref{eq:app-exceptional-relation} and \eqref{eq:app-analytic-identity-ranks}.
The factor $-3$ ensures a matching with the normalization of $F_4$ below.
 
Since $D$ is onto, we can choose $F_4$ such that
\begin{equation}\label{eq:fortuitous-preimage}
    DF_4=-iI_{\mathrm f}\,,
\end{equation}
where the explicit factor $-i$ comes from the $Q_0$ action and ensures that both $F_4$ and $I_{\mathrm f}$ have real (rational) coefficients in the multi-trace bases.
For fixed $I_{\mathrm f}$, the source $F_4$ is determined up to addition of a source closed in the free-trace algebra, and hence its cohomology class is determined up to addition of monotone classes.
We use this freedom to choose a representative adapted to the $Q_1$ action, as we will explain in Section~\ref{ssec:Q1-pairing}.
Since $\pi_7(I_{\mathrm f})=0$, while $I_{\mathrm f}$ is non-zero in the free-trace complex, $F_4$ is $Q_0$-closed for $\mathfrak{so}_7$ but not in the free-trace complex.

There is also a separate freedom in choosing the representative of the trace-relation class.
Indeed, $I_{\mathrm f}+Q_0J$, with $J\in\mathcal I_7^{\mathrm{src}}$, represents the same non-zero class in $\mathcal I_7^{\mathrm{tar}}/D\mathcal I_7^{\mathrm{src}}$.
This shift is compensated by $F_4\mapsto F_4-iJ$.
Since $J$ vanishes upon evaluation on $\mathfrak{so}_7$, the evaluated operator is unchanged, although its formal trace expression changes. 
This freedom therefore does not affect $[F_4]$, unlike adding a non-trivial monotone source class.
We keep the explicit relation \eqref{eq:main-explicit-If} in the construction below.

The choice of $I_{\mathrm f}$ ensures that $F_4$ defines a non-trivial and fortuitous $Q_0$-cohomology class \cite{Chang:2024zqi}.
To make this explicit, suppose for contradiction that, in the free-trace algebra,
\begin{equation}
    F_4=G+Q_0H+i_{\mathrm{src}}\,,
    \qquad Q_0G=0\,,
    \qquad i_{\mathrm{src}}\in\mathcal I_7^{\mathrm{src}}\,,
\end{equation}
where $G$ is closed in the free trace complex.
Applying $Q_0$ gives $-iI_{\mathrm f}=Di_{\mathrm{src}}$, contradicting $I_{\mathrm f}\notin D\mathcal I_7^{\mathrm{src}}$.
Thus the non-zero class of $I_{\mathrm f}$ in $\mathcal I_7^{\mathrm{tar}}/D\mathcal I_7^{\mathrm{src}}$ establishes both non-exactness and fortuity, as also follows from the general construction of \cite{Chang:2024zqi}.

\subsection{Quantum pairing}\label{ssec:Q1-pairing}
 
We now choose a preimage of $I_{\mathrm f}$ adapted to the $Q_1$ action and show that $[Q_1][F_4]$ is a non-zero multiple of $[\mathcal S^4]$. 
We then check that this image is unchanged by the addition of monotone classes to  $[F_4]$.
 
Since $Q_1$ annihilates $\mathcal C_\partial$, the $Q_1$ image of a source $F=F_\psi+F_\partial$, with $F_\psi\in\mathcal C_\psi$ and $F_\partial\in\mathcal C_\partial$, depends only on its $\psi\phi$ part $F_\psi$.
Closedness, on the other hand, involves both parts,
\begin{equation}\label{eq:closure-split}
    D_\psi F_\psi+D_\partial F_\partial=-iI_{\mathrm f}\,,
\end{equation}
with $D_\psi$ and $D_\partial$ the restrictions of $D$ to the two families.
We therefore choose $F_\psi$ first, as simply as possible, and then solve \eqref{eq:closure-split} for $F_\partial$.
Changing the representative of the trace-relation class can change which $F_\psi$ admit such a completion.
For the choice $I_{\mathrm f}$, a particularly simple choice for which $F_\partial$ exists is\footnote{Allowing a change of trace-relation representative, one can instead take $F'_\psi=-\tfrac15\mathcal S^3\sum_m\Tr(\psi_m\phi^m)$ and a corresponding derivative completion. Appendix~\ref{app:constr-pair-F4} gives this construction.}
\begin{equation}\label{eq:F4psi}
    F_{4,\psi}=\mathcal S^2\mathcal O_\psi\,,
    \qquad
    \mathcal O_\psi=\sum_{m=1}^3\Tr\bigl[\psi_m\bigl(\lambda_-\phi^m\lambda_+-\lambda_+\phi^m\lambda_-\bigr)\bigr]\,.
\end{equation}
The corresponding derivative part $F_{4,\partial}$ is given in Appendix~\ref{app:constr-pair-F4}.
Let us note that neither $F_{4,\psi}$ nor $F_{4,\partial}$ is separately $Q_0$-closed, even after evaluation on $\mathfrak{so}_7$ matrices. 
Being built from derivative words alone, $F_{4,\partial}$ is annihilated by $Q_1$, so that
\begin{equation}\label{eq:Q1F4-reduction}
    Q_1F_4=Q_1F_{4,\psi}=\mathcal S^2Q_1\mathcal O_\psi\,,
\end{equation}
where we used that $Q_1$ vanishes on any pair consisting of a gaugino $\lambda_\pm$ and another gaugino, a scalar $\phi^m$, or a fermion $\psi_m$ \cite{Budzik:2025zvu}.
Consequently, no pair involving a letter in $\mathcal S^2$ contributes to the $Q_1$ action.

Recall from Section~\ref{sec:nC-from-S} that the $Q_0$-cohomology of the zero-derivative gaugino sector $\mu=(4,4,0,0,0)$ is one dimensional and generated by $[\mathcal S^4]$, while the restriction $\rho_J$ to the four-root support \eqref{eq:so7-support} annihilates every $Q_0$-exact operator in this sector and keeps $\rho_J(\mathcal S^4)\neq0$.
We compute \eqref{eq:Q1F4-reduction} in the unrestricted theory and then apply $\rho_J$.
As the restricted sector contains a single Grassmann monomial, it suffices to demonstrate that
\begin{equation}\label{eq:Q1-detector-functional}
    \rho_J(Q_1F_4)=c\,\rho_J(\mathcal S^4)\,, \qquad c\neq 0\,,
\end{equation}
which implies $[Q_1][F_4]=c[\mathcal S^4]$ in the full $\mathfrak{so}_7$ cohomology.
To calculate this coefficient, we write the ordered product of the eight restricted Grassmann variables as $\Omega=x_1x_2x_3x_4y_1y_2y_3y_4$, with the roots ordered as in \eqref{eq:so7-support}.
Using $\Tr(E_{\alpha_r}E_{-\alpha_s})=2\delta_{rs}$ and the pair action \eqref{eq:Q1-psi-phi}, with further conventions specified in Appendix~\ref{app:constr-pair-F4}, we find
\begin{equation}\label{eq:app-direct-quantum-coefficients}
    \rho_J(Q_1F_4)=-4608\kappa_Q\Omega\,,\qquad
    \rho_J(\mathcal S^4)=384\Omega\,,
\end{equation}
where the second coefficient follows by explicit expansion.
It follows that
\begin{equation}\label{eq:Q1-F4-main-conclusion}
    [Q_1][F_4]=-12\kappa_Q[\mathcal S^4]\,,
\end{equation}
and therefore that $[\mathcal S^4]$ is $[Q_1]$-exact and is lifted at one loop, with the fortuitous class $[F_4]$ as a partner.

The calculation above establishes the pairing for a specific choice of $F_4$, or preimage of $-iI_{\mathrm f}$.
As explained in Section \ref{ssec:including-trace-relations}, this choice is not unique.
We now establish independence of the pairing under this choice.

Since two distinct preimages differ by a freely closed source $G\in\ker D$, it suffices to show that the $Q_1$-image of such $G$ is $Q_0$-exact.
For a basis $\{u_A\}$ of $\mathcal{C}_\infty^{\mathrm{src}}$ define the row-vector $q=(q_A)$ by
\begin{equation}\label{eq:Q1-source-row}
    \rho_J(Q_1u_A)=q_A\,\rho_J(\mathcal S^4)\,.
\end{equation}
Thus a source with coefficient (column) vector $f$ has coefficient $qf$ under $Q_1$, after restriction.
Computing this row and comparing it with $D$, we have verified in Appendix~\ref{app:constr-pair-F4} that
\begin{equation}\label{eq:Q1-free-kernel}
    q\big|_{\ker D}=0\,.
\end{equation}
Since the target cohomology is one dimensional and detected by $\rho_J$, it implies $[Q_1][G]=0$ for every freely closed source $G$ in this sector.
Every preimage $F_4'$ of $-iI_{\mathrm f}$ therefore satisfies
\begin{equation}\label{eq:app-preimage-independent-image}
    [Q_1][F_4']=[Q_1][F_4]=-12\kappa_Q[\mathcal S^4]\,,
\end{equation}
although its $Q_0$-cohomology class may differ from $[F_4]$ by a monotone class.
Moreover, every monotone class has a freely closed representative, so no monotone class in this source sector can have a non-zero $Q_1$ image proportional to $[\mathcal S^4]$.
It follows that the partner of $[\mathcal S^4]$ is necessarily fortuitous.\footnote{
We will provide a general argument for this observation in Section~\ref{sec:gen-patterns}.}
In addition, replacing $I_{\mathrm f}$ by $I_{\mathrm f}+Q_0J$ and $F_4$ by $F_4-iJ$, with $J$ a source trace identity, leaves the evaluated $\mathfrak{so}_7$ operator unchanged.
We conclude that the pairing depends neither on a choice of preimage of the trace relations, nor does it depend on the representative of the trace relation in $\mathcal I_7^{\mathrm{tar}}/D\mathcal I_7^{\mathrm{src}}$.

Finally, we find that $[F_4]$ has a non-zero Cartan restriction even modulo monotone classes.
We have checked this explicitly by restricting a basis of $\ker D$, \textit{i.e.}, the freely closed multi-traces, to the Cartan.
When we adjoin $\rho_C(F_4)$, this increases the rank of the resulting coefficient matrix by one which demonstrates its linear independence from the monotones.\footnote{The same property can be checked to hold for the BMN fortuitous partner of the CLCL monotone \cite{Gadde:2025yoa,Choi:2025bhi}, whose commutant descendant agrees with $[F_4]$ up to normalization and monotone classes.
This Coulomb visibility appears not to have been noted previously.}
We emphasize that the trace relation argument in \cite{Chang:2025mqp} does not rule out this possibility. 
Indeed, even though $I_{\mathrm f}$ vanishes on the Cartan, giving $\rho_C(Q_0F_4)=0$, this does not imply that $\rho_C(F_4)=0$.
Thus the pairing with $\mathcal S^4$ connects a Coulomb-visible fortuitous operator to a non-Coulomb monotone.
We discuss a possible connection between Coulomb-visible fortuitous classes and generalized monotones in the sense of \cite{Choi:2026faq} in Section~\ref{sec:discussion}.

\subsection{Higher powers and higher rank}\label{ssec:higher-rank-predictions}

The construction above suggests a systematic approach to the non-Coulomb powers $\mathcal S^m$ for $\mathfrak{so}_n$, with rank $r$ and dual Coxeter number $h=n-2$, in the range $r<m<h$.
It is natural to expect that these classes also become $[Q_1]$-exact.
We outline here how our calculations for $\mathfrak{so}_7$ generalize to higher rank.

To start, we note that a source under the $Q_1$ action for $\mathcal S^m$ has $2m$ letters and grading
\begin{equation}\label{eq:mum-sector}
    \mu_m=(m-1,m-1,1,1,1)\,,\qquad |\mu_m|=2m+1\,.
\end{equation}
It therefore again lies one letter below the bound, and the relevant free-trace singlet spaces have the schematic contents
\begin{equation}\label{eq:mum-letter-content}
\begin{aligned}
    \mathcal C_{\infty,m}^{\mathrm{src}}
    &:\quad \lambda^{2m-2}\psi_a\phi^a
    \ \oplus\ D_{\dot\alpha}\bigl(\lambda^{2m-3}\phi^1\phi^2\phi^3\bigr)\,,\\
    \mathcal C_{\infty,m}^{\mathrm{tar}}
    &:\quad \lambda^{2m-2}\phi^1\phi^2\phi^3\,.
\end{aligned}
\end{equation}
The $Q_0$ action again maps the source to a target with maximal letter content for the current grading, which itself is therefore annihilated by $Q_0$.
In the following, we denote the restriction of $Q_0$ to the source complex by $D_m$.

Let $\mathcal I_{n,m}^{\mathrm{src}}$ and $\mathcal I_{n,m}^{\mathrm{tar}}$ denote the spaces of trace relations in the above source and target spaces.
A target relation admits a preimage under $Q_0$ precisely when it lies in $\operatorname{im}D_m$.
The relations that produce fortuitous sources are therefore contained in
\begin{equation}\label{eq:higher-rank-relation-quotient}
    \frac{\mathcal I_{n,m}^{\mathrm{tar}}\cap\operatorname{im}D_m}
    {D_m\mathcal I_{n,m}^{\mathrm{src}}}\,.
\end{equation}
Each non-zero class in this quotient gives a non-exact, fortuitous source, determined up to the addition of monotone classes, by the same argument as for $F_4$.
When $D_m$ is onto, as in the $\mathfrak{so}_7$ example, the numerator is simply $\mathcal I_{n,m}^{\mathrm{tar}}$.
Appendices~\ref{app:finding-so7-trace-identities} and \ref{app:antisymmetrizer-ids} illustrate how to determine these spaces of trace relations.

We also note that if this quotient vanishes, $[\mathcal S^m]$ is protected at one loop.
Indeed, $[Q_1]$ annihilates powers of $\mathcal S$, and in fact every monotone, as shown in Section~\ref{ssec:rank-sources}, while the vanishing quotient excludes any fortuitous source that could make $[\mathcal S^m]$ $[Q_1]$-exact.
If the quotient is non-zero, it may contain several independent fortuitous directions.
Nevertheless, the pure-gaugino target cohomology is one-dimensional and generated by $[\mathcal S^m]$, so the induced $[Q_1]$ action has rank at most one.

We expect that the splitting into $\mathcal C_\psi$ and $\mathcal C_\partial$ in \eqref{eq:mum-letter-content} again simplifies the construction.
Since $Q_1$ annihilates $\mathcal C_\partial$, only the $\psi\phi$ part contributes to the pure-gaugino image.
A natural ansatz is thus
\begin{equation}
    F_m=F_{m,\psi}+F_{m,\partial}\,,
    \qquad
    F_{m,\psi}=\mathcal S^{m-2}\mathcal O_\psi\,,
\end{equation}
with $\mathcal O_\psi$ defined in \eqref{eq:F4psi}.
It remains to find a derivative completion $F_{m,\partial}$ such that $D_mF_m$ represents a non-zero class in \eqref{eq:higher-rank-relation-quotient}.
The pairing then follows from $Q_1\mathcal O_\psi=-12\kappa_Q\mathcal S^2$ modulo the pure-gaugino exact ideal, with the same coefficient at higher rank, since $\psi_m$ and $\phi^m$ are not adjacent in $\mathcal O_\psi$.
Thus, whenever the completion exists, we obtain
\begin{equation}
    [Q_1][F_m]=-12\kappa_Q[\mathcal S^m]\neq0\,,
    \qquad r<m<h\,.
\end{equation}
The lower bound follows from Section~\ref{ssec:rank-coulomb-images}, where we show that every $[Q_1]$ image has vanishing Cartan restriction, whereas $\mathcal S^m$ remains non-zero on the Cartan for $m\leq r$.

At fixed rank, a completion at one power also supplies completions for all higher non-zero powers.
Indeed, multiplication by $\mathcal S$ preserves $Q_0$-closedness, while the additional gauginos introduce no new $Q_1$ contractions.
Once $F_{m_0}$ has been constructed, we may therefore take
\begin{equation}\label{eq:higher-power-propagation}
    F_{m_0+k}=\mathcal S^kF_{m_0}\,,
    \qquad
    [Q_1][F_{m_0+k}]=-12\kappa_Q[\mathcal S^{m_0+k}]\,,
    \qquad 0\leq k<h-m_0\,.
\end{equation}
Thus it would suffice to find a derivative completion for the first non-Coulomb power, $m_0=r+1$, to establish the one-loop lifting of the entire non-Coulomb tower at that rank.

\section{General patterns of quantum lifting}\label{sec:gen-patterns}

In the preceding section, we showed that the non-Coulomb monotone $\mathcal S^4$ pairs with a Coulomb-visible fortuitous source under $[Q_1]$.
We now show that every non-zero one-loop pairing in the multi-trace sector necessarily has a fortuitous source and a non-Coulomb target, \textit{i.e.}, a target with vanishing Cartan restriction.
In particular, Coulomb monotones survive at this order, whereas non-Coulomb monotones can lift only by becoming $[Q_1]$-exact.

In more detail, in Section~\ref{ssec:rank-sources}, we show that the trace joining contribution to $Q_1$ vanishes after Cartan restriction.
Together with the classical graviton description and the rank independence of the joining action, this implies that $[Q_1]$ annihilates every monotone.
In Section~\ref{ssec:rank-coulomb-images}, we show that every $Q_0$-closed multi-trace operator becomes cohomologous to a multi-graviton expression after multiplication by a sufficiently high power of a scalar discriminant.
Applying $Q_1$ to this relation and using the vanishing of its multi-graviton image on the Cartan then excludes Coulomb targets.

We consider the classical families
\begin{equation}\label{eq:rank-classical-families}
    G_n=SU(n+1),\qquad SO(2n+1),\qquad USp(2n),\qquad SO(2n),
\end{equation}
where $n$ denotes the rank, and work with their complexified gauge algebras.
Throughout, we restrict to the multi-trace sector.\footnote{For $SO(2n)$, this is the reflection-even, or $O(2n)$-invariant, sector and excludes reflection-odd Pfaffian-type operators. Both $Q_0$ and $Q_1$ preserve this sector.}
As in the previous sections, $\rho_C$ restricts all letters to the Cartan subalgebra.

\subsection{\texorpdfstring{$Q_1$}{Q1}-closedness of monotones}\label{ssec:rank-sources}
\label{ssec:rank-cartan-action}

The one-loop supercharge acts on pairs of letters, including their covariant derivatives \cite{Budzik:2025zvu}.
Contractions between different traces join the two trace factors, and we first consider this part of $Q_1$.
It suffices to take a double trace $UV$, with $U,V$ single-trace operators, since $Q_1$ acts on only two letters at a time.
A contraction taking a letter $X$ from $U$ and a letter $Y$ from $V$ contributes a term of the form
\begin{equation}\label{eq:rank-cross-contraction}
    f^{a}{}_{cd}f^{bc}{}_{e}\,L^dM^e\,
    \frac{\partial U}{\partial X^a}
    \frac{\partial V}{\partial Y^b},
\end{equation}
up to a numerical coefficient, where $L,M$ are the output letters.
We refer to Section~\ref{ssec:gaugino-chiral-ring} for our Lie algebra conventions.

Cartan restriction sets the output indices $d,e$ equal to Cartan indices.
On the other hand, the differentiated traces vanish unless $a,b$ are Cartan indices.
Indeed, differentiating $U$ with respect to $X^a$ gives a sum of traces in which one occurrence of $X$ is replaced by $T_a$.
Since all remaining letters are diagonal after restriction, each trace vanishes if $T_a$ is off-diagonal.
The same argument applies to the derivative of $V$.

With $a,d$ Cartan indices, the first structure constant vanishes, and similarly for the second.
Every joining contribution therefore vanishes after Cartan restriction.
Since $\rho_C$ preserves products, we obtain
\begin{equation}\label{eq:rank-cartan-product-rule}
    \rho_C\bigl(Q_1(UV)\bigr)
    =\rho_C(Q_1U)\rho_C(V)
    +(-1)^{|U|}\rho_C(U)\rho_C(Q_1V),
\end{equation}
for arbitrary single traces $U,V$, whether or not they are $Q_0$-closed, where $|U|$ is the fermion parity of $U$.
The same formula extends to arbitrary multi-traces.
Thus the $Q_1$ action obeys a graded Leibniz rule after Cartan restriction.

As shown in \cite{Budzik:2025zvu}, the $Q_1$ image of a single-trace graviton is $Q_0$-exact.
This follows from the fact that $Q_1$ annihilates the scalar primary, and the graded commutators of $Q_1$ with the classical commutant generators are $Q_0$-exact on $Q_0$-closed operators, extending the statement to the descendants.
Since the classical transformations consist of (anti-)commutators, $\rho_CQ_0=0$.
Every exact operator therefore vanishes on the Cartan, so that \eqref{eq:rank-cartan-product-rule} yields
\begin{equation}\label{eq:rank-multi-graviton-cartan}
    \rho_C(Q_1G)=0
\end{equation}
for every multi-graviton expression $G$ at finite rank.

To strengthen \eqref{eq:rank-multi-graviton-cartan} to $[Q_1]$-closedness of monotone classes, it remains to show that the joining contribution is $Q_0$-exact.
For single-trace graviton representatives $U,V$, we write
\begin{equation}\label{eq:rank-joining-contribution}
    \mathcal J(U,V)
    \equiv Q_1(UV)-(Q_1U)V-(-1)^{|U|}U(Q_1V)\,.
\end{equation}
Since $Q_0U=Q_0V=0$, the graded Leibniz rule for $Q_0$ and $\{Q_0,Q_1\}=0$ imply $Q_0\mathcal J(U,V)=0$.
Its Cartan restriction vanishes by \eqref{eq:rank-cartan-product-rule}.
Moreover, each contraction joins the two input traces into a single trace without producing a separate closed color loop.
The resulting expression has rank-independent coefficients, so $\mathcal J(U,V)$ can be regarded as a fixed single-trace expression in the free trace complex.

We can therefore test its exactness using the single-trace cohomology before trace relations are imposed.
This cohomology has a basis of graviton classes, as reviewed in Section~\ref{sec:rev-class-coh}.
Restricting all letters to the same Cartan subalgebra gives polynomials in their bosonic and fermionic diagonal components.
At fixed grading and sufficiently large rank, every non-zero single-trace cohomology class has a non-zero Cartan restriction, as can be checked from the linear independence of the restricted graviton representatives.
Since $\mathcal J(U,V)$ is closed but has vanishing Cartan restriction at every rank, it must therefore be exact in this complex:
\begin{equation}\label{eq:rank-joining-primitive}
    \mathcal J(U,V)=Q_0H
\end{equation}
for some single-trace expression $H$.
Although we used the linear independence of the Cartan-restricted graviton basis at sufficiently large rank, the resulting identity holds before trace relations are imposed and therefore remains valid at arbitrary rank.

For a general product of gravitons, the $Q_1$ action consists of single-trace actions and joining terms of the form \eqref{eq:rank-joining-contribution}.
Since both contributions are $Q_0$-exact and multiplication by the remaining closed factors preserves exactness, we conclude that
\begin{equation}\label{eq:rank-obstruction-conclusion}
    [Q_1][G]=0
\end{equation}
for every monotone class $[G]$ at every rank.
This extends the single-trace result of \cite{Budzik:2025zvu} to the multi-trace sector.
Every non-zero one-loop pairing must therefore have a fortuitous source.
Its target may nevertheless be monotone, as $\mathcal S^4$ illustrates.

Since monotone classes are $[Q_1]$-closed, they can lift at this order only by becoming $[Q_1]$-exact.
This imposes a simple restriction on their grading.
$Q_1$ shifts $\mu$ by $(1,1,-1,-1,-1)$, as shown for the $\psi\phi$ entry in Section~\ref{ssec:constructing-source} (see \cite{Budzik:2025zvu} for the general case).
Since every operator has $n_\pm\geq0$, every non-zero class in the image of $[Q_1]$ must have $n_+\geq1$ and $n_-\geq1$.
Consequently, any monotone class with $n_+=0$ or $n_-=0$ survives the first quantum correction.

In particular, BMN operators have grading $(0,0,n_1,n_2,n_3)$ and therefore cannot represent $[Q_1]$-exact classes.
Every non-zero BMN monotone is thus protected at this order, including the non-Coulomb examples recently constructed in
\cite{Kim:2026bhc,Fortuity:2026summary}.
The same argument shows that fortuitous operators in the BMN sector can only lift by failing to be $[Q_1]$-closed.

\subsection{Non-Coulombness of \texorpdfstring{$Q_1$}{Q1} targets}\label{ssec:rank-coulomb-images}

We now show that the $[Q_1]$ image of any $Q_0$-cohomology class has vanishing Cartan restriction.
We work at fixed rank and take a multi-trace operator $F$ with $Q_0F=0$, allowing its Cartan restriction to be non-zero.
The $Q_0$-closedness of $F$ may be due to a trace relation at the fixed finite rank, and hence $F$ may be fortuitous.
We do not assume that its Cartan restriction vanishes or can be removed by subtracting monotones.
Indeed, the example of $F_4$ shows that this need not be possible.
The key observation, which we will prove below, is that multiplication by a sufficiently high power of a scalar discriminant $\Delta$ makes $F$ cohomologous to a multi-graviton expression,
\begin{equation}\label{eq:preamb-discrimanant-dressing}
    \Delta^kF=G+Q_0H\,,
\end{equation}
where $G$ is a multi-graviton and $H$ is some multi-trace expression.
We will use this identity, together with the vanishing of $\rho_C(Q_1G)$ established in the previous subsection, to show that $Q_1F$ has vanishing Cartan restriction.

To establish this identity, we temporarily work with a scalar whose eigenvalues are distinct and allow division by their differences.
This allows us, through non-linear field redefinitions, to organize all off-diagonal fields into $Q_0$ pairs.
After also replacing the diagonal variables by closed combinations, these pairs decouple and can be removed from the cohomology by the standard argument for \emph{contractible pairs}, reviewed in Appendix~2.B of \cite{Barnich:2000zw}.\footnote{Related non-linear field redefinitions eliminate an adjoint chiral multiplet made massive by a superpotential deformation in \cite{Choi:2025pqr}.
Here the scalar eigenvalue differences play the role of mass parameters, but the eigenvalues remain variables in the undeformed theory.
Field redefinitions also enter the construction of fortuitous $SU(N)$ BMN classes with an $SU(2)$ nonabelian core in \cite{deMelloKoch:2026BMNEveryN}.
The generalized monotone construction of \cite{Choi:2026faq} likewise allows inverse scalar letters, but requires a uniform construction as the rank varies.
We emphasize that our rational description is at fixed rank, and discuss these comparisons further in Section~\ref{sec:discussion}.}
After removing the pairs, the remaining expression is built from gravitons with coefficients rational in the scalar eigenvalues.
Clearing the denominators then gives the identity \eqref{eq:preamb-discrimanant-dressing}, valid also when eigenvalues coincide.

\subsubsection*{Pairing off-diagonal letters}

We write $\Phi=\phi^1(0)$ and, where its eigenvalues are distinct, choose a diagonal gauge,
\begin{equation}\label{eq:Phi-diag-gauge}
    \Phi=\operatorname{diag}(\zeta_1,\ldots,\zeta_m),
    \qquad \zeta_a\neq\zeta_b\quad(a\neq b),
\end{equation}
where $m$ is the dimension of the defining representation of $\mathfrak g$ and the eigenvalues obey the relevant constraints.
We emphasize that the $\zeta_a$ remain variables, not fixed vacuum expectation values.
We keep all other letters, including the non-zero modes of $\phi^1(z)$, unrestricted.

To identify the off-diagonal $Q_0$ pairs, we select the generating-field transformations in \eqref{eq:superalgebra-gener-fields} that contain a commutator with $\Phi$.
In this gauge, such a commutator singles out an off-diagonal letter multiplied by a non-zero eigenvalue difference:
\begin{equation}
    [\Phi,Y]_{ab}=(\zeta_a-\zeta_b)Y_{ab}\,,
    \qquad a\neq b.
\end{equation}
Since $Q_0\Phi=0$, the eigenvalues and, by the Leibniz rule, their inverse differences are $Q_0$-closed.
We can therefore factor out the eigenvalue difference in each of the selected $Q_0$ transformations and absorb the remaining terms into a redefined field.
Denoting these fields by hats, we obtain the exact pairings
\begin{equation}\label{eq:rank-generating-pairs}
    \begin{aligned}
        Q_0(\phi^1(z))_{ab}
            &=i(\zeta_a-\zeta_b)\widehat\lambda_{ab}(z),\\
        Q_0(\psi_3(z))_{ab}
            &=-2i(\zeta_a-\zeta_b)\widehat\phi^{\,2}_{ab}(z),\\
        Q_0(\psi_2(z))_{ab}
            &=2i(\zeta_a-\zeta_b)\widehat\phi^{\,3}_{ab}(z),\\
        Q_0f_{ab}(z)
            &=i(\zeta_a-\zeta_b)\widehat\psi_{1,ab}(z),
        \qquad a\neq b.
    \end{aligned}
\end{equation}
For the first pairing, the required redefinition is
\begin{equation}\label{eq:rank-corrected-generating-fields}
    \widehat\lambda_{ab}(z)
    =\lambda_{ab}(z)
    -\frac{[\lambda(z),\phi^1(z)-\Phi]_{ab}}{\zeta_a-\zeta_b}.
\end{equation}
This pairing starts at positive degree in $z$, since $(\phi^1(0))_{ab}=0$ for $a\neq b$ in the chosen gauge and $\lambda(0)=0$.
The other three redefinitions are
\begin{equation}\label{eq:rank-corrected-generating-fields-2}
    \begin{aligned}
        \widehat\phi^{\,2}_{ab}(z)
        &=(\phi^2(z))_{ab}
        +\frac{[\phi^1(z)-\Phi,\phi^2(z)]_{ab}}{\zeta_a-\zeta_b}
        +\frac{\{\lambda(z),\psi_3(z)\}_{ab}}{2(\zeta_a-\zeta_b)},\\[3pt]
        \widehat\phi^{\,3}_{ab}(z)
        &=(\phi^3(z))_{ab}
        +\frac{[\phi^1(z)-\Phi,\phi^3(z)]_{ab}}{\zeta_a-\zeta_b}
        -\frac{\{\lambda(z),\psi_2(z)\}_{ab}}{2(\zeta_a-\zeta_b)},\\[3pt]
        \widehat\psi_{1,ab}(z)
        &=(\psi_1(z))_{ab}
        +\frac{[\phi^1(z)-\Phi,\psi_1(z)]_{ab}}{\zeta_a-\zeta_b}\\
        &\quad+\frac{[\phi^2(z),\psi_2(z)]_{ab}+[\phi^3(z),\psi_3(z)]_{ab}-[\lambda(z),f(z)]_{ab}}{\zeta_a-\zeta_b}.
    \end{aligned}
\end{equation}
Nilpotence, together with $Q_0\zeta_a=0$, implies that all four hatted fields are $Q_0$-closed.
The four pairings thus cover every off-diagonal letter, provided that the change of variables is invertible.
To see the latter, keep the diagonal components and unredefined fields fixed.
We then recover the original off-diagonal fields order by order in $z$: first $\lambda$, then $\phi^2,\phi^3$, and finally $\psi_1$.
Corrections involving the field being recovered multiply it by $\phi^1(z)-\Phi$, which has no constant term, and therefore depend only on its already recovered lower-order coefficients.
All other fields are fixed or recovered earlier in this order, so the change of variables is invertible in finitely many steps at any fixed $z$-order.

\subsubsection*{Closed diagonal variables}

The off-diagonal letters now form $Q_0$ pairs, but they still enter the transformations of the diagonal variables.
For example, $Q_0(\psi_1)_{aa}=-2i[\phi^2,\phi^3]_{aa}$ depends on off-diagonal scalar components.
To decouple the pairs completely, we construct $Q_0$-closed replacements for the diagonal variables.
We require these to agree with the original variables on the Cartan, so that the redefinitions preserve the Cartan restriction of every operator.

We proceed in increasing weight
\begin{equation}\label{eq:rank-localization-weight}
    W=n_++n_-+n_2+n_3\,.
\end{equation}
Since $Q_0$ preserves the $\mu$-grading, it also preserves $W$.
Every independent letter except the undifferentiated $\Phi$ has positive weight, while the eigenvalues and their inverse differences have weight zero.

For each positive-weight letter $Y$, we choose a closed graviton representative $\mathcal G_{d,Y}$ containing $\Tr(\Phi^{d-1}Y)$.
Every term in $\mathcal G_{d,Y}$ contains $d$ letters and has total weight $W(Y)$.
At fixed letter number and $\mu$-grading, any term with $d-1$ copies of $\Phi$ is therefore proportional to $\Tr(\Phi^{d-1}Y)$.
The other terms, which ensure closedness of the full graviton, therefore contain fewer copies of $\Phi$, distributing the total weight $W(Y)$ among at least two positive-weight letters, each of smaller weight than $Y$.
Its Cartan restriction is thus
\begin{equation}\label{eq:rank-graviton-leading-moment}
    \rho_C(\mathcal G_{d,Y})
    =\sum_{a=1}^{m}\zeta_a^{d-1}Y_{aa}+R_{d,Y}\,,
\end{equation}
where $R_{d,Y}$ involves lower-weight diagonal letters and $\Phi$ eigenvalues.

We construct the closed diagonal variables in increasing weight.
We start with letters $Y$ of the smallest positive weight, for which $R_{d,Y}=0$, and define
\begin{equation*}
    \sum_a\zeta_a^{d-1}\widetilde Y_a=\mathcal G_{d,Y}.
\end{equation*}
For $SU(m)$, the values $d=2,\ldots,m$, together with tracelessness, determine the independent $\widetilde Y_a$.
At the smallest positive weight, these simply coincide with the original diagonal components, $\widetilde Y_a=Y_{aa}$.
For the orthogonal and symplectic families of rank $n$, the values $d=2,4,\ldots,2n$ suffice, using the paired eigenvalues $\pm\zeta_i$.
These systems are invertible when the $\Phi$ eigenvalues are distinct, and their inverses require only eigenvalue differences in the denominators.
Since the graviton expressions and eigenvalues are closed, the resulting variables are closed and agree with the original diagonal components on the Cartan.

We now proceed to letters $Y$ of higher weight.
Every lower-weight letter $X$ appearing in $R_{d,Y}$ already has a closed diagonal counterpart $\widetilde X_a$.
Replacing each $X_{aa}$ by $\widetilde X_a$ gives a closed expression $\widetilde R_{d,Y}$ with the same Cartan restriction.
We then define
\begin{equation}\label{eq:rank-closed-diagonal-definition}
    \sum_a\zeta_a^{d-1}\widetilde Y_a
    =\mathcal G_{d,Y}-\widetilde R_{d,Y}\,.
\end{equation}
The same linear systems determine the new variables.
Their right-hand sides are again closed, while on the Cartan the subtraction removes precisely $R_{d,Y}$.
Thus, at each weight,
\begin{equation}\label{eq:rank-closed-diagonal-properties}
    Q_0\widetilde Y_a=0,
    \qquad \rho_C(\widetilde Y_a)=Y_{aa}\,.
\end{equation}
The corrections $\widetilde Y_a-Y_{aa}$ may contain off-diagonal fields, but involve only lower-weight letters.
As with the off-diagonal redefinitions, we can therefore recover the original variables one weight at a time, with finitely many steps for any fixed expression.
The new diagonal variables are built from gravitons and $\Phi$ eigenvalues, with only eigenvalue differences in the denominators.

In the new variables, $Q_0$ acts only on the off-diagonal pairs and annihilates the corrected diagonal variables $\widetilde Y_a$.
The decoupled pairs can therefore be removed from the cohomology by subtracting a $Q_0$-exact term (as in Appendix 2.B of \cite{Barnich:2000zw}).
The remainder of any closed expression is obtained from its Cartan restriction by replacing each diagonal entry $Y_{aa}$ with its closed counterpart $\widetilde Y_a$.
Thus, once inverse eigenvalue differences are allowed, the cohomology is entirely described by the corrected diagonal variables, or equivalently by multi-graviton expressions with coefficients rational in the $\Phi$ eigenvalues.
After imposing residual gauge invariance, these coefficients can be written as rational functions of scalar traces.

We now return to the original polynomial multi-trace operator $F$, with $Q_0F=0$.
The change of variables leaves closed diagonal variables and decoupled off-diagonal $Q_0$ pairs.
The contractible-pair argument therefore shows that $F$ is cohomologous to its expression with both members of every pair set to zero, while keeping the corrected diagonal variables fixed.
Equivalently, there exists a rational operator $\widetilde H$ such that
\begin{equation}\label{eq:rank-rational-graviton-representative}
    F=\widetilde G+Q_0\widetilde H\,,
\end{equation}
where $\widetilde G$ is the remainder after the pairs are set to zero.
Because the corrected diagonal variables agree with the original ones on the Cartan, $\widetilde G$ is obtained from $\rho_C(F)$ by replacing each diagonal variable with its closed counterpart.
It follows that $\widetilde G$ is a finite sum of graviton products with coefficients rational in the $\Phi$ eigenvalues.
The equality holds in the enlarged space allowing inverse eigenvalue differences.
In particular, even though $F$ is polynomial, $\widetilde G$ and $\widetilde H$ separately may have poles where eigenvalues coincide.

Since $F$ is invariant and $Q_0$ commutes with the residual gauge symmetry, we may average $\widetilde G$ and $\widetilde H$ over these symmetries without changing the equality, including the additional reflection in the $SO(2n)$ trace sector.
The graviton products in $\widetilde G$ are themselves gauge invariant, so averaging acts only on their coefficients, which can then be written as rational functions of traces of powers of $\Phi$.
We emphasize that this rational description does not imply that $[F]$ has a multi-graviton representative in the original polynomial cohomology.
In particular, a Coulomb-visible fortuitous class can have a non-zero ``rational multi-graviton'' representative without admitting a polynomial one.
If $F$ is instead a non-Coulomb class, then $\rho_C(F)=0$ and hence $\widetilde G=0$ by construction.
Thus $F=Q_0\widetilde H$ is exact in the enlarged space, but remains non-exact in the original polynomial cohomology.

\subsubsection*{Scalar dressing and the non-Coulomb \texorpdfstring{$Q_1$}{Q1} image}

Let $\Delta$ be the discriminant of the characteristic polynomial of $\Phi$.
In terms of its eigenvalues,
\begin{equation}\label{eq:rank-localization-discriminant-cartan}
    \Delta=\prod_{a<b}(\zeta_a-\zeta_b)^2.
\end{equation}
This is a gauge-invariant polynomial in scalar traces and vanishes when eigenvalues coincide.
The finitely many denominators in both terms on the right-hand side of \eqref{eq:rank-rational-graviton-representative} can be cleared by a sufficiently high power of $\Delta$.
Since $Q_0\Delta=0$, we obtain a polynomial identity at the original fixed rank,
\begin{equation}\label{eq:rank-localization-clearing}
    \Delta^kF=G+Q_0H\,,
    \qquad Q_0F=0,
\end{equation}
where $G=\Delta^k\widetilde G$ and $H=\Delta^k\widetilde H$ are polynomial.
The coefficients of the graviton products in $G$ are now polynomials in scalar traces.
These traces are themselves gravitons, so $G$ is a polynomial multi-graviton expression.

If $\rho_C(F)=0$, we may take $\widetilde G=0$ as above and the statement reduces to
\begin{equation}\label{eq:rank-noncoulomb-annihilation}
    Q_0F=0,\quad \rho_C(F)=0
    \quad\Longrightarrow\quad
    \Delta^kF=Q_0H
\end{equation}
for sufficiently large $k$, whether $[F]$ is monotone or fortuitous.
For a Coulomb-visible class $\rho_C(F)\neq 0$, however, we have
\begin{equation}\label{eq:rank-visible-remainder}
    \rho_C(G)=\rho_C(\Delta)^k\rho_C(F)\neq0\,,
\end{equation}
since multiplication by a non-zero scalar polynomial cannot annihilate it.
Since $G$ has a non-vanishing Cartan restriction, it is a non-exact multi-graviton.
It follows that a discriminant dressing of the Coulomb-visible fortuitous operator $F_4$, for example, would make it monotone rather than zero.

We now apply $Q_1$ to \eqref{eq:rank-localization-clearing} and restrict to the Cartan.
The multi-graviton term obeys $\rho_C(Q_1G)=0$ by \eqref{eq:rank-multi-graviton-cartan}, while the exact term contributes zero because $\{Q_0,Q_1\}=0$ and $\rho_CQ_0=0$.
Since $\Delta^k$ contains only the scalar $\Phi$, its $Q_1$ image vanishes.
The restricted product rule \eqref{eq:rank-cartan-product-rule} therefore gives
\begin{equation}\label{eq:rank-cartan-discriminant}
    0=\rho_C\bigl(Q_1(\Delta^kF)\bigr)
     =\rho_C(\Delta)^k\,\rho_C(Q_1F)\, .
\end{equation}
As before, multiplication by the scalar $\rho_C(\Delta)^k\neq 0$ cannot annihilate $\rho_C(Q_1F)$, so this establishes for every polynomial multi-trace operator $F$ that
\begin{equation}\label{eq:rank-coulomb-image}
    Q_0F=0
    \quad\Longrightarrow\quad
    \rho_C(Q_1F)=0.
\end{equation}
Since exact changes of representative have zero Cartan restriction, every non-zero class in the image of $[Q_1]$ is non-Coulomb.
We stress that distinct eigenvalues were required only for the intermediate construction.
The resulting equations \eqref{eq:rank-localization-clearing} and \eqref{eq:rank-coulomb-image} are polynomial identities and are valid even when the eigenvalues coincide.

Combining this with the result of Section~\ref{ssec:rank-sources}, we conclude that every non-zero one-loop pairing in the multi-trace sector has a fortuitous source and a non-Coulomb target.
A Coulomb monotone is $[Q_1]$-closed and cannot be $[Q_1]$-exact, so it survives at this perturbative order.
Note that Coulomb visibility of an operator does not guarantee survival, since it can lift as a source.
Indeed, $F_4$ is Coulomb-visible and lifts as a source, with the non-Coulomb monotone $\mathcal{S}^4$ as its target.

\subsubsection*{Consequences for scalar dressings of descendants}

The discriminant relation also constrains scalar dressings of descendants.
A classical commutant generator $\mathcal D$ obeys the graded Leibniz rule and graded-commutes with $Q_0$, so it induces a graded derivation on $Q_0$-cohomology \cite{Budzik:2025zvu}.
It also preserves graviton multiplets and hence maps multi-graviton expressions to multi-graviton expressions.
Applying $\mathcal D$ to $[\Delta^kF]=[G]$, multiplying by $\Delta$ and using the original relation gives
\begin{equation}\label{eq:rank-descendant-dressing}
    [\Delta^{k+1}\mathcal DF]
    =
    [\Delta\,\mathcal DG-k(\mathcal D\Delta)G]\,.
\end{equation}
The right-hand side is a multi-graviton class, since $\Delta$ is itself a polynomial in scalar gravitons.
Repeating this argument for a descendant obtained with $s$ classical commutant generators gives
\begin{equation*}
    [\Delta^{k+s}\mathcal D_s\cdots\mathcal D_1F]=[G_s]\,,
\end{equation*}
where $G_s$ is a polynomial multi-graviton expression built from $\Delta$, $G$ and their commutant descendants.
For a non-Coulomb class, we may choose $G=0$, in which case every $G_s$ vanishes and the dressed descendants are $Q_0$-exact.

Under the proposed identification of fortuitous states with black hole microstates \cite{Chang:2024zqi}, this gives a partial no-hair result.
A fixed fortuitous class, or any fixed descendant, cannot generate an infinite tower of fortuitous classes through repeated scalar discriminant dressing.
After sufficiently many dressings, the resulting class is either multi-graviton or zero, depending on whether the original class was Coulomb visible or not.
This is reminiscent of the partial no-hair result of \cite{Choi:2023znd}, where products of the $SU(2)$ BMN threshold class with the scalar bilinears $\Tr(\phi^m\phi^n)$ were shown to be $Q_0$-exact after a single multiplication.
Our statement applies throughout the multi-trace sector of the classical families, including derivative letters, but allows a non-zero multi-graviton remainder and may require several powers of $\Delta$.
Whether similar restrictions hold for other graviton dressings remains an open question.

\newpage

\section{Conclusion and discussion}\label{sec:discussion}

We have studied non-Coulomb cohomology through the powers of the gaugino bilinear in the $SO(n)$ family, whose non-Coulomb character has a simple origin in Grassmann statistics.
To establish their non-exactness, we developed commuting root restrictions that retain information lost under Cartan restriction.
Using trace relations in the $SO(7)$ theory, we then constructed a fortuitous class $[F_4]$ that pairs with $\mathcal S^4$ under the one-loop supercharge $Q_1$.
We also observed that the class $[F_4]$ has a non-vanishing Cartan restriction, which remains modulo the addition of monotones, which demonstrates that the strict concept of fortuity does not imply non-Coulombness.\footnote{We comment on a stronger notion of fortuity, which would imply non-Coulombness, below.}
The identification of $\mathcal S^4$ as a superconformal descendant of the CLCL monotone connects this pairing to its previously established lifting \cite{Chang:2025mqp,Choi:2025bhi} and provides a transparent explanation of its non-Coulomb character.

Beyond this example, we have shown that every non-zero $[Q_1]$ pairing in the multi-trace sector of the finite rank classical gauge-group families has a fortuitous source and a non-Coulomb target, so Coulomb monotones survive at this perturbative order. 
Using a simple grading argument, our result also excludes the lifting of non-Coulomb BMN monotones, including those recently discovered for $SU(n)$ \cite{Kim:2026bhc,Fortuity:2026summary}.
The same argument shows that fortuitous operators in the BMN sector can only lift by failing to be $[Q_1]$-closed.
The non-Coulombness of $Q_1$ targets follows from \eqref{eq:rank-localization-clearing}.
At fixed rank, multiplication by a sufficiently high power of a scalar discriminant makes any classical class in the multi-trace sector cohomologous to a multi-graviton expression.
For a class with vanishing Cartan restriction, this expression can be chosen to vanish, giving a general partial no-hair result.
These results leave open several questions about the construction, quantum lifting and dressing of non-Coulomb classes.

\subsection*{Higher-rank quantum pairings}

The most immediate question is whether the pairing of $\mathcal S^4$ extends to the entire non-Coulomb gaugino bilinear tower at higher ranks, including possibly also their derivatives. 
As described in Section~\ref{ssec:higher-rank-predictions}, the remaining step for the powers of $\mathcal{S}$ (without derivatives) involves constructing the derivative completion $F_{m,\partial}$ of the proposed $F_{m,\psi}$ $Q_1$-source.
The latter already has the required $Q_1$ image, while the derivative terms are required to ensure $Q_0$-closedness through a trace relation.

A related question concerns the number of independent fortuitous sources in the relevant sector.
For $SO(7)$, the trace-relations quotient \eqref{eq:higher-rank-relation-quotient} is one-dimensional, so the fortuitous source class is unique up to addition of monotone classes.
It remains to determine whether this persists at higher rank.
The one-dimensionality of the pure-gaugino target cohomology implies that one can choose a basis in which at most one fortuitous class lifts as a source under $[Q_1]$.
Any additional classes must be annihilated by $[Q_1]$, but could still lift by being $[Q_1]$-exact, as targets of other pairings.

\subsection*{Coulomb-visible fortuity}

The class $F_4$ we constructed is clearly fortuitous, yet it has a non-zero Cartan restriction which cannot be cancelled by adding monotone classes.
On the other hand, the identity \eqref{eq:rank-rational-graviton-representative} expresses its class as a non-zero ``rational multi-graviton'', even if in ordinary cohomology it remains fortuitous.
In the conifold theory, there exist families of operators that are fortuitous in the ordinary polynomial description but admit a uniform construction across ranks using inverse scalar letters and baryonic factors \cite{Choi:2026faq}.
This motivates their interpretation as \emph{generalized monotones}, rather than \emph{strongly fortuitous} classes.
Our rational multi-graviton description of $F_4$ suggests a similar generalized monotone interpretation, which is further supported by the common ansatz $F_{m,\psi}=\mathcal S^{m-2}\mathcal O_\psi$ at higher powers and ranks in Section~\ref{ssec:higher-rank-predictions}.
The remaining step is to establish a uniform construction of the corresponding derivative completions $F_{m,\partial}$.

In contrast, non-Coulomb fortuitous classes disappear entirely in the rational graviton description, suggesting that they are natural candidates for \emph{strong fortuity}.
We leave a precise formulation of this distinction in the purely adjoint $\mathcal N=4$ theories considered here to future work.

\subsection*{Other non-Coulomb classes}

Non-Coulomb monotones also occur beyond the orthogonal theories.
Recent work has identified such classes in the BMN sectors of the $SU(3)$ and $SU(4)$ theories \cite{Kim:2026bhc}, as confirmed by and extended to $SO(5)$, $SO(7)$ and $USp(6)$ gauge groups \cite{Fortuity:2026summary}.
For unitary gauge algebras, $h-1=r$, so these examples cannot arise from the non-Coulomb tower built from $\mathcal S^m$.
They must instead involve other mechanisms by which a nontrivial multi-graviton class can vanish under Cartan restriction. 
The $SU(3)_R$ singlet in \cite{Kim:2026bhc} at $j=24$ in the $SU(3)$ theory is a particularly transparent example.
Its representative contains six scalars $\phi^n$ and three fermions $\psi_m$, so the $R$-singlet contraction necessarily includes an antisymmetrization of three scalar flavor indices.
On the $SU(3)$ Cartan, the three scalar flavor vectors formed from the diagonal entries sum to zero by tracelessness.
Their antisymmetric triple product therefore vanishes, as in the box-product argument of \cite{Gaikwad:2025ugk}.
This provides an alternative mechanism to the Grassmann-statistics explanation of the non-Coulomb powers of $\mathcal S$.
In both cases, the gauge rank limits the independent Cartan variables, while either the global-symmetry contraction or fermion
statistics forces the restricted operator to vanish.

The relation between $\mathcal S^4$ and the CLCL monotone shows that such a direct explanation of non-Coulombness can be more accessible in other members of the superconformal multiplet.
Cartan restriction commutes with the classical commutant action, so the non-Coulomb property passes to the entire multiplet.
It would be useful to identify similarly simple components of other non-Coulomb multiplets, allowing components outside the BMN truncation even when the original example lies within it.

Charge conjugation gives a further constraint in the unitary theories.
The authors of \cite{Kim:2026bhc} conjecture that fortuitous classes surviving quantum corrections have charge-conjugation parity opposite to gravitons in the corresponding charge sector.
For classical representatives containing $L$ letters, the graviton parity is $(-1)^L$ (see also \cite{Fortuity:2026summary}).
On the Cartan, charge conjugation changes the sign of every letter, so an operator with parity $(-1)^{L+1}$ necessarily has vanishing Cartan restriction.
The converse does not hold, as non-Coulomb monotones themselves demonstrate. 

Since $Q_1$ preserves both letter number and charge-conjugation parity, our source--target restrictions for the $Q_1$ action apply separately in the two parity sectors.
A pairing between a fortuitous source and a non-Coulomb monotone can only occur when the source has the graviton parity. 
In the opposite-parity sector, every non-zero pairing must join two fortuitous classes.
Pairing with non-Coulomb monotones thus provides one possible mechanism for removing fortuitous classes with graviton parity.
Our results do not require every such class to participate in a pairing, however, and opposite parity does not ensure survival. 
It remains to determine whether all fortuitous classes with graviton parity lift at one loop, or whether some survive this order and require a higher-order analysis.

More generally, it would be interesting to determine whether our restrictions on quantum pairings extend beyond one loop, and in particular whether Coulomb monotones remain protected.
Higher-loop corrections to the supercharge can act on more than two letters \cite{Budzik:2025zvu}, so the trace-joining argument requires further analysis.
A natural starting point is to investigate whether the higher-order contractions still vanish upon Cartan restriction, and whether the discriminant identity can then constrain the images to be non-Coulomb.

\subsection*{Restrictions on graviton dressings}

The discriminant relation \eqref{eq:rank-localization-clearing} shows that multiplying any classical class $F$ by a sufficiently high power of $\Delta$ gives a multi-graviton class.
For non-Coulomb classes this class is zero, whereas for Coulomb-visible classes such as $F_4$ it is non-zero.
This relation applies equally to any non-zero scalar combination $\Phi=v_m\phi^m$, yielding a family of discriminants with this property.
A natural next step is to determine the smallest power required for a given $F$ and find bounds in terms of the rank and its charges, starting with explicit constructions at low rank.

A related construction in the $SU(N)$ BMN sector \cite{deMelloKoch:2026BMNEveryN}, motivated by Higgsing, also uses rational expressions and denominator clearing.
There, $\Phi$ is also taken to be diagonal, but the off-diagonal components of the other letters within a chosen $2\times2$ block are not eliminated.
These form an $SU(2)$ non-abelian core, following the picture of \cite{deMelloKoch:2024pcs}, and only eigenvalue differences separating this block from the other colors enter the denominators.
Clearing these denominators then produces non-trivial fortuitous BMN classes for every $SU(N)$ with $N\geq3$.
We instead remove all off-diagonal $Q_0$ pairs by allowing for division by all eigenvalue differences of $\Phi$ and then clear the denominators with powers of the full scalar discriminant $\Delta$.
Sufficiently many such dressings therefore leave a multi-graviton class or zero, rather than a fortuitous class.

This distinction raises the question of which graviton dressings lead to exactness and which support infinite fortuitous towers.
An answer would sharpen the restrictions on graviton hair suggested by \cite{Choi:2023znd} and our discriminant hair.

\begin{acknowledgments}

We thank Chi-Ming Chang, Eunwoo Lee and Robert de Mello Koch  for valuable discussions from which this project originated, and Robert de Mello Koch for early collaboration on this project.
We are also grateful to Seok Kim and Siyul Lee for discussions, to Siyul Lee for collaboration on related work, and to Seunggyu Kim, Jehyun Lee, Siyul Lee, and Jaewon Song for bringing related results to our attention.
TK is supported by a Simons Foundation fellowship through the Targeted Grant to Instituto Balseiro. KM is supported by the South African Research Chairs Initiative of the Department of Science, Technology, and Innovation and the National Research Foundation Grant 78554. 
SvL acknowledges support from the National Research Foundation of South Africa (NRF) through the Research Development Grants for Y-Rated Researchers programme (Grant No. CSRP250405308980). 
Any opinion, finding, conclusion or recommendation expressed in this publication is that of the authors, and the NRF accepts no liability in this regard.
TK gratefully acknowledges the hospitality of the Mandelstam Institute for Theoretical Physics at the University of the Witwatersrand, where a part of this work was carried out during an academic visit.

During the preparation of this manuscript, the authors used LLMs to assist with editing, refining technical arguments and programming. Every mathematical and physical statement, calculation, and citation in the final manuscript was independently checked by the authors who take full responsibility for every aspect of this work.

\end{acknowledgments}


\begin{appendix}

\section{Lie algebra conventions}\label{app:lie-algebra-review}

This appendix reviews the aspects of Lie algebras used in the main text and fixes conventions.
Most of this material can be found in \cite[Ch.~13]{DiFrancesco:1997nk} and \cite{Humphreys:1980dw}.

\subsection{Root systems and the Weyl group}\label{app:root-system}

For a complex simple Lie algebra $\mathfrak g$ of dimension $d$ and rank $r$, fix a Cartan subalgebra $\mathfrak h\subset\mathfrak g$ and decompose
\begin{equation}
    \mathfrak g=\mathfrak h\oplus\bigoplus_{\alpha\in \Phi}\mathfrak g_\alpha \, ,
    \qquad
    \mathfrak g_\alpha=\bigl\{x\in\mathfrak g\,:\,[H,x]=\alpha(H)\,x\ \ \forall H\in\mathfrak h\bigr\} \, ,
    \label{eq:cartan-decomposition}
\end{equation}
where the root system $\Phi\subset\mathfrak h^{*}$ is the finite set of non-zero $\alpha$ for which $\mathfrak g_\alpha\neq0$.
Each root space is one dimensional, so that $|\Phi|=d-r$.
A choice of positive roots $\Phi_+$, equivalently of simple roots $\Pi=\{\alpha_1,\dots,\alpha_r\}$, fixes the decomposition
\begin{equation}
    \mathfrak g=\mathfrak h\oplus\mathfrak n^{+}\oplus\mathfrak n^{-} \, ,
    \qquad
    \mathfrak n^{\pm}=\bigoplus_{\alpha\in \Phi_{\pm}}\mathfrak g_\alpha \, ,
    \label{eq:triangular}
\end{equation}
and every root is an integer combination of simple roots whose coefficients are either all non-negative or all non-positive.

Let $\langle\,\cdot\,,\,\cdot\,\rangle: \mathfrak g \times \mathfrak g\longrightarrow\mathbb C $ be a non-degenerate bilinear invariant form.
It annihilates $\mathfrak g_\alpha\times\mathfrak g_\beta$ unless $\alpha+\beta=0$, pairs $\mathfrak g_\alpha$ with $\mathfrak g_{-\alpha}$, and identifies each root $\alpha$ with the element $t_\alpha\in\mathfrak h$ determined by
\begin{equation}
    \langle t_\alpha,H\rangle=\alpha(H) \, ,\qquad \forall\, H\in\mathfrak h \, .
    \label{eq:form-dual}
\end{equation}
The induced bilinear form on the roots is $(\alpha,\beta)=\langle t_\alpha,t_\beta\rangle$.
We normalize $\langle\,\cdot\,,\,\cdot\,\rangle$ so that long roots have length squared two, in terms of which the Killing form reads
\begin{equation}
    \operatorname{Tr}_{\rm ad}\bigl(\operatorname{ad}X\,\operatorname{ad}Y\bigr)=2h^{\vee}\,\langle X,Y\rangle \, ,
    \qquad X,Y\in\mathfrak g \, .
    \label{eq:form-normalisation}
\end{equation}
Using this normalization, the real span $\mathfrak h^*_{\mathbb R}$ of the roots is an $r$-dimensional Euclidean space with inner product $(\cdot,\cdot)$.

\begin{figure}[t]
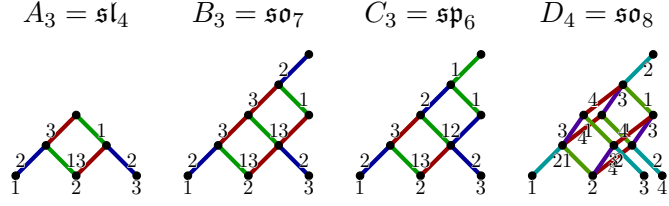

\centering
\begin{tabular}{cccc}
 $A_3=\mathfrak{sl}_{4}$ & $B_3=\mathfrak{so}_{7}$ & $C_3=\mathfrak{sp}_{6}$ & $D_4=\mathfrak{so}_{8}$\\[6pt]
\hasse[three D=false, edge length=.8cm]{A}{3} &
\hasse[three D=false, edge length=.8cm]{B}{3} &
\hasse[three D=false, edge length=.8cm]{C}{3} &
\hasse[three D=false, edge length=.8cm]{D}{4}
\end{tabular}
\caption{Hasse diagrams of the posets of positive roots: each node is a positive root, each edge adds the simple root labelling it, and the top node is the highest root $\theta$.
The height of $\theta$, the sum of its coefficients in simple roots, is ${\rm ht}(\theta)=h_{\rm Cox}-1$, and along any upward path from a simple root $\alpha_j$ to $\theta$ the label $i$ occurs $a_i-\delta_{ij}$ times.}
\label{fig:hasse}
\end{figure}

The co-root of $\alpha$ is $\alpha^{\vee}=2t_\alpha/(\alpha,\alpha)$, and the Weyl group $W$ is the group of symmetries of $\Phi$ generated by the reflections
\begin{equation}
    s_\alpha(x)=x-x(\alpha^{\vee})\,\alpha \, ,\qquad \alpha\in \Phi \, ,\quad x\in\mathfrak h^* \, .
    \label{eq:weyl-reflections}
\end{equation}
Writing the highest root and its co-root as $\theta=\sum_i a_i\alpha_i$ and $\theta^{\vee}=\sum_i a_i^{\vee}\alpha_i^{\vee}$, with $a_i$ the marks and $a_i^{\vee}$ the co-marks, the Coxeter and dual Coxeter numbers are
\begin{equation}
    h_{\rm Cox}=1+\sum_i a_i \, ,
    \qquad
    h^{\vee}=1+\sum_i a_i^{\vee} \, .
    \label{eq:coxeter-numbers}
\end{equation}
Since the highest root is long, $\theta^{\vee}=t_\theta=\sum_i a_i\,t_{\alpha_i}$, and comparing with the expansion of $\theta^\vee$ in simple co-roots gives
\begin{equation}
    a_i^{\vee}=\tfrac12 (\alpha_i,\alpha_i)\,a_i \, ,
    \label{eq:comark-from-mark}
\end{equation}
so that the co-mark of a long simple root equals its mark, while that of a short simple root is half of it, or a third for $G_2$.
The two numbers $h_{\rm Cox}$ and $h^{\vee}$ therefore agree for simply laced algebras but not in general.
For example, for $B_3$ the marks are $(1,2,2)$ and $\alpha_3$ is short, so the co-marks are $(1,2,1)$ and $h_{\rm Cox}=6$ while $h^{\vee}=5$.
Throughout the main text $h$ denotes the dual Coxeter number $h^{\vee}$.
See Figure~\ref{fig:hasse} for the Hasse diagrams of a few Lie algebras, which record the above data in a useful form.

For $A_r$ we use $e_1,\ldots,e_{r+1}$ with $(e_i,e_j)=\delta_{ij}$ and restrict to the hyperplane orthogonal to their sum. 
For $B_r$ and $D_r$ we use $r$ such orthonormal vectors, while for $C_r$ we take $(e_i,e_j)=\tfrac12\delta_{ij}$. 
In every case the long roots then have length squared two, and the coordinates below are the coefficients in the $e_i$ basis.
The classical root systems are then
\begin{equation}
\begin{aligned}
    A_{r}=\mathfrak{sl}_{r+1} \ &: \ e_i-e_j \, , \qquad && 1\leq i\neq j\leq r+1 \, ,\\
    B_{r}=\mathfrak{so}_{2r+1} \ &: \ \pm e_i\pm e_j \, ,\quad \pm e_i \, , && 1\leq i<j\leq r \, ,\\
    C_{r}=\mathfrak{sp}_{2r} \ &: \ \pm e_i\pm e_j \, ,\quad \pm2e_i \, , && 1\leq i<j\leq r \, ,\\
    D_{r}=\mathfrak{so}_{2r} \ &: \ \pm e_i\pm e_j \, , && 1\leq i<j\leq r \, ,
\end{aligned}
    \label{eq:classical-roots}
\end{equation}
where the signs are chosen independently, and for the single index roots, $i$ runs over $1,2,\dots,r$.
We call the coefficients $c_i$ in $\alpha=\sum_i c_i e_i$ the coordinates of a root.

The Weyl group \eqref{eq:weyl-reflections} acts on these coordinates in a simple way.
The reflection in $e_i-e_j$ exchanges $e_i$ and $e_j$, the reflection in $e_i$ or $2e_i$ flips the sign $e_i\to-e_i$, and the reflection in $e_i+e_j$ exchanges $e_i$ and $e_j$ while flipping both signs.
Hence $W(A_r)$ is the group of permutations of $e_1,\dots,e_{r+1}$, of order $(r+1)!$, and $W(B_r)=W(C_r)$ is the group of all signed permutations of $e_1,\dots,e_{r}$, of order $2^r r!$.
Since $D_r$ has no roots $e_i$ or $2e_i$, sign flips only occur in pairs, and $W(D_r)$ is the group of signed permutations with an even number of sign flips, of order $2^{r-1}r!$.
A single sign flip is nevertheless a symmetry of the $D_r$ root system, realized by an element of $O(2r)$ with determinant $-1$ rather than by the Weyl group.
This distinction is used in Section~\ref{ssec:comm-suports-son}.

\subsection{Cartan--Weyl basis}\label{app:cw-basis}

For the support restrictions of the main text it is convenient to rescale the generators so that the invariant form is the identity on $\mathfrak h$ and the root vectors are paired to unity.
A normalized Cartan--Weyl basis consists of an orthonormal basis $H_1,\dots,H_r$ of the real span $\mathfrak h_{\mathbb R}$ of the $t_\alpha$ together with root vectors $E_\alpha$, $\alpha\in \Phi$, rescaled so that
\begin{equation}
    \langle H_i,H_j\rangle=\delta_{ij} \, ,\qquad
    \langle E_\alpha,E_\beta\rangle=\delta_{\alpha+\beta,0} \, ,\qquad
    \langle H_i,E_\alpha\rangle=0 \, .
    \label{eq:cw-normalisation}
\end{equation}
Writing $T_a=(H_i,E_\alpha)$ for the resulting basis of $\mathfrak g$, the dual basis with respect to $\langle\cdot,\cdot\rangle$ is read off immediately from \eqref{eq:cw-normalisation},
\begin{equation}
    \langle T_a,T^b\rangle=\delta_a^{\,b} \, ,
    \qquad
    H^i=H_i \, ,
    \qquad
    E^\alpha=E_{-\alpha} \, .
    \label{eq:cw-dual-basis}
\end{equation}
It is this basis and dual basis that enter the paired support restrictions of the main text.
Note that $H_i$ denotes an orthonormal generator of $\mathfrak h$ and \emph{not} a simple co-root $\alpha_i^{\vee}$, as it would in a Chevalley basis.

In this basis the brackets are
\begin{equation}
\begin{gathered}
    [H_i,H_j]=0 \, ,
    \qquad
    [H_i,E_\alpha]=\alpha(H_i)\,E_\alpha \, ,\\[6pt]
    [E_\alpha,E_\beta]=
    \begin{cases}
        N_{\alpha\beta}\,E_{\alpha+\beta} \, , & \alpha+\beta\in \Phi \, ,\\[2pt]
        t_\alpha \, , & \beta=-\alpha \, ,\\[2pt]
        0 \, , & \text{otherwise} \, ,
    \end{cases}
\end{gathered}
    \label{eq:cw-brackets}
\end{equation}
with $t_\alpha$ as in \eqref{eq:form-dual} and with structure constants $N_{\alpha\beta}\neq0$ whenever $\alpha+\beta\in \Phi$.
The feature of \eqref{eq:cw-brackets} that is used repeatedly in the main text is
\begin{equation}
    [E_\alpha,E_\beta]=0
    \qquad\Longleftrightarrow\qquad
    \alpha+\beta\notin \Phi\cup\{0\} \, .
    \label{eq:commuting-roots}
\end{equation}
The search for a set of mutually commuting root vectors thus becomes a combinatorial question about sums of roots.

\subsection{Explicit basis for \texorpdfstring{$\mathfrak{so}_n$}{so(n)}}\label{app:son}

Let $\mathfrak g=\mathfrak{so}_n$ with $n\geq5$, so that the defining representation has dimension $n$ and
\begin{equation}
    r={\rm rk}(\mathfrak g)=\Bigl\lfloor \frac n2\Bigr\rfloor \, ,
    \qquad
    \dim\mathfrak g=\tfrac12 n(n-1) \, ,
\end{equation}
with $\mathfrak{so}_{2r+1}=B_r$ and $\mathfrak{so}_{2r}=D_r$.
We realize $\mathfrak g$ by choosing the symmetric bilinear form $\Xi$ (see \eqref{eq:so(n)-bil-form}) in a basis $v_1,\dots,v_n$ of the defining representation in which the symmetric form pairs $v_{2i-1}$ with $v_{2i}$,
\begin{equation}
    \Xi=\sum_{i=1}^{r}\bigl(\mathsf E_{2i-1,2i}+\mathsf E_{2i,2i-1}\bigr)+(n-2r)\,\mathsf E_{nn} \, ,
    \label{eq:so-split-form}
\end{equation}
where $\mathsf E_{ab}$ is the matrix unit with a single entry $1$ in row $a$ and column $b$, and the last term, present only for $n$ odd, pairs $v_n$ with itself.
The Cartan subalgebra then consists of the diagonal matrices in $\mathfrak g$, and we take the orthonormal generators of \eqref{eq:cw-normalisation} to be
\begin{equation}
    H_i=\mathsf E_{2i-1,2i-1}-\mathsf E_{2i,2i} \, ,
    \qquad i=1,\dots,r \, .
    \label{eq:so-cartan}
\end{equation}
Let $e_1,\dots,e_r\in\mathfrak h^{*}$ be the dual basis,
\begin{equation}
    e_j(H_i)=\delta_{ij} \, ,
    \qquad\text{so that}\qquad
    (e_i,e_j)=\delta_{ij} \, ,
    \label{eq:e-basis}
\end{equation}
and $v_{2i-1}$, $v_{2i}$ and, for $n$ odd, $v_n$ carry the weights $e_i$, $-e_i$ and $0$.
To the root systems of $B_r$ and $D_r$ in \eqref{eq:classical-roots} we assign the root vectors
\begin{equation}
\begin{gathered}
    E_{e_i-e_j}=\mathsf E_{2i-1,2j-1}-\mathsf E_{2j,2i} \, ,
    \qquad
    E_{e_i+e_j}=\mathsf E_{2i-1,2j}-\mathsf E_{2j-1,2i} \, ,
    \\
    E_{e_i}=\mathsf E_{2i-1,n}-\mathsf E_{n,2i} \, ,
\end{gathered}
    \label{eq:so-root-vectors}
\end{equation}
together with $E_{-\alpha}=E_\alpha^{T}$, where $E_{e_i}$ occurs only for $n$ odd.
In the defining representation $\Tr(XY)=2\langle X,Y\rangle$, so that \eqref{eq:so-cartan} and \eqref{eq:so-root-vectors} form a normalized Cartan--Weyl basis, with $[E_\alpha,E_{-\alpha}]=t_\alpha$ and $\Tr(E_\alpha E_{-\alpha})=2$, which is the normalization used in Section~\ref{ssec:examples}.

A convenient set of simple roots is $\alpha_i=e_i-e_{i+1}$ for $i<r$, together with $\alpha_r=e_r$ for $B_r$ and $\alpha_r=e_{r-1}+e_r$ for $D_r$.
The positive roots are then $e_i\pm e_j$ with $i<j$, together with $e_i$ for $B_r$.
The highest root and its co-root are
\begin{equation}
    \theta=e_1+e_2 \, ,
    \qquad
    (\theta,\theta)=2 \, ,
    \qquad
    \theta^{\vee}=t_\theta=H_1+H_2 \, ,
    \label{eq:so-highest-root}
\end{equation}
and the dual Coxeter number takes the uniform value
\begin{equation}
    h\equiv h^{\vee}=n-2
    \label{eq:so-dual-coxeter}
\end{equation}
for both series.

\section{Review of Etingof's proof of the CDSW conjecture}\label{app:etinghof-proof}

This appendix reviews the structure of Etingof's proof of the CDSW conjecture and its relation to the restriction method of Section~\ref{sec:nC-from-S}.
The common starting point is to impose the pure relations $\{\lambda_+,\lambda_+\}=\{\lambda_-,\lambda_-\}=0$ in the derivative-free gaugino ring, and only then impose the mixed relation $\{\lambda_+,\lambda_-\}=0$.
Etingof and Kac introduced this approach and proved the $G_2$ case \cite{Etingof:2003dd}, while Etingof subsequently developed it into an algebraic proof for the classical Lie algebras \cite{Etingof:2003fy}.

In the notation of Section~\ref{ssec:gaugino-chiral-ring}, the classical chiral ring is $R^{\mathfrak g}/I^{\mathfrak g}$, and the nilpotence statement of the CDSW conjecture is \eqref{eq:cdsw-conj}.
The algebra $R$ factorizes as $R=\Lambda[x]\otimes\Lambda[y]$, where $\Lambda[x]=\bigoplus_{n_+} R_{(n_+,0)}$ is generated by $x^a$, corresponding to the components of $\lambda_+$, and $\Lambda[y]=\bigoplus_{n_-} R_{(0,n_-)}$ is generated by $y_a$, the components of $\lambda_-$.
The pure relations each lie in one of these factors, and we define the corresponding quotients
\begin{equation}\label{eq:B-definition}
    B_+=\bigoplus_{n_+} R_{(n_+,0)}/I_{(n_+,0)} \, ,
    \qquad
    B_-=\bigoplus_{n_-} R_{(0,n_-)}/I_{(0,n_-)} \, .
\end{equation}
Here $I_{(n_+,0)}$ is the degree-$p$ part of the ideal generated by the components of $\{\lambda_+,\lambda_+\}$ in $\Lambda[x]$, and similarly for $I_{(0,n_-)}$ in $\Lambda[y]$.
Thus $B_+$ is the algebra of polynomials in $x^a$ subject to $\lambda_+^2=0$, and $B_-$ is the analogous algebra in $y_a$.
The quadratic generators of the pure ideals transform in the adjoint, so the $\mathfrak g$ action preserves each ideal and therefore descends to the quotients $B_\pm$.
Imposing the pure relations on $R$ gives $B_+\otimes B_-$.
Let us denote the remaining ideal inside $B_+\otimes B_-$ by $\bar I$, which is generated by the components of the mixed relation.
Projecting onto $\mathfrak{g}$-invariants maps each ideal into itself and therefore commutes with these quotients, so that
\begin{equation}\label{eq:E-quotient}
    R^{\mathfrak g}/I^{\mathfrak g}\simeq E/\bar I^{\mathfrak g} \, ,
    \qquad
    E=\bigl(B_+\otimes B_-\bigr)^{\mathfrak g} \, .
\end{equation}
The argument therefore first determines $B_\pm$, then their invariants $E$, and finally the effect of the mixed relation.

\subsubsection*{Pure relations}

Kostant's theorem \cite{Kostant_2004,Kostant:abelian} describes the full $\mathfrak g$-module structure of $B_\pm$ in terms of abelian ideals of a Borel subalgebra.
Let $\mathfrak b$ be the Borel subalgebra spanned by the $H_i$ and the $E_\alpha$ with $\alpha>0$.
An abelian ideal of $\mathfrak b$ is an abelian subalgebra $\mathfrak a\subset\mathfrak b$ with $[\mathfrak b,\mathfrak a]\subseteq\mathfrak a$, and it is spanned by the $E_\alpha$ for a set of positive roots, which we also denote by $\mathfrak a$.
Each such $\mathfrak a$ supplies the two monomials
\begin{equation}\label{eq:hw-vector}
    v^+_{\mathfrak a}=\prod_{\alpha\in\mathfrak a}x^\alpha\in B_+ \, ,
    \qquad
    v^-_{\mathfrak a}=\prod_{\alpha\in\mathfrak a}y_\alpha\in B_- \, .
\end{equation}
Under the induced action of $\mathfrak{g}$, $x^\alpha$ has weight $-\alpha$ and $y_\alpha$ has weight $\alpha$.
Because $\mathfrak a$ is an ideal of $\mathfrak b$, $v^-_{\mathfrak a}$ is a highest-weight vector of weight $\sum_{\alpha\in\mathfrak a}\alpha$ and $v^+_{\mathfrak a}$ a lowest-weight vector of the opposite weight.
Indeed, $E_\beta$ with $\beta>0$ acts on $y_\alpha$ by a term proportional to $y_{\alpha+\beta}$, and $E_{-\beta}$ acts on $x^\alpha$ by a term proportional to $x^{\alpha+\beta}$.
If $\alpha+\beta$ is not a root, this term is absent, otherwise the ideal property puts $\alpha+\beta$ in $\mathfrak a$, so the product contains a repeated Grassmann variable and vanishes.

Kostant's theorem states that these vectors account for everything \cite{Kostant_2004,Kostant:abelian},
\begin{equation}\label{eq:B-as-Va}
    B_-\simeq\bigoplus_{\mathfrak a}V_{\mathfrak a} \, ,
    \qquad
    B_+\simeq\bigoplus_{\mathfrak a}V_{\mathfrak a}^* \, ,
\end{equation}
where $V_{\mathfrak a}$ is the irreducible module generated by $v^-_{\mathfrak a}$ and the decomposition is multiplicity free.
The sum runs over the $2^r$ abelian ideals of $\mathfrak b$, including the trivial ideal \cite{Kostant:1998abelian,Cellini:2000ad}.

\subsubsection*{Invariants}

Since \eqref{eq:B-as-Va} is multiplicity free, Schur's lemma leaves exactly one invariant for each abelian ideal,
\begin{equation}\label{eq:E-decomposition}
    E=\bigoplus_{\mathfrak a}\bigl(V_{\mathfrak a}^*\otimes V_{\mathfrak a}\bigr)^{\mathfrak g} \, ,
    \qquad
    \dim E=2^r \, ,
\end{equation}
namely the canonical invariant in $V_{\mathfrak a}^*\otimes V_{\mathfrak a}$, of bidegree $(|\mathfrak a|,|\mathfrak a|)$.
Hence $E$ is concentrated on the diagonal, with top bidegree $(\max|\mathfrak a|,\max|\mathfrak a|)$, where the largest abelian ideal of $\mathfrak b$ has the dimension of the largest abelian subalgebra of $\mathfrak g$, as tabulated for the classical algebras in Table~\ref{tab:classical-lie-data}.
For the abelian ideal $\mathbb C E_\theta$ with $\theta$ the highest root, the highest-weight vector is $y_\theta$, of weight $\theta$, so that $V_{\mathbb C E_\theta}$ is the adjoint, realized by the span of the $y_a$, and its dual by the span of the $x^a$.
The canonical invariant pairing the two is $x^ay_a$, proportional to the gaugino bilinear $\mathcal S=\Tr(\lambda_+\lambda_-)$.
It follows that $\mathcal S^p$ vanishes in $E$ for $p>\max|\mathfrak a|$, before the mixed relation is imposed.

\subsubsection*{Mixed relation}

In bidegree $(1,1)$, $\bar I$ is spanned by the components of $\{\lambda_+,\lambda_-\}$, which transform in the adjoint and therefore contain no invariant.
Thus $\bar I^{\mathfrak g}_{(1,1)}=0$ and $\mathcal S$ is non-trivial.
Determining the mixed ideal in higher bidegrees requires an understanding of the ring structure of $E$, rather than just its module decomposition.

For the classical algebras, Etingof constructs algebra generators of $E$ from the basic invariant polynomials of $\mathfrak g$ \cite{Etingof:2003fy}.
The generator associated with the quadratic invariant is proportional to $\mathcal S$, while the remaining generators generate $\bar I^{\mathfrak g}$.
A comparison of graded dimensions shows that there are no relations among these generators below bidegree $(h,h)$, which ensures that $\mathcal S^{h-1}$ survives the quotient.
Together with the explicit relation $\mathcal S^h=0$ established in \cite{Witten:2003ye}, this gives
\begin{equation}
    E/\bar I^{\mathfrak g}\simeq\mathbb C[\mathcal S]/(\mathcal S^h) \, .
\end{equation}
When $\max|\mathfrak a|>h-1$, the mixed quotient therefore imposes a stronger truncation than the degree bound supplied by the pure relations.

\subsubsection*{Comparison with the restriction method}

The restriction method also separates the pure and mixed relations.
In its first step, Section~\ref{ssec:strong-comm-vanish-ideal}, we solve for the pure relations by restricting the gauginos to take value in a commuting support $J$, satisfying \eqref{eq:lower-lower-comm}.
This restriction $\rho_J$ annihilates $\{\lambda_+,\lambda_+\}$, so that $\prod_{a\in J}x^a$ is non-zero in $B_+$ by \eqref{eq:restriction-contrapositive}, and likewise $\prod_{a\in J}y_a$ in $B_-$.
Note that these supports are only required to be abelian, and do not have to be ideals of $\mathfrak b$ as well.
The remaining exact expressions restrict to the alternating multi-traces $\mathcal T_\nu$, with at least one factor $\mathcal T_m$ for $m>1$.
These are precisely the generators of the mixed invariant ideal $\bar I^{\mathfrak g}$.
Rather than reconstructing $E$ and its quotient by $\bar I^{\mathfrak g}$, we compare their restrictions with that of $\mathcal S^{h-1}$ on a family of supports.
Whenever this separates $\mathcal S^{h-1}$ from the residual ideal, it demonstrates the non-triviality of the class without determining the full invariant algebra.

\subsection{Example: \texorpdfstring{$\mathfrak{sl}_3$}{sl3}}

Here $r=2$ and $h=3$.
Writing $\alpha_1,\alpha_2$ for the simple roots $e_1-e_2$ and $e_2-e_3$, and $\theta=\alpha_1+\alpha_2$, the four abelian ideals of $\mathfrak b$ are
\begin{equation}
    0\,,\qquad \mathbb C E_\theta\,,\qquad
    \operatorname{span}\{E_{\alpha_1},E_\theta\}\,,\qquad
    \operatorname{span}\{E_{\alpha_2},E_\theta\}\,.
\end{equation}
Their highest weights are $0$, $\theta$, $2\alpha_1+\alpha_2$ and $\alpha_1+2\alpha_2$, giving the trivial, adjoint and two dual ten-dimensional representations.
Thus \eqref{eq:B-as-Va} gives
\begin{equation}
    B_\pm\simeq\mathbf 1_{(0)}\oplus\mathbf 8_{(1)}
    \oplus\bigl(\mathbf{10}\oplus\overline{\mathbf{10}}\bigr)_{(2)}\,,
\end{equation}
where the subscripts indicate degree.
In particular, $B_\pm$ vanish above degree two, so $\mathcal S^3=0$ already after imposing the pure relations.

Taking invariants gives a four-dimensional $E$, with one invariant in each of bidegrees $(0,0)$ and $(1,1)$ and two in bidegree $(2,2)$.
For $\mathfrak{sl}_3$, the basic invariant polynomials have degrees two and three.
Etingof's construction associates to them algebra generators of $E$ in bidegrees $(1,1)$ and $(2,2)$, which can be chosen as
\begin{equation}
    \mathcal S\,,\qquad
    \mathcal T_2=\Tr\bigl((\lambda_+\lambda_-)^2\bigr)
    \,.
\end{equation}
Since these generate $E$ and its $(2,2)$ component is two-dimensional, $\mathcal S^2$ and $\mathcal T_2$ must be linearly independent.
Thus a basis is $1,\mathcal S,\mathcal S^2,\mathcal T_2$.

The mixed invariant ideal is generated by the generator associated with the cubic invariant, namely $\mathcal T_2$ \cite{Etingof:2003fy}.
Since $E$ vanishes above bidegree $(2,2)$, this ideal is simply
\begin{equation}
    \bar I^{\mathfrak{sl}_3}=\mathbb C\,\mathcal T_2\,.
\end{equation}
Consequently, $\mathcal S^2$ survives the mixed quotient, while $\mathcal S^3=0$ already follows from the pure relations, since $B_\pm$ have no components above degree two.
Thus, one finds
\begin{equation}
    E/\bar I^{\mathfrak{sl}_3}
    \simeq\mathbb C[\mathcal S]/(\mathcal S^3)\,.
\end{equation}
The mixed relation therefore removes one of the two invariants in bidegree $(2,2)$, but does not affect the nilpotence bound of $\mathcal{S}$ deduced from the pure relations.

\subsection{Example: \texorpdfstring{$\mathfrak{so}_7$}{so7}}

Here $r=3$ and $h=5$.
The eight abelian ideals are given by
\begin{equation}
\begin{aligned}
    \mathfrak a_0&=\varnothing,
    &\qquad \mathfrak a_1&=\{e_1+e_2\},\\
    \mathfrak a_2&=\{e_1+e_2,e_1+e_3\},
    &\qquad \mathfrak a_3^{(1)}&=\{e_1+e_2,e_1+e_3,e_2+e_3\},\\
    \mathfrak a_3^{(2)}&=\{e_1+e_2,e_1+e_3,e_1\},
    &\qquad \mathfrak a_4^{(1)}&=\{e_1+e_2,e_1+e_3,e_1-e_3,e_1\},\\
    \mathfrak a_4^{(2)}&=\{e_1+e_2,e_1+e_3,e_2+e_3,e_1\},
    &\qquad \mathfrak a_5&=\{e_1,e_1\pm e_2,e_1\pm e_3\}.
\end{aligned}
\end{equation}
where the subscript indicates the dimension and the simple roots are as in Appendix \ref{app:son}.
Consequently, we find
\begin{equation}
\begin{aligned}
    B_\pm\simeq{}&
    \mathbf1_{(0)}\oplus\mathbf{21}_{(1)}
    \oplus\mathbf{189}_{(2)}
    \oplus(\mathbf{294}\oplus\mathbf{616})_{(3)}
    \oplus(\mathbf{819}\oplus\mathbf{1386})_{(4)}\oplus\mathbf{378}_{(5)}\,,
\end{aligned}
\end{equation}
where the subscripts indicate degree.
We see that $B_\pm$ vanish above degree five, so the pure relations already imply $\mathcal S^6=0$.

Taking invariants gives an eight-dimensional $E$, with graded dimensions
\begin{equation}
    \sum_{p\geq0}\dim E_{(p,p)}\,t^p
    =1+t+t^2+2t^3+2t^4+t^5\,.
\end{equation}
The basic invariant polynomials of $\mathfrak{so}_7$ have degrees two, four and six.
Etingof's corresponding generators of $E$ can be chosen as
\begin{equation}
    \mathcal S\,,\qquad \mathcal T_3\,,\qquad \mathcal T_5\,,
\end{equation}
of bidegrees $(1,1)$, $(3,3)$ and $(5,5)$, respectively \cite{Etingof:2003fy}.
In particular, the basis of the two-dimensional $(4,4)$ comprises of $\mathcal S^4$ and $\mathcal S\mathcal T_3$.

The mixed invariant ideal is generated by $\mathcal T_3$ and $\mathcal T_5$.
For $h=5$, Etingof's proposition 2.12 gives a relation in $E$ of the form
\begin{equation}
    c\,\mathcal S^5+a\,\mathcal S^2\mathcal T_3
    +b\,\mathcal T_5=0\,,
    \qquad c\neq0
\end{equation}
and his proposition 2.6 states that this is the first such relation.
It follows that $\mathcal S^5$ vanishes in the mixed quotient.
Moreover, in bidegree $(4,4)$, the ideal removes $\mathcal S\mathcal T_3$ but not the independent invariant $\mathcal S^4$.
Consequently,
\begin{equation}
    E/\bar I^{\mathfrak{so}_7}
    \simeq\mathbb C[\mathcal S]/(\mathcal S^5)\,.
\end{equation}
Here, we see that the mixed relation reduces the nilpotence bound further than the pure relations, in contrast with the $\mathfrak{sl}_3$ example.

\section{Commuting supports in \texorpdfstring{$\mathfrak{so}_n$}{so(n)}}\label{app:comm-supports}

In this appendix, we provide a general construction of commuting subsets of cardinality $\geq h-1$, as used in Section~\ref{ssec:comm-suports-son}, for $\mathfrak{so}_n$ with $n\geq7$, $r=\lfloor n/2\rfloor$ and $h=n-2$.
Note that these subsets are not strongly commuting, \textit{i.e.}, $[T_a,T_b]=[T^a,T^b]=0$ for any pair $a,b$ in our sets, but $[T_a,T^b]\neq 0$ in general, where we work in the Cartan-Weyl basis of the Lie algebra (see Appendix \ref{app:lie-algebra-review}).

To this end, we recall from \eqref{eq:classical-roots} that the roots of $\mathfrak{so}_{n}$ are $\pm e_{i}\pm e_{j}$, $i<j\le r$, together with $\pm e_{i}$ when $n$ is odd, so that every root has at most two non-vanishing coordinates, each equal to $\pm1$.
Now, fix a nonempty subset $K\subseteq\{1,\dots,r\}$ of size $k$ and let $H_{K}\in\mathfrak{h}$ be the Cartan element for which
\begin{equation}
  e_{j}(H_{K})=\begin{cases}1,& j\in K,\\ 0,& j\notin K.\end{cases}
  \label{eq:K-cartan}
\end{equation}
The adjoint action of $H_{K}$ grades $\mathfrak{g}$,
\begin{equation}
  \mathfrak{g}=\bigoplus_{a=-2}^{2}\mathfrak{g}_{a} \, ,\qquad
  \mathfrak{g}_{a}=\{X\in\mathfrak{g}\,:\,[H_{K},X]=aX\} \, ,
  \label{eq:K-decomposition}
\end{equation}
where only these degrees occur, since $\alpha(H_K)\in\{0,\pm1,\pm2\}$ for every root $\alpha$.
As we now show, abelian subalgebras are naturally built from the positive graded pieces.
We first carry this out for general $k$, and then return to $k=2$, which is the simplest case and admits a more familiar description related to Heisenberg subalgebras of $\mathfrak{g}$.

\subsubsection*{Level \texorpdfstring{$k$}{k} construction}

The positive graded pieces are
\begin{equation}
  \mathfrak{g}_{2}=\operatorname{span}\bigl\{E_{e_{i}+e_{j}}\bigr\}_{i<j\in K} \, ,
  \qquad
  \mathfrak{g}_{1}=\operatorname{span}\bigl\{E_{e_{i}\pm e_{l}}\,,\;E_{e_{i}}\bigr\}_{i\in K,\,l\notin K} \, ,
  \label{eq:degree-pieces}
\end{equation}
where the short-root vectors $E_{e_i}$ occur only for odd $n$.
Hence $\mathfrak{g}_{2}$ has dimension $\binom{k}{2}$ and $\mathfrak{g}_1$ has dimension $k(n-2k)$.
Indeed, $\mathfrak g_1$ contains the $2k(r-k)$ roots $e_i\pm e_l$ with $i\in K$ and $l\notin K$, and for odd $n$ the $k$ short roots $e_i$, giving a total of $k(2(r-k)+[\,n\text{ odd}\,])=k(n-2k)$, where $[\,n\text{ odd}\,]$ is $1$ for odd $n$ and $0$ otherwise, so that $n=2r+[\,n\text{ odd}\,]$.

The Lie bracket adds degrees,
\begin{equation}\label{eq:additive-bracket}
  [\mathfrak g_a,\mathfrak g_b]\subseteq \mathfrak g_{a+b}\,,
\end{equation}
so that, since there are no roots of degree $3$ or $4$,
\begin{equation}
  [\mathfrak{g}_{2},\mathfrak{g}_{2}]=[\mathfrak{g}_{2},\mathfrak{g}_{1}]=0 \, ,\qquad
  [\mathfrak{g}_{1},\mathfrak{g}_{1}]\subseteq\mathfrak{g}_{2} \, .
  \label{eq:two-step}
\end{equation}
We may therefore keep all of $\mathfrak g_2$, and only have to choose a set $L$ of degree-one roots whose root vectors commute, after which we set
\begin{equation}
  J_K(L) = \{e_{i}+e_{j}\}_{i<j\in K}\;\cup\;L \,.
  \label{eq:graded-support}
\end{equation}
Such sets $L$ are easily described.
First, if $\alpha=e_{i}\pm e_{l}$ and $\beta=e_{j}\pm e_{l'}$ with $i,j\in K$ and $l\neq l' \notin K$, then $\alpha+\beta$ is not a root, since it has at least three non-vanishing coordinates or a coordinate equal to $\pm2$.
The same holds if $\beta=e_{j}$ is short.
So $L$ can be decomposed into the blocks
\begin{equation}
  L=\bigcup_{l\notin K}L_{l}\;\cup\;L_{0} \, ,\qquad
  L_{l}\subseteq\{e_{i}\pm e_{l}: i\in K\} \, ,\qquad
  L_{0}\subseteq\{e_{i}: i\in K\} \, ,
  \label{eq:blocks}
\end{equation}
where $L_0=\varnothing$ for even $n$.
Within a block, $(e_{i}+\varepsilon e_{l})+(e_{j}+\varepsilon'e_{l})=e_{i}+e_{j}+(\varepsilon+\varepsilon')e_{l}$ is a root precisely when $i\neq j$ and $\varepsilon'=-\varepsilon$.
Hence $L_{l}$ is abelian if and only if one of the following holds:
\begin{enumerate}
    \item all roots of $L_l$ carry the same sign of $e_l$, as for instance
    \begin{equation}
        L_l=\{\,e_i+e_l,\;e_j+e_l\,\}\quad\text{with } i\neq j\in K \, ;
        \label{eq:different-index-pair}
    \end{equation}
    \item $L_l$ contains both signs, in which case it is the pair
    \begin{equation}
        L_{l}=\{\,e_{i}+e_{l},\;e_{i}-e_{l}\,\}\quad\text{for a single }i\in K \, .
        \label{eq:same-index-pair}
    \end{equation}
\end{enumerate}
On the other hand, $L_{0}$ can contain at most one root, since $e_i+e_j$ is a root for $i\neq j$.
For the family used in the main text, we keep two roots from each block $L_l$ and, when $n$ is odd, one short root.

It is convenient to name the possibilities for the choice of two roots per block.
We say that a block $L_{l}$ is of \emph{type $i$} when it is the pair \eqref{eq:same-index-pair}, and of \emph{type $ij$} when it consists of $e_{i}+\varepsilon e_{l}$ and $e_{j}+\varepsilon e_{l}$ with $i\neq j$ and a common sign $\varepsilon$, as in \eqref{eq:different-index-pair}.
Similarly, we say $L_{0}=\{e_{i}\}$ is of type $i$.
A choice of $L$ is then specified by a type for each of the $r-k$ blocks, together with a type for $L_{0}$ when $n$ is odd.
The cardinality of the resulting set is
\begin{equation}
  |J_{K}(L)| \;=\; \binom{k}{2}+2(r-k)+[\,n\text{ odd}\,] \;=\; \binom{k}{2}+(n-2k) \, ,
  \label{eq:graded-cardinality}
\end{equation}
so that
\begin{equation}
  |J_{K}(L)|-(h-1)\;=\;\frac{(k-2)(k-3)}{2} \,.
  \label{eq:graded-excess}
\end{equation}
Such a support therefore always satisfies $|J_{K}(L)|\geq h-1$, as required.
For $k=2$ and $k=3$ it is a minimal support, while for $k=1$ and $k>3$ the set $J_{K}(L)$ serves as the larger support from which the subsets in \eqref{eq:chosen-subsets} are drawn, each of which is again commuting.
The extreme case $k=r$ reproduces the two cardinalities entering the $\max|\mathfrak{a}|$ column of Table~\ref{tab:classical-lie-data}.\footnote{The other extreme value is $k=1$, where there is no degree-two part and $J_{K}(L)$ consists of the roots $e_{1}\pm e_{l}$ with $l=2,\dots,r$, together with $e_{1}$ for odd $n$, of cardinality $n-2=h$. For odd $n$ this is the remaining entry $2r-1$ of the table.}
Indeed, in this case there are no blocks, and $J_{K}(L)$ consists of the roots $e_{i}+e_{j}$ with $i<j$, together with one short root for odd $n$, of cardinality $r(r-1)/2$ for even $n$ and $r(r-1)/2+1$ for odd $n$.
The intermediate values of $k$ interpolate between the two.

The data specifying $L$ is conveniently represented by the graphs of Figure~\ref{fig:multigraphs}.
A block of type $i$ is drawn as a loop at the vertex $i$, a block of type $ij$ as an edge joining $i$ and $j$, and $L_{0}=\{e_{i}\}$ marks the vertex $i$.
Each of the $r-k$ blocks thus contributes one loop or one edge, which carries its own index $l\notin K$, and the roots $\{e_{i}+e_{j}\}_{i<j\in K}$ of \eqref{eq:graded-support}, being present for every choice of $L$, are left undrawn.
The common sign of a block of type $ij$ is not recorded either, since it can be made positive by the sign flip $e_l\mapsto-e_l$, which is realized by an element of $O(n)$ and leaves the trace invariants unchanged (Appendix~\ref{app:root-system}).

\subsubsection*{Level two and the highest root}

After a permutation of the coordinates, $K=\{1,2\}$, so that $H_{K}=\theta^{\vee}$ is the coroot of the highest root $\theta=e_{1}+e_{2}$ of \eqref{eq:so-highest-root}, and $\mathfrak{g}_{2}=\mathbb CE_\theta$ is one dimensional.
The elements $\theta^\vee$ and $E_{\pm \theta}$ generate an $\mathfrak{sl}_2$ subalgebra inside $\mathfrak g_{-2}\oplus\mathfrak g_0\oplus\mathfrak g_2$.
Since $\theta^\vee$ acts on $\mathfrak g_{\pm1}$ with eigenvalue $\pm1$, these two graded pieces combine into two-dimensional $\mathfrak{sl}_2$ modules.
As $\mathfrak g_2$ is one dimensional, the bracket $[\mathfrak g_1,\mathfrak g_1]\subseteq\mathfrak g_2$ of \eqref{eq:two-step} defines a skew-symmetric form on $\mathfrak g_1$,
\begin{equation}
  [X,Y]=\omega_\theta(X,Y)E_\theta\,,
  \qquad X,Y\in\mathfrak g_1\,.
\end{equation}
This form is non-degenerate, since the roots in $\mathfrak g_1$ appear in complementary pairs
\begin{equation}
  \{\alpha\,,\theta-\alpha\}\,, \qquad [E_\alpha,E_{\theta-\alpha}]=N_{\alpha,\theta-\alpha}E_\theta\,,\quad N_{\alpha,\theta-\alpha}\neq0\,,
\end{equation}
while the root vectors of any other two degree-one roots commute.
This implies that $\mathfrak g_1\oplus\mathfrak g_2$ is a Heisenberg algebra with center $\mathfrak g_2$.
A polarization of this Heisenberg algebra consists of the center together with a maximal subspace of $\mathfrak g_1$ on which $\omega_\theta$ vanishes, and for a polarization spanned by root vectors this amounts to choosing exactly one root from each complementary pair.
If $L$ denotes such a choice, then
\begin{equation}
    J_\theta(L)=\{\theta\}\cup L
\end{equation}
corresponds to a commuting subset.
Indeed, the only possible non-zero brackets among the root vectors in $\mathfrak g_1$ come from choosing both members of one complementary pair, and this has been avoided, while $E_\theta$ commutes with all of $\mathfrak g_1$ by \eqref{eq:two-step}.
This is nothing but $J_{K}(L)$ of \eqref{eq:graded-support} at $k=2$: the complementary pairs are $\{e_1\pm e_l,\,e_2\mp e_l\}$ and, for $n$ odd, $\{e_1,e_2\}$, so that choosing one root from each pair gives two roots in every block $L_l$ and, for odd $n$, one root in $L_0$.
The cardinality of $J_\theta(L)$ is therefore $h-1$, as it must be by \eqref{eq:graded-excess}.
This is consistent with and related to the dimension $\dim\mathcal O_{\min}=2(h-1)$ of the minimal nilpotent orbit of $\mathfrak g$ \cite{WangMinimalNilpotentOrbit}.
Namely, the highest-root grading gives $\dim\mathcal O_{\min}=\dim\mathfrak g_1+2$, so that the $\dim\mathfrak g_1=2(h-2)$ roots of degree one form $h-2$ complementary pairs, and adding $\theta$ gives $|J_\theta(L)|=h-1$.

\section{Superconformal primaries from the gaugino bilinear}\label{app:clcl-monotone}

We derive the relation \eqref{eq:clcl-descendant} by acting with six commutant superconformal charges on $\mathcal S^4$ to obtain the CLCL primary of \cite{Chang:2025mqp,Choi:2025bhi}:
\begin{equation}\label{eq:app-clcl-descendant}
    \epsilon_{m_1n_1p_1}\epsilon_{m_2n_2p_2}
    \bigl(\bar S^{m_1}_{\dot\alpha}\bar S^{m_2\dot\alpha}
    \bar S^{n_1}_{\dot\beta}\bar S^{n_2\dot\beta}
    \bar S^{p_1}_{\dot\gamma}\bar S^{p_2\dot\gamma}\bigr)(\mathcal{S}^4)
    \;\propto\; M_1-3M_2 \, .
\end{equation}
We also compute the analogous primaries corresponding to higher powers of $\mathcal S$.

We use the classical supercharge action on local operators at the origin.
The supercharges (anti-)commuting with our choice of the classical supercharge $Q_0$ in the main text\footnote{See Section \ref{sec:rev-class-coh} for our notation and conventions.} generate the gaugino from each scalar \cite{Grant:2008sk},
\begin{equation}\label{eq:SQlambda-1}
    \lambda_{\dot\beta}=-[\bar Q_{n\dot\beta},\phi^n]\,,
\end{equation}
with no sum over $n$.
Taking $n=m$, the super-Jacobi identity gives
\begin{equation}\label{eq:SQlambda-2}
    \{\bar S^{m\dot\alpha},\lambda_{\dot\beta}\}
    =-[\{\bar S^{m\dot\alpha},\bar Q_{m\dot\beta}\},\phi^m]
    +\{\bar Q_{m\dot\beta},[\bar S^{m\dot\alpha},\phi^m]\}\,.
\end{equation}
The second term vanishes because the scalar is a superconformal primary.
Normalizing the first anticommutator by $[\{\bar S^{m\dot\alpha},\bar Q_{m\dot\beta}\},\phi^m]=\delta^{\dot\alpha}_{\dot\beta}\phi^m$, we obtain
\begin{equation}\label{eq:S-action-lambda}
    \{\bar S^{m\dot\alpha},\lambda_{\dot\beta}\}
    =-\delta^{\dot\alpha}_{\dot\beta}\phi^m\,,
    \qquad
    [\bar S^{m\dot\alpha},\phi^n]=0\,.
\end{equation}
The action of the $\bar{S}^{m\dot\alpha}$ satisfies a graded Leibniz rule and is component by component in the gauge index, so it simply replaces letters inside traces, without splitting or joining those traces.
Applying all six charges exhausts the non-trivial superconformal action on these letters, so the resulting operator, when non-zero, is a classical superconformal primary.

For $\mathcal S^4$, the six charges act on six distinct gauginos, since they annihilate the scalars.
The two untouched gauginos either lie in the same trace or in two different traces, so only two structures can appear:
\begin{equation}\label{eq:M1M2}
\begin{aligned}
    M_1&=\epsilon_{m_1n_1p_1}\epsilon_{m_2n_2p_2}\,
    u^{m_1m_2}u^{n_1n_2}u^{p_1p_2}\,
    \Tr(\lambda_{\dot\alpha}\lambda^{\dot\alpha})\,,\\
    M_2&=\epsilon_{m_1n_1p_1}\epsilon_{m_2n_2p_2}\,
    \Tr(\phi^{m_1}\lambda_{\dot\alpha})
    \Tr(\phi^{m_2}\lambda^{\dot\alpha})\,
    u^{n_1n_2}u^{p_1p_2}\,,
\end{aligned}
\end{equation}
where $u^{mn}=\Tr(\phi^m\phi^n)$, following the conventions of \cite{Choi:2025bhi}.

We note that the componentwise action \eqref{eq:S-action-lambda} commutes with restricting all letters to the same Cartan subalgebra.
Since $\mathcal S^4$ vanishes on the Cartan of $\mathfrak{so}_7$, the resulting primary must vanish there as well.
One easily checks that on the Cartan,\footnote{For an explicit representation, see Appendix \ref{app:son}.} $M_1=3M_2$, with $M_{1,2}$ non-vanishing, so the result has to be proportional to $M_1-3M_2$.
We confirm this relative coefficient by explicit action with the $\bar{S}^{m\dot\alpha}$ below.
For now, we note that since all classical commutant transformations preserve Cartan-valued fields, the non-Coulomb property holds for the entire superconformal multiplet.

To identify the resulting class with the CLCL monotone, note that the basis of \cite{Choi:2025bhi} contains a third structure,
\begin{equation}\label{eq:M3}
    M_3=\epsilon_{m_1n_1p_1}\epsilon_{m_2n_2p_2}\,
    w^{m_1m_2}u^{n_1n_2}u^{p_1p_2}\,,
    \qquad
    w^{mn}=\Tr\bigl(\lambda_{\dot\alpha}\lambda^{\dot\alpha}
    \phi^{(m}\phi^{n)}\bigr)\,.
\end{equation}
Its length-four trace cannot be produced by the six charges acting on the four length-two traces of $\mathcal S^4$.
On the $\mathfrak{so}_7$ Cartan, the three structures obey $M_1=3M_2=6M_3$, so their non-Coulomb combinations form a two-dimensional space.
Within this space, \cite{Choi:2025bhi} identifies $M_1+2M_2-10M_3$ as $Q_0$-exact and represents the non-trivial monotone by $\mathcal O_{\mathrm m}=M_1-M_2-4M_3$.
Since
\begin{equation}\label{eq:M3-free}
    M_1-3M_2
    =\tfrac53\mathcal O_{\mathrm m}
    -\tfrac23\bigl(M_1+2M_2-10M_3\bigr)\,,
\end{equation}
our construction gives the CLCL class (up to normalization), establishing \eqref{eq:app-clcl-descendant}.
The non-zero $M_3$ coefficient of the exact direction makes this the unique $M_3$-free representative of the resulting class within the $M_{1,2,3}$ space.

\subsubsection*{Higher powers of \texorpdfstring{$\mathcal{S}$}{S}}

For $\mathcal S^5$ of $\mathfrak{so}_8$, the six superconformal charges leave four gauginos.
These can lie in two bilinears, in one bilinear and two mixed traces, or in four mixed traces.
The corresponding structures are then
\begin{equation}\label{eq:W123}
\begin{gathered}
    W_1=M_1\,\Tr(\lambda_{\dot\beta}\lambda^{\dot\beta})\,,
    \qquad
    W_2=M_2\,\Tr(\lambda_{\dot\beta}\lambda^{\dot\beta})\,,\\[4pt]
    W_3=\epsilon_{m_1n_1p_1}\epsilon_{m_2n_2p_2}\,
    u^{m_1m_2}\,
    \Tr(\phi^{n_1}\lambda_{\dot\beta})
    \Tr(\phi^{n_2}\lambda^{\dot\beta})\,
    \Tr(\phi^{p_1}\lambda_{\dot\gamma})
    \Tr(\phi^{p_2}\lambda^{\dot\gamma})\,,
\end{gathered}
\end{equation}
where $M_{1,2}$ are the structures from before.
On the rank-four Cartan, the three structures obey $W_1=6W_2-4W_3$ so the unique combination that vanishes on the Cartan, up to normalization, is
\begin{equation}\label{eq:Sbar6ons5}
    W_1-6W_2+4W_3\,.
\end{equation}
Below we verify this by acting with the $\bar{S}^{m\dot\alpha}$ on $\mathcal S^5$.
For higher powers, just Cartan vanishing need not determine a unique combination.
For example, for $\mathcal S^6$, the four analogous structures instead obey two independent relations on the rank-four Cartan of $\mathfrak{so}_9$, and the primary can only be determined through explicit action with the $\bar{S}^{m\dot\alpha}$.

To this end, we introduce six Grassmann parameters $\xi_{n\dot\alpha}$.
The six-charge product in \eqref{eq:app-clcl-descendant} is a non-zero multiple of the coefficient of the ordered product of all six parameters in $\exp(\xi_{n\dot\alpha}\bar S^{n\dot\alpha})$.
The exponential acts on each gaugino as $\lambda_{\dot\alpha}\mapsto\lambda_{\dot\alpha}-\xi_{n\dot\alpha}\phi^n$, with no further terms because the charges annihilate the scalars.
Since the charges satisfy a graded Leibniz rule, their exponential acts on a product by making this substitution in every gaugino factor.
We then collect the coefficient of the six Grassmann parameters.
For $m\geq3$, the resulting primary is proportional to
\begin{equation}\label{eq:partner-coefficient}
    [\xi^6]\left\{
    \Tr\Bigl[
        (\lambda_{\dot\alpha}-\xi_{n\dot\alpha}\phi^n)
        (\lambda^{\dot\alpha}-\xi_p{}^{\dot\alpha}\phi^p)
    \Bigr]\right\}^{m}\,,
\end{equation}
where $[\xi^6]$ extracts the coefficient of the ordered product of all six parameters.
We use the spinor-contracted gaugino bilinear for convenience, which differs from our definition of $\mathcal S$ by a factor of $2$.

Let $u$ denote the $3\times3$ scalar Gram matrix, with entries $u^{mn}=\Tr(\phi^m\phi^n)$ introduced above.
Assuming first that $\det u\neq0$, we separate the part of the gaugino orthogonal to the three scalars with respect to the trace form:
\begin{equation}\label{eq:lambda-perp}
    \lambda^\perp_{\dot\alpha}
    =\lambda_{\dot\alpha}
    -(u^{-1})_{np}\Tr(\phi^p\lambda_{\dot\alpha})\,\phi^n\,,
    \qquad
    \Tr(\phi^n\lambda^\perp_{\dot\alpha})=0\,.
\end{equation}
Shifting the Grassmann parameters by expressions independent of $\xi$ does not change their top coefficient.
We can therefore absorb the component of $\lambda_{\dot\alpha}$ along the scalars into $\xi_{n\dot\alpha}$, after which the bilinear in \eqref{eq:partner-coefficient} becomes
\begin{equation}
\Tr(\lambda^\perp_{\dot\alpha}\lambda^{\perp\dot\alpha})
    +u^{np}\xi_{n\dot\alpha}\xi_p{}^{\dot\alpha}\,.
\end{equation}
Exactly three factors of the second term supply the six parameters, with coefficient proportional to $\det u$.
Thus the primary obtained from $\mathcal S^m$ is proportional to
\begin{equation}\label{eq:partner-closed}
    \det(u)\,
    \Bigl[\Tr\bigl(\lambda^\perp_{\dot\alpha}
    \lambda^{\perp\dot\alpha}\bigr)\Bigr]^{m-3}\,,
    \qquad m\geq3\,.
\end{equation}
Although $\lambda^\perp$ uses $u^{-1}$, the full expression equals the polynomial \eqref{eq:partner-coefficient} up to a non-zero numerical factor.
Its apparent denominators therefore cancel, and this polynomial expression extends the result to $\det u=0$.

For $m=4$, \eqref{eq:partner-closed} can be seen to equal $\tfrac16(M_1-3M_2)$, and for $m=5$ it equals $\tfrac16(W_1-6W_2+4W_3)$.
The first expression is non-zero, as can be seen by taking the three scalars along three orthogonal generators of $\mathfrak{so}_7$ and the two gaugino components along a fourth.
Then both $\det u$ and the orthogonal gaugino bilinear are non-zero, confirming the non-zero coefficient in \eqref{eq:app-clcl-descendant}.

The same formula also explains the non-Coulomb character.
On a rank-$r$ Cartan with $r\geq3$, three generic scalars occupy three directions, leaving only $r-3$ directions for the orthogonal gauginos.
Each remaining direction supplies one pair of Grassmann variables, so their bilinear vanishes when raised to a power greater than $r-3$.
Thus \eqref{eq:partner-closed}, which contains $m-3$ such factors, vanishes on the Cartan for $m>r$, just as $\mathcal S^m$ does.

The resulting primaries are $Q_0$-closed because they contain only closed letters. Whenever $\mathcal S^m$ is non-exact, its primary is too, since the corresponding commutant supercharges recover the non-zero descendant $[\mathcal S^m]$, and these charges preserve $Q_0$-exactness. 
The quantum pairing of the higher-rank classes remains to be established, as outlined in Section~\ref{ssec:higher-rank-predictions}.

\section{Trace relations and the explicit quantum pairing}\label{app:trace-relations}

In this appendix, we construct the $\mathfrak{so}_7$ trace relation $I_{\mathrm f}$ used in Section~\ref{ssec:constructing-source}, give an explicit operator $F_4$ satisfying $Q_0F_4=-iI_{\mathrm f}$, and provide details of the $Q_1$ pairing of $[F_4]$ with $[\mathcal{S}^4]$, as established in Section~\ref{ssec:Q1-pairing}.

\subsection{Finding the \texorpdfstring{$\mathfrak{so}_7$}{so(7)} trace relations}\label{app:finding-so7-trace-identities}

We first explain how the trace relations are found numerically.
Choose multi-trace bases $\{u_A\}_{A=1}^{312}$ and $\{v_B\}_{B=1}^{85}$ in the $SU(2)_R$ Lorentz and $SU(3)_R$ R-symmetry singlet sector\footnote{See \eqref{eq:app-spin-singlet} and \eqref{eq:app-flavor-projections} for explicit singlet projectors.} of the source and target spaces of Section~\ref{ssec:constructing-source}, whose letter content we repeat here for convenience,
\begin{equation}\label{eq:app-mu4-letter-content}
\begin{aligned}
    \mathcal C_\infty^{\mathrm{src}}&:\quad \lambda^6\psi_m\phi^m\ \oplus\ D_{\dot\alpha}\bigl(\lambda^5\phi^1\phi^2\phi^3\bigr)\,,
    \\
    \mathcal C_\infty^{\mathrm{tar}}&:\quad \lambda^6\phi^1\phi^2\phi^3\,.
\end{aligned}
\end{equation}
The bases incorporate (graded) cyclicity and the orthogonal transpose relations \eqref{eq:so(n)-bil-form}, but not yet $\mathfrak{so}_7$ trace relations.
We find the latter by evaluating both bases on randomly sampled $\mathfrak{so}_7$ matrices, keeping track of the Grassmann nature of fermionic letters, and identifying candidate trace relations as the null spaces of the resulting evaluation matrices.

Let us first note that a singlet basis element contains in general many multi-trace terms with different letter contents.
For evaluation on $\mathfrak{so}_7$ matrices, we separate the fixed letter-content components $\nu$, writing $u_A=\sum_\nu u_{A,\nu}$ and similarly for the target.
Terms with different letter contents cannot cancel in a trace relation, so a relation among the singlets is equivalent to
\begin{equation}
    \sum_A z_Au_A=0
    \quad\Longleftrightarrow\quad
    \sum_A z_Au_{A,\nu}=0
    \quad\text{for every }\nu\,.
\end{equation}
The evaluation matrix constructed below collects the conditions for each $\nu$ separately, so that its null space consists of the coefficient vectors $(z_A)$ that satisfy these conditions simultaneously.

For each fixed letter content $\nu$, bosonic letters are evaluated on independent random elements of $\mathfrak{so}_7$.
A fermionic letter $\chi$ occurring $r$ times in $\nu$ is instead written as
\begin{equation}\label{eq:app-grassmann-evaluation}
    \chi=\sum_{a=1}^{r}\chi^a M_a\,,
\end{equation}
with $M_a$ independent random elements of $\mathfrak{so}_7$ and $\chi^a$ independent Grassmann variables, with a separate set for each fermionic letter.
We extract the coefficient of the ordered product of all Grassmann variables introduced for this letter content.
Using the same sampled assignment for fixed $\nu$, we place the resulting coefficient of $u_{A,\nu}$ in column $A$ of one row of the evaluation matrix $E_{\mathrm{src}}$, and assemble the entire row.
If $u_A$ has no component with content $\nu$, the entry is $0$.
Further samples and different $\nu$ provide additional rows.
The target matrix $E_{\mathrm{tar}}$ is constructed in the same way.

Rows are accumulated until the observed rank stabilizes.\footnote{The row reduction is carried out over the finite field with prime $p=1{,}000{,}003$.
Choosing a prime $p$ that divides none of the denominators of the evaluation-matrix entries and clearing all denominators, a non-zero minor modulo $p$ implies a non-zero minor over $\mathbb{Q}$.
The modulo $p$ rank therefore gives lower bounds on the exact rank.}
A source $\sum_A z_Au_A$ passes every sampled test precisely when $E_{\mathrm{src}}z=0$, and similarly for the target.
We therefore compute the two null spaces, whose elements give candidate trace relations.
These candidates are then tested on further independent sampled assignments.
We find
\begin{equation}\label{eq:app-evaluation-ranks}
    \operatorname{rank}E_{\mathrm{src}}=309\,,
    \qquad
    \operatorname{rank}E_{\mathrm{tar}}=81\,,
\end{equation}
so that the numerical null spaces have dimensions $312-309=3$ and $85-81=4$.
The numerical ranks give upper bounds on the dimensions of the genuine trace-relation spaces:
\begin{equation}\label{eq:app-relation-upper-bound}
    \dim\mathcal I_7^{\mathrm{src}}\leq3\,,
    \qquad
    \dim\mathcal I_7^{\mathrm{tar}}\leq4\,.
\end{equation}
The exact trace relations constructed in the next subsection give three independent source relations and four independent target relations.
These relations supply lower bounds on the dimensions, so that the spaces have dimensions $3$ and $4$ exactly, and the explicit relations below span them.

\subsection{Antisymmetrizer expressions for the relation}\label{app:antisymmetrizer-ids}

We now construct the three source and four target trace relations explicitly.
The vanishing of the relations will follow from the antisymmetrization of eight vector indices in seven dimensions.
We use the standard antisymmetrization relations of orthogonal invariant theory, as reviewed in \cite[Section~3.1]{LehrerZhang2012}.
We confirm the independence of the resulting relations by comparing their coefficients in the formal multi-trace bases $\{u_A\}_{A=1}^{312}$ and $\{v_B\}_{B=1}^{85}$, respectively.
As such, they prove the saturation of the bounds in Appendix~\ref{app:finding-so7-trace-identities}.

We work in a basis of the seven-dimensional defining representation of $\mathfrak{so}_7$ in which the preserved symmetric bilinear form is $\delta_{ij}$, so adjoint-valued matrices obey $X^T=-X$.
We use the supercharge conventions of Section~\ref{ssec:constructing-source} and write
\begin{equation}\label{eq:app-normalized-letters}
    \lambda_{ab}\equiv D_{(a}\lambda_{b)}=\tfrac12\bigl(D_a\lambda_b+D_b\lambda_a\bigr)\,,\qquad a,b\in\{+,-\}\,.
\end{equation}
The relevant $Q_0$ actions are
\begin{equation}\label{eq:app-Q0-letters}
\begin{aligned}
    \{Q_0,\psi_m\}&=-i\epsilon_{mnp}[\phi^n,\phi^p]\,,\\
    [Q_0,D_a\phi^m]&=-i[\lambda_a,\phi^m]\,,\qquad
    \{Q_0,\lambda_{ab}\}=-i\{\lambda_a,\lambda_b\}\,.
\end{aligned}
\end{equation}
The letters $\lambda_a$ and $\phi^m$ are $Q_0$-closed, and $Q_0$ acts on products by the graded Leibniz rule.

Define the unnormalized eight-index antisymmetrizer by
\begin{equation}
    \Delta^{i_1\cdots i_8}_{j_1\cdots j_8}\equiv\sum_{\sigma\in S_8}\operatorname{sgn}(\sigma)\delta^{i_1}_{j_{\sigma(1)}}\cdots\delta^{i_8}_{j_{\sigma(8)}}\,.
\end{equation}
It is identically zero when the vector indices take seven values.
For matrices $A_1,\ldots,A_8$, not necessarily antisymmetric, define
\begin{equation}\label{eq:app-B-contraction}
\begin{aligned}
    \mathcal B(A_1,\ldots,A_8)\equiv{}&\Delta^{i_1\cdots i_8}_{j_1\cdots j_8}(A_1)_{i_1i_2}(A_2)_{i_3i_4}(A_3)_{j_1j_2}(A_4)_{j_3j_4}\\
    &\times(A_5)_{i_5j_5}(A_6)_{i_6j_6}(A_7)_{i_7j_7}(A_8)_{i_8j_8}\,.
\end{aligned}
\end{equation}
The arguments of $\mathcal B$ fall into three groups according to their index contractions: $A_1,A_2$ each carry two $i$ indices, $A_3,A_4$ each carry two $j$ indices, and $A_5,\ldots,A_8$ each carry one index of each kind.
This particular contraction pattern turns out to be a convenient choice.
The matrix entries are to be multiplied in the displayed order to keep track of Grassmann signs for fermionic matrices.
Expanding $\Delta$ contracts the matrix indices into products of traces.
Any transposes introduced by the index contractions are reduced using the antisymmetry of the constituent letters.
The vanishing of $\mathcal B$ at matrix size seven follows from $\Delta^{i_1\cdots i_8}_{j_1\cdots j_8}=0$.

To keep the expressions compact, we define singlet projectors for the global $SU(2)_R$ right-handed Lorentz symmetry and $SU(3)_R$ R-symmetry.
For $SU(2)_R$, let $\mathsf L_+$ and $\mathsf L_-$ raise and lower spinor indices, with
\begin{equation}
    \mathsf L_+\lambda_-=\lambda_+\,,\qquad \mathsf L_-\lambda_+=\lambda_-\,,\qquad
    \mathsf L_+\lambda_+=\mathsf L_-\lambda_-=0\,.
\end{equation}
They act in the same way on each dotted derivative index, annihilate $\phi^m$ and $\psi_m$, and act on products by the Leibniz rule.
The expressions to be projected below have six spinor indices and weight zero, \textit{i.e.}, an equal number of $+$ and $-$ indices.
The Lorentz-singlet projector is then given by
\begin{equation}\label{eq:app-spin-singlet}
    \mathscr P_0\equiv 1-\tfrac12\mathsf L_-\mathsf L_++\tfrac1{12}\mathsf L_-^2\mathsf L_+^2-\tfrac1{144}\mathsf L_-^3\mathsf L_+^3\,.
\end{equation}
It acts as the identity on spin zero and vanishes on the weight-zero states of spins one, two and three.

For the $SU(3)_R$ singlet projection, we antisymmetrize the three scalar labels and sum over the scalar--chiralino pair:\footnote{Note that these are the only letter combinations appearing in our bases \eqref{eq:app-mu4-letter-content}.}
\begin{equation}\label{eq:app-flavor-projections}
\begin{aligned}
    \mathscr A_\epsilon X(\phi^1,\phi^2,\phi^3)&=\tfrac1{6}\sum_{\sigma\in S_3}\operatorname{sgn}(\sigma)X(\phi^{\sigma(1)},\phi^{\sigma(2)},\phi^{\sigma(3)})\,,\\
    \mathscr A_\delta X(\psi_1,\phi^1)&=\tfrac1{3}\sum_{m=1}^3X(\psi_m,\phi^m)\,.
\end{aligned}
\end{equation}
These operations relabel letters without changing their positions in a trace.
Note that the supercharge $Q_0$ respects these singlet conditions.

To construct explicit representatives of the numerical relations, we assign the required letters to the slots of \eqref{eq:app-B-contraction}.
For an eight-letter source, each letter occupies one slot, while for a nine-letter target one slot contains a product of two letters.
For each assignment, we expand $\Delta$, apply the singlet operations, and collect the coefficients of the canonical free-trace words, keeping track of transpose and Grassmann signs.
These formal coefficient vectors are then compared with the numerical relation spaces, and we keep searching for independent expressions until we span them.
Note that the resulting expressions are not unique, but simply form a basis for the spaces.
We also note that the symmetries of $\mathcal{B}$ constrain the non-trivial letter-to-slot assignments.
Indeed, since $\mathcal B$ is symmetric under exchanges of bosonic arguments within each of the groups $\{1,2\}$, $\{3,4\}$ and $\{5,6,7,8\}$, placing two (undifferentiated) scalar arguments in the same group makes the flavor singlet projection vanish.

For the three source letter families in \eqref{eq:app-mu4-letter-content} containing $\psi$, $D\phi$ and $D\lambda$, respectively, the search gives
\begin{equation}\label{eq:app-source-identities}
\begin{aligned}
    J_\psi&\equiv \mathscr P_0\mathscr A_\delta\mathcal B\bigl(\lambda_+,\lambda_-,\lambda_+,\lambda_-,\lambda_+,\lambda_-,\psi_1,\phi^1\bigr)\,,\\
    J_{D\phi}&\equiv \mathscr P_0\mathscr A_\epsilon\mathcal B\bigl(\lambda_+,\lambda_-,\lambda_-,\phi^1,\lambda_-,\lambda_+,\phi^2,D_+\phi^3\bigr)\,,\\
    J_{D\lambda}&\equiv \mathscr P_0\mathscr A_\epsilon\mathcal B\bigl(\lambda_+,\phi^1,\lambda_-,\phi^2,\lambda_+,\lambda_-,\lambda_{+-},\phi^3\bigr)\,.
\end{aligned}
\end{equation}
Each expression vanishes for $\mathfrak{so}_7$ by \eqref{eq:app-B-contraction}.
However, their free-trace expansions are non-zero and have disjoint letter contents, so they are linearly independent in the source space.
In addition, these expressions are chosen such that their $Q_0$ images are linearly independent in the target space as well.
A fourth, independent target identity can be written as
\begin{equation}\label{eq:app-exceptional-relation}
    I_{\mathrm f}\equiv -3\mathscr P_0\mathscr A_\epsilon\mathcal B\bigl(\lambda_+,\lambda_-,\lambda_-,\phi^1,\lambda_+,\lambda_-,\lambda_+\phi^2,\phi^3\bigr)\,.
\end{equation}
Note that the seventh argument contains two letters, so that this is indeed a nine-letter target singlet, and the prefactor is chosen in order to match the normalization of $F_4$, to be constructed in Appendix \ref{app:constr-pair-F4}.
Again, $I_{\mathrm f}$ is guaranteed to vanish on $\mathfrak{so}_7$ matrices by \eqref{eq:app-B-contraction}.

We have thus found that
\begin{equation}\label{eq:app-analytic-identity-ranks}
\begin{aligned}
    \dim\operatorname{span}\{J_\psi,J_{D\phi},J_{D\lambda}\}&=3\,,\\
    \dim\operatorname{span}\{Q_0J_\psi,Q_0J_{D\phi},Q_0J_{D\lambda},I_{\mathrm f}\}&=4\,.
\end{aligned}
\end{equation}
Combined with the numerical upper bounds, these dimensions prove that the three $J$'s form a basis of the source identity space, while their three $Q_0$ images together with $I_{\mathrm f}$ form a basis of the target identity space.
In particular, the three independent images $Q_0J_\psi$, $Q_0J_{D\phi}$ and $Q_0J_{D\lambda}$ span $Q_0\mathcal I_7^{\mathrm{src}}$, while $I_{\mathrm f}$ supplies the remaining target direction.

Since the target expressions saturate the letter bound, they are $Q_0$-closed, and the quotient $\mathcal I_7^{\mathrm{tar}}/Q_0\mathcal I_7^{\mathrm{src}}$ is one-dimensional, generated by $[I_{\mathrm f}]$. Surjectivity of $D$ supplies a source satisfying $Q_0F_4=-iI_{\mathrm f}$. A change of relation representative is compensated by
\begin{equation}\label{eq:app-relation-representative-freedom}
    I_{\mathrm f}'=I_{\mathrm f}+Q_0J,\qquad
    F_4'=F_4-iJ,\qquad J\in\mathcal I_7^{\mathrm{src}}.
\end{equation}
This leaves the evaluated $\mathfrak{so}_7$ operator unchanged. An independently chosen preimage can additionally differ by a monotone class. As shown in Section~\ref{ssec:constructing-source}, the non-zero quotient class ensures that $[F_4]$ is non-exact and fortuitous.

\subsection{Constructing and pairing \texorpdfstring{$F_4$}{F4}}\label{app:constr-pair-F4}

In this subsection, we work out an explicit expression for $F_4$ in the form described schematically in Section~\ref{ssec:Q1-pairing}. 
We then act with $Q_1$ to establish its pairing with $[\mathcal{S}^4]$, and check the preimage independence of this pairing.
Throughout, $I_{\mathrm f}$ denotes the fixed relation \eqref{eq:app-exceptional-relation}.

\subsubsection*{Derivative completion for \texorpdfstring{$F_4$}{F4}}

Recall from Section~\ref{ssec:Q1-pairing} the splitting $\mathcal C_\infty^{\mathrm{src}}=\mathcal C_\psi\oplus\mathcal C_\partial$ and the corresponding blocks $D_\psi,D_\partial$ of $Q_0$.
For a chosen $F_\psi$, the derivative completion is found by solving
\begin{equation}\label{eq:app-closure-split}
    D_\partial F_\partial=-iI_{\mathrm f}-D_\psi F_\psi\,.
\end{equation}
We tested singlet-projected multi-trace candidates for $F_\psi$, solving this linear equation for their overall coefficient and $F_\partial$.
An admissible choice is
\begin{equation}\label{eq:app-short-Fpsi}
    F_{4,\psi}=6\,\mathscr P_0\mathscr A_\delta\Bigl[\mathcal S^2\Tr(\psi_1\lambda_-\phi^1\lambda_+)\Bigr]\,.
\end{equation}
Since $\mathcal S^2$ is already a Lorentz singlet, the projections give
\begin{equation}\label{eq:app-Fpsi-expanded}
    F_{4,\psi}=\mathcal S^2\mathcal O_\psi\,,\qquad
    \mathcal O_\psi=\sum_{m=1}^3\Tr\bigl[\psi_m\bigl(\lambda_-\phi^m\lambda_+-\lambda_+\phi^m\lambda_-\bigr)\bigr]\,.
\end{equation}
A corresponding solution for the derivative part is
\begin{equation}\label{eq:app-explicit-F4}
    F_{4,\partial}=\mathscr P_0\mathscr A_\epsilon\sum_{j=1}^{60}c_jW_j\,,
\end{equation}
with the coefficients and words given in Table~\ref{tab:F4-derivative-seeds}.
Expanding the singlet projections and comparing every canonical multi-trace coefficient verifies
\begin{equation}\label{eq:app-explicit-F4-closure}
    Q_0F_4=Q_0F_{4,\psi}+Q_0F_{4,\partial}=-iI_{\mathrm f}\,,
\end{equation}
as an identity in the free-trace algebra.
Thus $F_4$ is $Q_0$-closed upon evaluation on $\mathfrak{so}_7$, although neither part is separately closed there.

\subsubsection*{\texorpdfstring{$Q_1$}{Q1} contraction}

Only the $\psi\phi$ pair contributes to the $Q_1$ action on this source \cite{Budzik:2025zvu}:
\begin{equation}\label{eq:app-Q1-psi-phi}
    Q_1\bigl((\psi_m)^a(\phi^n)^b\bigr)=-\kappa_Q\,\delta_m^{\,n}\,f^{a}{}_{cd}f^{bc}{}_{e}(\lambda_{\dot\alpha})^d(\lambda^{\dot\alpha})^e\,.
\end{equation}
Here we use the conventions of Section~\ref{ssec:gaugino-chiral-ring} and Appendix~\ref{app:cw-basis}: $\langle X,Y\rangle=\tfrac12\Tr(XY)$ for $\mathfrak{so}_7$, $f_{abc}=\langle[T_a,T_b],T_c\rangle$, and color indices are raised with the inverse of $g_{ab}=\langle T_a,T_b\rangle$.
Our spinor convention is $\lambda^+=\lambda_-$ and $\lambda^-=-\lambda_+$.
Since $Q_1$ annihilates $\mathcal C_\partial$ and no pair involving a gaugino contributes, the calculation reduces to
\begin{equation}\label{eq:app-quantum-short-source}
    Q_1F_4=\mathcal S^2Q_1\mathcal O_\psi\,.
\end{equation}
All color contractions in the $Q_1$ action have to be performed in the full Lie algebra before restricting the output gauginos to the four-root support $J=\{\alpha_1,\ldots,\alpha_4\}$ of \eqref{eq:so7-support}:
\begin{equation}\label{eq:app-output-support}
    \rho_J(\lambda_+)=\sum_{r=1}^4x_rE_{\alpha_r}\,,\qquad
    \rho_J(\lambda_-)=\sum_{r=1}^4y_rE_{-\alpha_r}\,.
\end{equation}
We order the roots as in that equation and normalize them by $\Tr(E_{\alpha_r}E_{-\alpha_s})=2\delta_{rs}$.
Extracting the coefficient of $\Omega=x_1x_2x_3x_4y_1y_2y_3y_4$ gives the values in \eqref{eq:app-direct-quantum-coefficients}, with ratio $-12\kappa_Q$.
Section~\ref{ssec:Q1-pairing} explains why this restricted coefficient determines the full cohomological pairing.

\subsubsection*{Preimage independence}

For fixed $I_{\mathrm f}$, any other preimage has the form $F_4'=F_4+G$ with $G\in\ker D$.
Since $G$ is closed in the free-trace algebra, any non-zero $Q_0$-cohomology class that it represents after evaluation on $\mathfrak{so}_7$ is monotone.
Different preimages can therefore represent distinct $Q_0$ cohomology classes, related by addition of a monotone class.
We now check that this freedom does not change their $Q_1$ image.

To this end, we compute the row $q$ of \eqref{eq:Q1-source-row} on the full $\mathcal C^{\mathrm{src}}_\infty$ basis.
Let $D^+$ be any right inverse of $D$, so that $DD^+=1$.
Then $1-D^+D$ projects onto $\ker D$, and explicit computation gives
\begin{equation}\label{eq:app-preimage-independence}
    q(1-D^+D)=0\,.
\end{equation}
Thus $q$ vanishes on every freely closed source, independently of the choice of right inverse.
Since $q$ detects the coefficient of the unique target class $[\mathcal S^4]$, this establishes
\begin{equation}
    [Q_1][G]=0\,,
    \qquad
    [Q_1][F_4+G]=[Q_1][F_4]\,.
\end{equation}
The pairing is consequently independent of the preimage, although its source is not a uniquely specified $Q_0$-cohomology class.

Furthermore, as already discussed around \eqref{eq:app-relation-representative-freedom}, changing the representative of the trace-relation class shifts $F_4$ by a source trace relation.
Unlike a monotone addition, this shift leaves the evaluated operator itself unchanged, and hence also leaves its $Q_1$ pairing unchanged.

\subsubsection*{An alternative choice of $F_\psi$}

Changing the representative of the trace-relation class can allow a simpler $F_\psi$.
For example, set $U\equiv\mathcal S^3\sum_m\Tr(\psi_m\phi^m)$ and solve
\begin{equation}\label{eq:app-alternative-completion}
    aD_\psi U+D_\partial V=-iI_{\mathrm f}+\sum_{t\in\{\psi,D\phi,D\lambda\}}b_tQ_0J_t
\end{equation}
for $a$, $V\in\mathcal C_\partial$ and the coefficients $b_t$.
An exact solution exists with $a=-1/5$, $b_\psi=3/20$ and $b_{D\phi}=b_{D\lambda}=0$.
It gives $F_4'=-U/5+V$ and $Q_0F_4'=-iI_{\mathrm f}'$, where
\begin{equation}\label{eq:app-alternative-antisymmetrizer}
\begin{aligned}
    I_{\mathrm f}'&=I_{\mathrm f}+\tfrac{3i}{20}Q_0J_\psi\\
    &=I_{\mathrm f}+\tfrac35\mathscr P_0\mathscr A_\epsilon\mathcal B\bigl(\lambda_+,\lambda_-,\lambda_+,\lambda_-,\lambda_+,\lambda_-,\phi^1\phi^2,\phi^3\bigr)\,,
\end{aligned}
\end{equation}
where the second line follows by acting on the only non-closed slot of $J_\psi$.

From \eqref{eq:app-relation-representative-freedom}, it is clear that the shifted relation $I_{\mathrm f}'$ admits the preimage
\begin{equation}
    \widetilde F_4=F_4+\tfrac3{20}J_\psi\,.
\end{equation}
Since $J_\psi$ vanishes on $\mathfrak{so}_7$, $\widetilde F_4$ and $F_4$ define the same evaluated operator.
However, the source $F_4'$ constructed above was obtained by independently solving the completion equation with the prescribed simpler $\psi\phi$ part.
It has the same classical image as $\widetilde F_4$, so
\begin{equation}
    F_4'=\widetilde F_4+G\,,
    \qquad Q_0G=0
\end{equation}
in the free-trace algebra.
Thus $[F_4']-[F_4]=[G]$, which is either zero or a monotone class.
The calculation \eqref{eq:app-preimage-independence} shows that $[Q_1][G]=0$, and hence
\begin{equation}
    [Q_1][F_4']=[Q_1][F_4]=-12\kappa_Q[\mathcal S^4]\,.
\end{equation}
Equality of the source classes would additionally require $G$ to be $Q_0$-exact at $\mathfrak{so}_7$.

\begin{longtable}{r r l}
\caption{Coefficients $c_j$ and words $W_j$ of $F_{4,\partial}$ in \eqref{eq:app-explicit-F4}}\label{tab:F4-derivative-seeds}\\
$j$&$c_j$&$W_j$\\ \hline
\endfirsthead
$j$&$c_j$&$W_j$\\ \hline
\endhead
1&$-48$&$\Tr\bigl(\phi^1\phi^2\phi^3\lambda_-\lambda_+\bigr)\Tr\bigl(\lambda_-\lambda_-\lambda_{++}\bigr)$\\
2&$288$&$\Tr\bigl(\phi^1\phi^2\lambda_-\bigr)\Tr\bigl(\phi^3\lambda_-\lambda_+\bigr)\Tr\bigl(\lambda_-\lambda_{++}\bigr)$\\
3&$-336$&$\Tr\bigl(\phi^1\phi^2\lambda_-\bigr)\Tr\bigl(\phi^3\lambda_-\lambda_+\lambda_{+-}\lambda_+\bigr)$\\
4&$528$&$\Tr\bigl(\phi^1\phi^2\lambda_-\bigr)\Tr\bigl(\phi^3\lambda_-\lambda_{+-}\lambda_+\lambda_+\bigr)$\\
5&$-60$&$\Tr\bigl(\phi^1\phi^2\lambda_-\bigr)\Tr\bigl(\phi^3\lambda_-\lambda_{++}\bigr)\Tr\bigl(\lambda_-\lambda_+\bigr)$\\
6&$-288$&$\Tr\bigl(\phi^1\phi^2\lambda_-\bigr)\Tr\bigl(\phi^3\lambda_+\lambda_-\lambda_+\lambda_{+-}\bigr)$\\
7&$-48$&$\Tr\bigl(\phi^1\phi^2\lambda_-\phi^3\lambda_-\lambda_{++}\bigr)\Tr\bigl(\lambda_-\lambda_+\bigr)$\\
8&$72$&$\Tr\bigl(\phi^1\phi^2\lambda_-\phi^3\lambda_+\lambda_-\lambda_{++}\lambda_-\bigr)$\\
9&$24$&$\Tr\bigl(\phi^1\phi^2\lambda_-\phi^3\lambda_+\lambda_+\lambda_{--}\lambda_+\bigr)$\\
10&$-96$&$\Tr\bigl(\phi^1\phi^2\lambda_-\phi^3\lambda_+\lambda_{+-}\lambda_-\lambda_+\bigr)$\\
11&$96$&$\Tr\bigl(\phi^1\phi^2\lambda_-\phi^3\lambda_{+-}\lambda_+\lambda_+\lambda_-\bigr)$\\
12&$-18$&$\Tr\bigl(\phi^1\phi^2\lambda_-\lambda_-\bigr)\Tr\bigl(\phi^3\lambda_{++}\bigr)\Tr\bigl(\lambda_-\lambda_+\bigr)$\\
13&$360$&$\Tr\bigl(\phi^1\phi^2\lambda_-\lambda_-\phi^3\lambda_-\lambda_+\lambda_{++}\bigr)$\\
14&$672$&$\Tr\bigl(\phi^1\phi^2\lambda_-\lambda_-\lambda_+\phi^3\lambda_+\lambda_{+-}\bigr)$\\
15&$48$&$\Tr\bigl(\phi^1\phi^2\lambda_-\lambda_-\lambda_+\phi^3\lambda_{+-}\lambda_+\bigr)$\\
16&$-36$&$\Tr\bigl(\phi^1\phi^2\lambda_-\lambda_-\lambda_+\lambda_-\bigr)\Tr\bigl(\lambda_+D_+\phi^3\bigr)$\\
17&$-72$&$\Tr\bigl(\phi^1\phi^2\lambda_-\lambda_-D_+\phi^3\lambda_+\bigr)\Tr\bigl(\lambda_-\lambda_+\bigr)$\\
18&$-72$&$\Tr\bigl(\phi^1\phi^2\lambda_-\lambda_-\lambda_{+-}\lambda_+\bigr)\Tr\bigl(\phi^3\lambda_+\bigr)$\\
19&$720$&$\Tr\bigl(\phi^1\phi^2\lambda_-\lambda_+\bigr)\Tr\bigl(\phi^3\lambda_-\lambda_+\lambda_{+-}\bigr)$\\
20&$-288$&$\Tr\bigl(\phi^1\phi^2\lambda_-\lambda_+\phi^3\lambda_-\lambda_+\lambda_{+-}\bigr)$\\
21&$-1008$&$\Tr\bigl(\phi^1\phi^2\lambda_-\lambda_+\phi^3\lambda_-\lambda_{+-}\lambda_+\bigr)$\\
22&$288$&$\Tr\bigl(\phi^1\phi^2\lambda_-\lambda_+\lambda_-\lambda_+D_+\phi^3\lambda_-\bigr)$\\
23&$-624$&$\Tr\bigl(\phi^1\phi^2\lambda_-\lambda_+\lambda_+\phi^3\lambda_{+-}\lambda_-\bigr)$\\
24&$144$&$\Tr\bigl(\phi^1\phi^2\lambda_-\lambda_+\lambda_+\lambda_-\phi^3\lambda_{+-}\bigr)$\\
25&$-324$&$\Tr\bigl(\phi^1\phi^2\lambda_-\lambda_+\lambda_{++}\lambda_-\bigr)\Tr\bigl(\phi^3\lambda_-\bigr)$\\
26&$-96$&$\Tr\bigl(\phi^1\phi^2\lambda_-D_+\phi^3\bigr)\Tr\bigl(\lambda_-\lambda_-\lambda_+\lambda_+\bigr)$\\
27&$-48$&$\Tr\bigl(\phi^1\phi^2\lambda_-\lambda_{+-}\bigr)\Tr\bigl(\phi^3\lambda_-\lambda_+\lambda_+\bigr)$\\
28&$120$&$\Tr\bigl(\phi^1\phi^2\lambda_{--}\bigr)\Tr\bigl(\phi^3\lambda_-\bigr)\Tr\bigl(\lambda_+\lambda_+\lambda_+\bigr)$\\
29&$288$&$\Tr\bigl(\phi^1\phi^2\lambda_{+-}\bigr)\Tr\bigl(\phi^3\lambda_-\lambda_-\lambda_+\lambda_+\bigr)$\\
30&$24$&$\Tr\bigl(\phi^1\lambda_-\bigr)\Tr\bigl(\phi^2\lambda_-\bigr)\Tr\bigl(\phi^3\lambda_-\bigr)\Tr\bigl(\lambda_+\lambda_{++}\bigr)$\\
31&$216$&$\Tr\bigl(\phi^1\lambda_-\bigr)\Tr\bigl(\phi^2\lambda_-\bigr)\Tr\bigl(\phi^3\lambda_-\lambda_{++}\lambda_+\bigr)$\\
32&$27$&$\Tr\bigl(\phi^1\lambda_-\bigr)\Tr\bigl(\phi^2\lambda_-\bigr)\Tr\bigl(\phi^3\lambda_{++}\bigr)\Tr\bigl(\lambda_-\lambda_+\bigr)$\\
33&$108$&$\Tr\bigl(\phi^1\lambda_-\bigr)\Tr\bigl(\phi^2\lambda_-\bigr)\Tr\bigl(\lambda_-\lambda_+\lambda_+D_+\phi^3\bigr)$\\
34&$-96$&$\Tr\bigl(\phi^1\lambda_-\bigr)\Tr\bigl(\phi^2\lambda_-\lambda_-\lambda_+\lambda_+D_+\phi^3\bigr)$\\
35&$-324$&$\Tr\bigl(\phi^1\lambda_-\bigr)\Tr\bigl(\phi^2\lambda_-\lambda_+\phi^3\lambda_-\lambda_{++}\bigr)$\\
36&$648$&$\Tr\bigl(\phi^1\lambda_-\bigr)\Tr\bigl(\phi^2\lambda_-\lambda_+\lambda_-D_+\phi^3\lambda_+\bigr)$\\
37&$336$&$\Tr\bigl(\phi^1\lambda_-\bigr)\Tr\bigl(\phi^2\lambda_-\lambda_+\lambda_+\lambda_-D_+\phi^3\bigr)$\\
38&$108$&$\Tr\bigl(\phi^1\lambda_-\bigr)\Tr\bigl(\phi^2\lambda_-D_+\phi^3\lambda_+\bigr)\Tr\bigl(\lambda_-\lambda_+\bigr)$\\
39&$-312$&$\Tr\bigl(\phi^1\lambda_-\bigr)\Tr\bigl(\phi^2D_-\phi^3\lambda_-\lambda_+\lambda_+\lambda_+\bigr)$\\
40&$204$&$\Tr\bigl(\phi^1\lambda_-\bigr)\Tr\bigl(\phi^2D_-\phi^3\lambda_+\lambda_+\bigr)\Tr\bigl(\lambda_-\lambda_+\bigr)$\\
41&$168$&$\Tr\bigl(\phi^1\lambda_-\bigr)\Tr\bigl(\phi^2\lambda_{--}\lambda_+\phi^3\lambda_+\lambda_+\bigr)$\\
42&$-528$&$\Tr\bigl(\phi^1\lambda_-\bigr)\Tr\bigl(\phi^2\lambda_+\lambda_-\lambda_+D_-\phi^3\lambda_+\bigr)$\\
43&$-54$&$\Tr\bigl(\phi^1\lambda_-\bigr)\Tr\bigl(\phi^2\lambda_+\lambda_+\bigr)\Tr\bigl(\lambda_-\lambda_-D_+\phi^3\bigr)$\\
44&$24$&$\Tr\bigl(\phi^1\lambda_-\bigr)\Tr\bigl(\phi^2D_+\phi^3\bigr)\Tr\bigl(\lambda_-\lambda_-\lambda_+\lambda_+\bigr)$\\
45&$-6$&$\Tr\bigl(\phi^1\lambda_-\bigr)\Tr\bigl(\phi^2D_+\phi^3\bigr)\Tr\bigl(\lambda_-\lambda_+\bigr)\Tr\bigl(\lambda_-\lambda_+\bigr)$\\
46&$-96$&$\Tr\bigl(\phi^1\lambda_-\phi^2\lambda_-D_-\phi^3\bigr)\Tr\bigl(\lambda_+\lambda_+\lambda_+\bigr)$\\
47&$-324$&$\Tr\bigl(\phi^1\lambda_-\phi^2\lambda_-\lambda_+\lambda_-\bigr)\Tr\bigl(\phi^3\lambda_{++}\bigr)$\\
48&$192$&$\Tr\bigl(\phi^1\lambda_-\phi^2\lambda_-D_+\phi^3\lambda_+\bigr)\Tr\bigl(\lambda_-\lambda_+\bigr)$\\
49&$432$&$\Tr\bigl(\phi^1\lambda_-\phi^2D_-\phi^3\lambda_-\lambda_+\lambda_+\lambda_+\bigr)$\\
50&$-432$&$\Tr\bigl(\phi^1\lambda_-\phi^2\lambda_+\phi^3\lambda_-\lambda_{+-}\lambda_+\bigr)$\\
51&$408$&$\Tr\bigl(\phi^1\lambda_-\phi^2\lambda_+\lambda_-D_+\phi^3\bigr)\Tr\bigl(\lambda_-\lambda_+\bigr)$\\
52&$-216$&$\Tr\bigl(\phi^1\lambda_-\phi^2\lambda_+\lambda_{--}\phi^3\lambda_+\lambda_+\bigr)$\\
53&$432$&$\Tr\bigl(\phi^1\lambda_-\phi^2\lambda_+\lambda_+\lambda_-\lambda_-D_+\phi^3\bigr)$\\
54&$-72$&$\Tr\bigl(\phi^1\lambda_-\lambda_-\bigr)\Tr\bigl(\phi^2\lambda_-\lambda_+\bigr)\Tr\bigl(\lambda_+D_+\phi^3\bigr)$\\
55&$648$&$\Tr\bigl(\phi^1\lambda_-\lambda_-\bigr)\Tr\bigl(\phi^2\lambda_-\lambda_+\lambda_+D_+\phi^3\bigr)$\\
56&$432$&$\Tr\bigl(\phi^1\lambda_-\lambda_-\bigr)\Tr\bigl(\phi^2\lambda_+D_-\phi^3\lambda_+\lambda_+\bigr)$\\
57&$-144$&$\Tr\bigl(\phi^1\lambda_-\lambda_-\phi^2\lambda_-\lambda_+\phi^3\lambda_{++}\bigr)$\\
58&$288$&$\Tr\bigl(\phi^1\lambda_-\lambda_-\phi^2\lambda_+\lambda_-\phi^3\lambda_{++}\bigr)$\\
59&$-1296$&$\Tr\bigl(\phi^1\lambda_-\lambda_+\phi^2D_-\phi^3\lambda_+\lambda_+\lambda_-\bigr)$\\
60&$-384$&$\Tr\bigl(\phi^1\lambda_-\lambda_+\lambda_-\lambda_-\lambda_+\phi^2D_+\phi^3\bigr)$\\
\end{longtable}

\end{appendix}

\begingroup
\interlinepenalty=10000
\bibliographystyle{JHEP}
\bibliography{refs}
\endgroup

\end{document}